\documentclass[a4paper,11pt]{article}
\pdfoutput=1 

\usepackage{jhepp} 
\usepackage[T1]{fontenc} 
\usepackage{pgfplots} 
\usepackage{tikz}
\usetikzlibrary{external}
\usepackage{capt-of}
 \usepackage{slashed}
 \usepackage{hhline}
 \usepackage{comment}
 \usepackage{framed}
 \usepackage{braket}
 \usepackage{subcaption}
\title{\boldmath Axion-like particles as mediators for dark matter: beyond freeze-out}
\newcommand{\mca}[1]{\mathcal{#1}}
\newcommand{\mrm}[1]{\mathrm{#1}}

\newcommand{\tr}{\operatorname{tr}}
\newcommand{\into}{\ensuremath{\,\rightarrow\,}}
\renewcommand{\d}{\operatorname{d}\!}
\newcommand{\hc}{+\,\mathrm{h.c.}}
\newcommand{\Tp}{T^\prime}

\newcommand{\vev}[1]{\langle #1 \rangle}

\newcommand{\U}[1]{\ensuremath{\mathrm{U}(#1)}}

\author[a]{ A. Bharucha,}
\author[b]{F. Br{\"u}mmer,}
\author[c]{N. Desai}
\author[a]{and S. Mutzel}

\affiliation[a]{Aix Marseille Univ, Universit\'{e} de Toulon, CNRS, CPT, IPhU, Marseille, France}
\affiliation[b]{LUPM, UMR5299, Universit\'{e} de Montpellier,\\ 34095 Montpellier, France}
\affiliation[c]{Department of Theoretical Physics, Tata Institute of Fundamental Research\\Mumbai, India}
\emailAdd{aoife.bharucha@cpt.univ-mrs.fr}
\emailAdd{felix.bruemmer@umontpellier.fr}
\emailAdd{desai@theory.tifr.res.in}
\emailAdd{sophie.mutzel@cpt.univ-mrs.fr}

\abstract{We consider an axion-like particle (ALP) coupled to Standard Model (SM) fermions as a mediator between the SM and a fermionic dark matter (DM) particle. We explore the case where the ALP-SM and/or the ALP-DM couplings are too small to allow for DM generation via standard freeze-out. DM is therefore thermally decoupled from the visible sector and must be generated through either freeze-in or decoupled freeze-out (DFO). In the DFO regime, we present an improved approach to obtain the relic density by solving a set of three stiff coupled Boltzmann equations, one of which describes the energy transfer from the SM to the dark sector.  Having determined the region of parameter space where the correct relic density is obtained, we revisit experimental constraints from electron beam dump experiments, rare $B$ and $K$ decays, exotic Higgs decays at the LHC, astrophysics, dark matter searches and cosmology. In particular, for our specific ALP scenario we (re)calculate and improve beam dump, flavour and supernova constraints. Throughout our calculation we implement state-of-the-art chiral perturbation theory results for the ALP partial decay width to hadrons. We find that while the DFO region, which predicts extremely small ALP-fermion couplings, can probably only be constrained by cosmological observables, the freeze-in region covers a wide area of parameter space that may be accessible to other more direct probes. Some of this parameter space is already excluded, but a significant part should be accessible to future collider experiments.
}

\begin{document} 
\maketitle
\flushbottom

\section{Introduction}
\label{sec:intro}
The last two decades have seen a surge in the theoretical exploration of dark matter (DM) models, and brought a wealth of new data to constrain them. The possibility that dark matter directly couples to Standard Model (SM) states via renormalizable interactions is by now severely constrained. For example, LHC data and direct detection experiments rule out vast parts of the parameter space for standard electroweak-scale WIMPs produced via thermal freeze-out \cite{limitsDM, colliderreview}. This has led to a paradigm shift towards various other possibilities for dark matter candidates, with masses above 1 TeV or below 10 GeV, and with other production mechanisms. In particular, models where the DM-SM interactions are mediated by additional particles have attracted much attention.

Among such models are those with a fermionic dark matter candidate and a pseudoscalar mediator $a$, interacting with the SM via a Lagrangian of the form
\begin{equation}
{\cal L}= a\sum_{f} C_f\,\frac{m_f}{f_a}\,\bar f i\gamma_5 f+a\, C_\chi\,\frac{m_{\chi}}{f_a}\,\bar\chi i\gamma_5\chi+\ldots\label{eq:pseudscalarmediator}
\end{equation}
Here $f$ stands for any SM fermion, and $\chi$ is the DM particle. One possible motivation for this type of interaction is that it allows the efficient suppression of direct detection cross sections while still allowing for large enough DM-SM interactions to create the observed relic abundance via thermal freeze-out. However, freeze-out is of course by no means the only mechanism for dark matter production. In fact, one of the oldest dark matter candidates is itself a pseudoscalar state which is extremely light and extremely weakly coupled, namely a non-thermally produced QCD axion. At intermediate values of the couplings, dark matter could be produced by the freeze-in mechanism: Assuming that its abundance is initially zero after reheating, dark matter will be gradually produced by scattering processes in the thermal plasma without ever reaching thermal equilibrium, until these processes decouple and the DM abundance remains constant \cite{FreezeIn, Dawn}.

In this paper we study a dark matter model with interactions of the type of eq.~\eqref{eq:pseudscalarmediator} but for a range of couplings which are too small to give the correct relic density through freeze-out. Our motivation is unrelated to direct detection constraints (which are irrelevant for such small couplings) but are rather guided by top-down reasoning: If there exists a light pseudoscalar $a$, it might well be an axion-like particle (ALP), by which we mean the pseudo-Goldstone boson of an approximate $\U{1}_{\rm PQ}$ global symmetry which is spontaneously broken at a high scale $f_a$; this provides a reason for it to be light. It is natural to have $a$ emerge from an extended Higgs sector in some UV completion of the SM, which could explain the flavour-preserving couplings to SM states of eq.~\eqref{eq:pseudscalarmediator}. In this case, besides these effectively renormalizable couplings, one also expects dimension-5 couplings of the form
\begin{equation}
 {\cal L}= a\sum_{f} C_f\,\frac{y_f}{\sqrt{2}f_a}\,h\,\bar f i\gamma_5 f+\ldots\label{eq:pseudscalarmediator-dim5}
\end{equation}
where $h$ is the SM Higgs boson, which will turn out to be important for the dark matter abundance in parts of the parameter space. 

If there are additional fermions $\chi$ charged under $\U{1}_{\rm PQ}$, they would also couple to the ALP. If there are no additional fermions charged under both $\U{1}_{\rm PQ}$ and under the Standard Model, there exists a field basis in which no dimension-5 couplings $aF\widetilde F$ between the ALP and the SM gauge bosons are induced above the electroweak scale. The only relevant couplings between the SM and the ALP are then those of eqs.~\eqref{eq:pseudscalarmediator} and \eqref{eq:pseudscalarmediator-dim5}. Moreover, since there is no evidence for new physics close to the electroweak scale, the scale $f_a$ should be large, $f_a\gg$ TeV. Hence the ALP couplings to fermions may be too small to allow for dark matter production via freeze-out, and other production mechanisms should be studied.

It is tempting to try to identify $a$ in this scenario with a (DFSZ-like) QCD axion itself. However, if a QCD axion were massive enough to be short-lived on cosmological scales (thus mediating the SM-DM interactions, rather than being part of the dark matter sector itself), it would be subject to multiple astrophysical and laboratory constraints, see e.g.~\cite{DiLuzio:2020wdo} for an overview. While it may be possible to circumvent these constraints by extensive model-building, in this paper we will not attempt to do so but rather assume that the $a$ mass is mainly due to some explicitly $\U{1}_{\rm PQ}$-breaking effect other than the anomaly. The usual QCD axion relation between $m_a$ and $f_a$ therefore no longer holds, and $a$ does not contribute to solving the strong CP problem.

To summarise our findings, we will show that the observed dark matter abundance can be produced in this model by several distinct mechanisms, depending on the values of the ALP-SM and ALP-DM couplings as well as on the ALP and DM masses. It can be produced via ALP-mediated freeze-in from the scattering of SM particles, via freeze-in from the scattering of ALPs, via freeze-out of ALP-DM interactions which kept the DM in equilibrium with the ALP, or finally via standard freeze-out. We will numerically study a number of example scenarios with regard to the most relevant cosmological, astrophysical and collider constraints.

\section{Model}
\label{sec:model}
\subsection{Particle content, interactions and relevant processes}

Our model contains a single Dirac fermion $\chi$ and a pseudoscalar ALP $a$. The Lagrangian is
\begin{align}\nonumber
\mathcal{L} =&\;{\cal L}_{\rm SM}+\frac{1}{2} \partial_\mu a \partial^\mu a +\bar\chi (i\partial\!\!\!\slash-m_\chi)\chi+ i\sum_f  \frac{C_f}{f_a} \left(m_f+\frac{y_f}{\sqrt{2}}h\right)\,a\bar f \gamma_5 f \\
& - \frac{1}{2}m^{2}_{a} a^2 + i\frac{ C_\chi}{f_a}\,m_\chi\, a\bar \chi \gamma_5 \chi\,, \label{eq:lagrangian}
\end{align}
where $f$ is any Standard Model fermion with mass $m_f$ and Yukawa coupling $y_f$ and $h$ is the Higgs boson. This Lagrangian is valid below the electroweak symmetry breaking scale; see Appendix \ref{app:model} for some details on a possible UV completion.
We have not included any coupling of the ALP to gauge bosons via $a\,\tr F\widetilde F$ terms. Including them would be perfectly possible but would lead to a proliferation of parameters, so we restrict our analysis to more minimal models where there are no heavy fermions charged under both the SM gauge group and $\U{1}_{\rm PQ}$. Nevertheless, effective $a\gamma\gamma$ and $agg$ vertices are induced at one loop by (finite) SM fermion triangle graphs. We will often refer to the hidden sector (HS), by which we mean the ensemble of ALPs and DM particles.
For future convenience, we define the effective axion couplings by
\begin{equation}\label{eq:gaff}
 g_{aff}=\frac{C_f}{f_a}\,,\qquad g_{a\chi\chi}=\frac{C_\chi}{f_a}\,,
\end{equation}
where we will later refer to $g_{a\chi\chi}$ as the hidden sector coupling, and to  $g_{aff}$ as the connector coupling. In this model, dark matter is stabilized by a global $\U{1}$ $\chi$-number symmetry which also ensures that it can only be produced in $\chi\bar\chi$ pairs. A variant with a Majorana dark matter candidate stabilized by a $\mathbb{Z}_2$ symmetry could also be viable.

From the above Lagrangian we can deduce the various means by which:
\begin{itemize}
    \item the dark matter can interact with Standard Model particles, e.g.~through interactions with fermions mediated by the pseudoscalar, $f\bar f \leftrightarrow\chi\bar \chi$ or $f\bar f \leftrightarrow h\chi\bar \chi$, or via a loop diagram with gauge bosons, $\gamma\gamma\leftrightarrow \chi \bar \chi$;
    \item the ALP can interact with the SM, e.g.~via $q \bar q\leftrightarrow g a$, $q g\leftrightarrow q a$, or the (inverse) decay $f\bar f\leftrightarrow a$, or $f\bar f\leftrightarrow ha$;
    \item the ALPs and the dark matter can interact, i.e.~$aa\leftrightarrow\chi\bar \chi$ or $\chi a\leftrightarrow\chi a$.
\end{itemize}
These different interactions, where those responsible for DM and ALP generation are summarised in table~\ref{tab:processes+couplings}, give rise to a rich phenomenology. Different DM generation mechanisms are at play in different regions of the parameter space. The study of these mechanisms was pioneered in a different model by \cite{4ways}, and more recently refined in \cite{Hambye:2019dwd}.
\renewcommand{\arraystretch}{1.2}
\begin{table}
\begin{center}
\begin{tabular}{|c|ccc|c|}
\hline
\textbf{Interaction} & \multicolumn{3}{|c|}{\textbf{Processes}} & \textbf{Scaling} \\
\hline
SM $\leftrightarrow $ ALPs & \hspace{.5cm} \parbox[c][2.2cm]{1.2cm}{\includegraphics[scale=0.5]{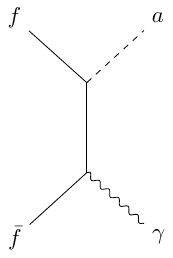}}&\parbox[c][2.5cm]{2cm}{\includegraphics[scale=0.5]{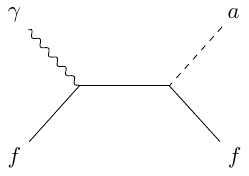}}&\parbox[c][2.5cm]{1.2cm}{\includegraphics[scale=0.5]{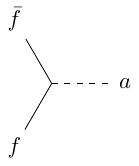}}\parbox[c][2.5cm]{2cm}{\includegraphics[scale=0.5]{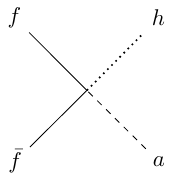}}&$g_{aff}^2$ \\
 \hline
SM $\leftrightarrow \chi$ &\multicolumn{3}{|c|}{ \parbox[c][2.5cm]{2cm}{\includegraphics[scale=0.5]{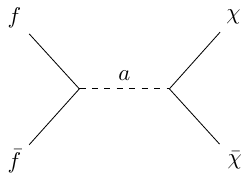}}
\quad\parbox[c][2.5cm]{2cm}{\includegraphics[scale=0.5]{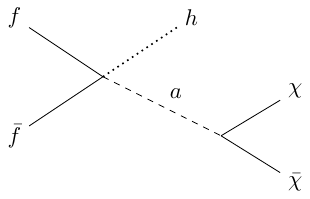}}} & $(g_{aff}\cdot g_{a\chi\chi})^2$\\
\hline
 ALPs $\leftrightarrow \chi$ & \multicolumn{3}{|c|}{\parbox[c][2.5cm]{2cm}{\includegraphics[scale=0.45]{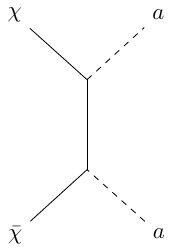}}
 }& $g_{a\chi\chi}^4$\\
\hline
\end{tabular}
\captionof{table}{Different sector interaction processes and their scaling with the hidden sector and connector couplings. Note that the diagrams $f\bar f\leftrightarrow\gamma a$ and $f\gamma\leftrightarrow fa$ shown are equivalent to those for $qq\leftrightarrow g a $, $q g\leftrightarrow qa$, $\bar{f} \gamma \leftrightarrow \bar{f} a$ and $\bar q g\leftrightarrow \bar qa$ and therefore the latter are not shown explicitly.}\label{tab:processes+couplings}
\end{center}
\end{table}
\renewcommand{\arraystretch}{1.}

As detailed in the next sections, the mechanisms at play in our model are the following\footnote{Note also that if the ALP is coupled to a strongly-interacting hidden sector, the DM can be produced via an additional, qualitatively different production mechanism, the so-called SIMP mechanism \cite{Hochberg_2018}.}:
\begin{itemize}
 \item For sufficiently small $g_{a\chi\chi}$ and $g_{a ff}$, and assuming that the HS relic density is negligible at reheating, dark matter is produced by freeze-in. Depending on the couplings, the dominant process may be direct freeze-in $f\bar f\into\chi\bar\chi$ mediated by an off-shell ALP, or ALP production from SM states followed by $aa\into\chi\bar\chi$ (where the ALP may or may not be in equilibrium with the SM). These processes are infrared dominated, since they are induced by (effectively) renormalizable operators. For large reheating temperatures, the dominant process for direct freeze-in will be $2\into 3$ scattering $f\bar f\into h\chi\bar\chi$, which is suppressed by phase space but ultraviolet-dominated, i.e.~sensitive to the reheating temperature.
 \item For intermediate $g_{aff}$, ALPs will be produced abundantly from the SM thermal bath but will not reach thermal equilibrium with the SM particles. However, if $g_{a\chi\chi}$ is sufficiently large, the hidden sector will form a separate thermal bath at a lower temperature, in which DM is produced by freeze-out. We study this mechanism of freeze-out from a thermally decoupled dark sector, or to put it simply, decoupled freeze-out (DFO), in detail.
 \item For larger $g_{aff}$ and  $g_{a\chi\chi}$, the entire dark sector will thermalize with the SM, and DM can be produced via standard freeze-out. Since we are interested in the small-coupling regime, and since the standard freeze-out phase is already well studied in the literature, we will not analyse it in detail.
\end{itemize}

\subsection{General Boltzmann equations}

In order to study the various DM generation mechanisms we start from the covariant form of the Boltzmann equation for the Friedmann-Lema\^{i}tre-Robertson-Walker metric, where the momenta of the particles have been integrated over. The SM particles are in equilibrium with the photon bath at temperature $T$, allowing us to replace the distributions of SM particles by the equilibrium distributions. We neglect quantum statistical factors, assuming Maxwell-Boltzmann statistics:
\begin{equation}
f_{\rm eq}(p,T)=e^{-E/T},
\end{equation}
such that the particles' number density, energy density and pressure equilibrium distributions are given by
\begin{align}
\nonumber n_{\rm{eq}}(T)=&\frac{g}{(2\pi)^3}\,\int d^3 p\,f_{\rm eq}(p,T) = \frac{g}{2 \pi ^2}\,m^2\,T K_2\left(m/T\right)\,,\\
\nonumber \rho_{\rm{eq}}(T)=&\frac{g}{(2\pi)^3}\,\int d^3 p\,f_{\rm eq}(p,T) E=\frac{g}{2 \pi ^2}\,m^2\,T \left(m K_1\left(m/T\right)+3 T K_2\left(m/T\right)\right)\,,\qquad\\
\label{eq:MB}P_{\rm{eq}}(T)=&\frac{g}{(2\pi)^3}\,\int d^3 p\ f_{\rm eq}(p,T) \frac{|p|^2}{3E}  = \frac{g }{2 \pi ^2}m^2 T^2 K_2\left(m/T\right)=Tn_{\rm{eq}}(T) \;,
\end{align}
respectively, for a given particle of mass $m$ and degrees of freedom $g$, where $K_1(x)$ and $K_2(x)$ are the first and second order Bessel functions.

More generally, the hidden sector particles may be in kinetic equilibrium at a temperature $T'$ which is different from $T$:
\begin{equation}
f(p,T')=\frac{n(T')}{n_{\rm eq}(T')}e^{-E/T'}\,.
\end{equation}
Kinetic equilibrium is not reached in the freeze-in phase, where the scattering among hidden sector particles is not efficient (see \cite{Belanger:2020npe} for a detailed analysis of freeze-in in a related model).

The Boltzmann equations governing the evolution of the $a$ and $\chi$ number densities are given by
\begin{align}
\nonumber\frac{\d n_{\chi}}{\d t}+3 H n_{\chi}=&\sum_f \left\langle\sigma_{\chi\bar \chi\rightarrow f \bar f}v\right\rangle\left(\left( n_{\chi}^{\rm{eq}}\right)^2-n_{\chi}^{2}\right)+\left\langle\sigma_{a a \rightarrow \chi\bar\chi } v\right\rangle n_{a}^{2}-\braket{\sigma_{\chi\bar\chi \rightarrow a a} v}n_{\chi}^2\\
\nonumber&+\sum_{i,j,k}\left(\braket{\sigma_{\chi\bar\chi \rightarrow ijk}v} \left(\left(n_\chi^{\rm eq}\right)^2-n_\chi^2\right)+\braket{\sigma_{\chi\bar\chi i\rightarrow jk}v^2} n_i^{\rm eq}\left(\left(n_\chi^{\rm eq}\right)^2-n_\chi^2\right)\right)\,,\\
\nonumber\frac{\d n_a}{\d t}+3 Hn_a=&\sum_{i,j,k}\left\langle\sigma_{i a \rightarrow j k} v\right\rangle\left(n_{a}^{\rm{eq}} n_{i}^{\rm{eq}}-n_{a} n_{i}^{\rm{eq}}\right)+ \braket{\Gamma_{a}}\left(n_{a}^{\rm{eq}}-n_{a}\right)\\
\label{eq:generalBoltzmann}&\,\,-\braket{\sigma_{a a \rightarrow \chi\bar\chi} v}n_{a}^2+\braket{\sigma_{\chi\bar\chi \rightarrow a a} v}n_{\chi}^2\,.
\end{align}
Here  $i,j,k$ are SM particles which are involved in the relevant processes, as listed in table~\ref{tab:processes+couplings}. For example, if all fermion couplings $C_f$ are of the same order and at temperatures where all SM particles are relativistic, the dominant contribution for ALP production comes from $i=g$, $j=\bar{k}=t$. The cross sections for the processes entering eq.~\eqref{eq:generalBoltzmann} are given in appendix~\ref{app:xsecs}. Most of the processes appearing in eq.~\eqref{eq:generalBoltzmann} are $2\to 2$ processes, and enter via the typical thermally averaged cross section  $\braket{\sigma v}$, defined by\footnote{We have not explicitly stated the temperature dependence of the thermally averaged cross sections in eq.~\eqref{eq:generalBoltzmann} as this depends on the production regime considered. For example, in the DFO mechanism the hidden sector interactions take place at the common temperature of the hidden sector which is different from that of the SM, see below in section \ref{sec:reannihilation}.}
\begin{align}
\braket{\sigma_{12\to 34}v}=\frac{C}{2\,T K_2(m_1/T)\,K_2(m_2/T)}\int_{s_{\rm{min}}}^\infty \sigma(s) \frac{F(m_1,m_2,s)^2}{m_1^2 m_2^2\sqrt{s}}\,K_1(\sqrt{s}/T)\,\d s\,\,,
\end{align}
where $\sigma(s)$ is the cross section for the process $12\to34$ as a function of the squared centre-of-mass energy $s$. Here $C$ is an additional factor of $1/2$ for the case of identical particles in the initial state ($C=1$ for non-identical initial state particles). We have further used the abbreviation
\begin{align}
F(m_1,m_2,s)=\frac{\sqrt{(s-(m_1+m_2)^2)(s-(m_1-m_2)^2)}}{2} 
\end{align} 
and the lower limit of the integral is given by
$s_{\rm{min}}= \text{max}\left((m_1+m_2)^2,(m_3+m_4)^2\right)\,$.
Note that, where appropriate, we have made use of detailed balance,
\begin{align}
\braket{\sigma_{i j \rightarrow k l} v}n_{i}^{\rm{eq}} n_{j}^{\rm{eq}}=\braket{\sigma_{k l \rightarrow i j} v} n_{k}^{\rm{eq}} n_{l}^{\rm{eq}} \; .
\end{align}
For the (inverse) decays which contribute, we define the thermally averaged decay rate to be 
\begin{equation}
\braket{\Gamma_a}=\Gamma_a \frac{K_1(m_a/T)}{K_2(m_a/T)} \; .
\end{equation}
where $\Gamma_a$ is the decay rate of the pseudoscalar $a$.
We provide details of the ALP decay rate in the various decay channels in appendix~\ref{app:decaywidths}. For the calculation of the DM relic density we restrict ourselves to (inverse) decay into fermions in the perturbative regime, see eq.~\eqref{eq:decayaxionleptons}.\footnote{For energies in the range $\sim 0.5 - 1.2$~GeV various hadronic decay channels open up. However, when we present results on the DM relic density, we are either in the perturbative regime or apply a hard cut on the decay width below $T\leq 600$~MeV (see also the discussion in section~\ref{sec:DFOresults} and in appendix~\ref{app:decaywidths}).} 

The Hubble rate in eq.~\eqref{eq:generalBoltzmann} is given in terms of the total energy density of the universe $\rho=\rho_{\rm{SM}}+\rho_{\rm{HS}}$, via
\begin{equation}\label{eq:Friedmann}
H = \left( \frac{8}{3} \pi G \rho \right)^{1/2} \; ,
\end{equation}
where $G$ is the gravitational constant. The energy density of the visible sector is
\begin{equation}\label{eq:rhoSM}
\rho_{\rm{SM}} = g_{*\rho,\rm{SM}}(T)\frac{\pi^2}{30}T^4 \; ,
\end{equation}
with $g_{*\rho,\rm{SM}}(T)$ the SM effective degrees of freedom in energy. 
The hidden sector energy density differs according to the production regime as we will explain in more detail in section \ref{sec:reannihilation}. For the SM effective degrees of freedom in energy and in entropy, denoted by $g_{*\rho}$ and $g_{*s}$, respectively, we adopt the recent results from an improved analysis, covering a wide range of temperatures~\cite{Saikawa:2018rcs}. 
This analysis is based on state-of-the-art results of perturbative and non-perturbative calculations of thermodynamic quantities in the SM. In particular, for an improved description of the QCD phase transition they adopt results from the ``Budapest-Wuppertal''-collaboration \cite{Borsanyi:2016ksw}  in $2 + 1 + 1$ flavour lattice QCD. We use the fit functions provided in this paper in our analysis. 

Note that by solving this set of differential equations we only calculate the abundance of $\chi$ particles and later set $n_{\rm{DM}}=n_\chi+n_{\bar{\chi}}=2n_\chi$. In addition, we point out that later in this section we will find that it is convenient to make a change of variables from $t$ to $z=m_\chi /T$, and the differential equations will be written as a function of $z$ instead of $t$.

\subsection{Freeze-in}\label{sec:FI}
The freeze-in mechanism can generate the observed DM abundance for the case that both $g_{aff}$ and $g_{a\chi\chi}$ are sufficiently small, assuming that the initial abundance of DM particles and ALPs is zero or negligible. Both $a$ and $\chi$ will be produced from interactions amongst SM particles in the thermal bath, but for small $g_{aff}$ and $g_{a\chi\chi}$ the DM will not thermalize, neither with the photons nor with the ALPs. The production stops once the temperature of the particles involved drops below the DM mass scale, since the abundances become Boltzmann-suppressed and the production rate becomes negligible compared to the Hubble expansion rate. The co-moving dark matter number density is effectively frozen from that point on, while the ALPs will eventually decay back to Standard Model states.

The interplay of the different sectors allows for three different freeze-in regimes, namely freeze-in of the DM directly from SM particles, freeze-in from the ALPs and sequential freeze-in~\cite{Hambye:2019dwd,Belanger:2020npe}. All three regimes are characterised by a value of $g_{a\chi\chi}$ which is too small for the hidden sector particles to establish thermal equilibrium among each other. Depending on the range in $g_{aff}$, the ALPs may or may not be in equilibrium with the SM, as described in the following.

\subsubsection{Freeze-in from SM particles}\label{sec:FISM}
In this regime, the DM is produced directly from the SM fermions via an $s$-channel ALP (see table~\ref{tab:processes+couplings}). In order to generate the correct relic density via this mechanism, one requires $g_{a\chi\chi}$ to be small, such that $\braket{\sigma_{aa \rightarrow \chi\bar\chi}v} \propto g_{a\chi\chi}^4$ is small enough for $aa \to \chi \bar\chi$ to be negligible compared to $f\bar f\to (h)\chi \bar \chi$. In this regime we find that ALPs are in thermal equilibrium with the SM, i.e.~$n_a = n_a^{\rm{eq}}(T)$.

In order to calculate the relic abundance we require the appropriate Boltzmann equation for this region, which we obtain by adapting eq.~\eqref{eq:generalBoltzmann} accordingly. We first define the dark matter yield $Y=Y_\chi$ as usual by
\begin{equation}
 Y=\frac{n_\chi}{s}
\end{equation}
where $s=s_{\rm{SM}}+s_{\rm{HS}}$ is the total entropy density of the universe. The entropy density in the freeze-in regime is given by
\begin{equation}
s = \frac{2 \pi^2}{45} g_{*s}(T) T^3 \; ,
\end{equation} 
with the effective degrees of freedom in entropy given by $g_{*s}(T)=g_{*s\rm{,SM}}(T)+g_{*s\rm{,a}}(T)$ and we can neglect the entropy density of the DM particles.

The ALP abundance plays no role in this freeze-in scenario, so we neglect all terms in eq.~\eqref{eq:generalBoltzmann} involving $n_a$. Since the DM is never in equilibrium, $Y\ll Y_{\rm eq}$, we can neglect the back-reaction as usual. Concerning the contribution to the DM yield from $2\into 2$ scattering, it is
convenient to introduce the reaction density for the interaction $ij\to kl$ \cite{4ways}
\begin{equation}
\gamma_{ij\to kl}=\langle\sigma_{ij\to kl}\,v\rangle n_i^{\rm eq}n_j^{\rm eq}
\end{equation}
in terms of which the Boltzmann equation eq.~\eqref{eq:generalBoltzmann} for the yield from freeze-in becomes
\begin{equation}
 \frac{\d Y}{\d T}=- \sum_f\frac{\gamma_{\chi\bar\chi \into f\bar f}}{ 3 H s^2}\frac{\d s}{\d T}+(2\,\leftrightarrow\,3\text{ terms})\,.
\end{equation}
Here we have replaced the $z$ dependence by a dependence on $T$.
The corresponding infrared-dominated contribution to the yield today at $T\approx 0$ is therefore obtained by integrating the reaction density,
\begin{equation}
Y_{0, {\rm IR}}=-\sum_f \int_{0}^\infty\frac{\gamma_{\chi\bar\chi \into f\bar f}}{3Hs^2} \frac{\d s}{\d T}\d T\,. \label{eq:freezein}
\end{equation}

An important point concerns the relative magnitude of the contributions from $2\into 2$ interactions and $2\into 3$ interactions. In a hypothetical model with only renormalizable couplings between the ALP and the SM fermions, as defined by eq.~\eqref{eq:pseudscalarmediator}, freeze-in would proceed only via $f\bar f\into \chi\bar\chi$ scattering, which is infrared dominated. However, as we stated previously, in realistic models where the ALP-fermion couplings are generated from the ALP mixing with the Standard Model Higgs, one also obtains the dimension-5 couplings of eq.~\eqref{eq:pseudscalarmediator-dim5}. These couplings induce  $2\into 3$ interactions which are phase-space suppressed, but nevertheless cannot be neglected in general,\footnote{Note on the other hand that the $f\bar f h\to\chi \bar \chi$ interactions included in eq.~\eqref{eq:generalBoltzmann} can safely be neglected.} because they are ultraviolet dominated and their contributions therefore scale as the reheating temperature $T_{\rm RH}$. For instance, if $T_{\rm RH}$  is large enough to neglect all masses of the involved particles, then the dark matter yield from such a process can be estimated as \cite{Elahi:2014fsa}
\begin{equation}\label{eq:maxTRH}
Y_{0,{\rm UV}}\approx \frac{135}{(2\pi)^9}\frac{1}{1.66\,g_*^{3/2}} g_{aff}^2 g_{a\chi\chi}^2\,y_f^2\,m_\chi^2\,T_{\rm RH}\,M_P\,. 
\end{equation}
A thorough analysis of ALP production from UV freeze-in, allowing for all possible dimension-5 couplings between the ALP and the SM to be present, was conducted in \cite{Salvio:2013iaa}, and models similar to ours have been studied in both their UV and IR freeze-in phases in \cite{Biswas:2019iqm}. We will eventually find that, for reheating temperatures $T_{\rm RH}\gtrsim 200$~GeV, UV freeze-in via $2\to 3$ scattering gives significant contributions to the dark matter yield; see appendix \ref{app:xsecs} for details of the computation.

\subsubsection{Freeze-in from ALPs}\label{sec:FIALP}
In this regime, the ALPs are in thermal equilibrium with the SM, and produce the DM via the $t$-channel process $aa\to\chi \bar\chi$.  Obtaining the observed relic density via this mechanism requires a larger $g_{a\chi\chi}$ compared to the case in section ~\ref{sec:FISM}, and a correspondingly smaller $g_{aff}$ to avoid overproduction.

Since the ALPs are in thermal equilibrium with the SM particles, this regime is conceptually very similar to the previous one. We just need to replace the reaction density $\gamma_{\chi \bar \chi \into f\bar f}$ in eq.~\eqref{eq:freezein} by the appropriate $\gamma_{\chi \bar \chi \into a a}$:
\begin{equation}\label{eq:freezeinfromALPs}
Y_0=- \int_{0}^\infty\frac{\gamma_{\chi \bar \chi \into a a}}{3H s^2} \frac{\d s}{\d T}\d T.
\end{equation}

In this regime, there are no direct sizeable UV-dominated contributions to the DM yield. However, UV-dominated $2\to 2$ processes such as $ f \bar f\to ah$ will contribute to the ALP abundance and will therefore reduce the value of  $g_{aff}$ at which the ALPs reach equilibrium with the SM. We determine this boundary by numerically solving the Boltzmann equation for the ALPs and checking whether $n_a\gtrsim n_a^{\rm eq}$. For IR-dominated contributions, the yield is essentially insensitive to the upper integration limit (provided it is chosen above the freeze-out temperature of the top quark). Since the UV-dominated contributions depend on the reheating temperature, the appropriate integration limit is at $z_{RH}=m_\chi/T_{RH}$. As always we assume that the universe has reheated into SM particles only, i.e.~$n_a(T_{RH}) \approx 0$. Note that ${\cal O}(1)$ corrections to the UV contributions can be expected, due to the uncertainty associated with the details of the reheating mechanism.

\subsubsection{Sequential freeze-in}\label{sec:seqFI}
In the sequential freeze-in regime, dark matter is again produced from ALPs via $aa\to\chi\bar\chi$. The difference with the previous regime is that the ALPs are not thermalized. Nevertheless, the ALP abundance is sufficient to obtain the correct DM relic density.

Calculating the DM relic abundance is non-trivial, since the ALPs are not even in kinetic equilibrium, and therefore the unintegrated Boltzmann equations governing the individual momentum modes should be employed for a precise quantitative analysis. This was done in a study dedicated to the freeze-in regime in ref.~\cite{Belanger:2020npe} for a related model. Due to time limitations, we choose to follow ref.~\cite{Hambye:2019dwd}, where a simplified analysis is proposed employing the integrated Boltzmann equations
\begin{align}
\nonumber
\frac{\d n_a}{\d t}+3 Hn_a=&\sum_{i,j,k} \gamma^{\rm eq}_{i a\to j k}+ \braket{\Gamma_{a}}n_{a}^{\rm{eq}}(T)\,,\\
\label{eq:seq-freezein}\frac{\d n_{\chi}}{\d t}+3 H n_{\chi} =&\,\gamma_{\chi\bar\chi\to aa}\,,
\end{align}
where $\gamma_{aa\to \chi\bar\chi}=\langle \sigma_{aa\to\chi\bar\chi}v\rangle(T)\,n_a^2$, and $n_a$ is determined from the first equation. The second term on the RHS of the first equation describes the production of the ALP via inverse decays. We have neglected the term $-\gamma_{\chi\bar\chi\to aa}$ in the first equation since $n_{a}^{\rm{eq}}(T)\gg n_a$. The temperature of the ALPs is taken to be the temperature $T$ of the SM bath.

This gives a rough estimate of the parameter values leading to the observed relic density; however we stress that this method is relying on several rather crude approximations. Specifically, we assume that the axions are in kinetic equilibrium in order to be able to define a temperature, and we then equate this temperature with that of the photons. It should be possible to obtain a more precise result following the method of ref.~\cite{Belanger:2020npe}, but this is beyond the scope of our work.

\subsection{Decoupled freeze-out (DFO)}\label{sec:reannihilation}

For a sufficiently large hidden sector coupling $g_{a\chi\chi}$, the interactions between the HS particles will be strong enough for them to thermalize. If the ALP-SM couplings $g_{aff}$ are sufficiently small, then the hidden sector will nevertheless be thermally decoupled from the SM thermal bath, at a temperature $\Tp \ll T$.
This regime is somewhat more complicated to analyse than the freeze-in regime, due to an increased number of interactions playing a role, and due to the non-trivial interplay between the hidden sector number densities and the temperature $\Tp$. We will therefore proceed to explain some of the technical aspects in more detail.

As in the case of freeze-in, we assume the initial number density of DM particles and of ALPs to be negligibly small. The DM and ALPs  will then freeze in until the abundances are sufficient to allow the particles to thermalize. This happens once the reaction rate for $a a \,\leftrightarrow\,\chi\bar\chi$ scattering exceeds the expansion rate of the universe, 
\begin{equation}
\Gamma_{a a\into\chi\bar\chi} \equiv \langle \sigma_{a a\into\chi\bar\chi}\, v\rangle n^\text{eq}_a(T^\prime) > H\,.
\label{gammaannihH}
\end{equation}
While $a a \,\leftrightarrow\,\chi\bar\chi$ is the dominant process for DM and ALP annihilation and production, the DM and ALP number densities will approximately track the equilibrium values $n^{\rm eq}_\chi(\Tp)$ and $n^{\rm eq}_a(\Tp)$, respectively. 
After $\Tp$ falls below the mass of the $\chi$ and $a$, we allow for the possibility that $n_\chi$ and $n_a$ deviate from the equilibrium distributions. The number densities $n_\chi$ and $n_a$ and the hidden-sector temperature $\Tp$ are then given by the solution of a system of three coupled differential equations, as explained in the following.

The temperature $T'$ of the hidden sector is calculated from the hidden-sector energy density $\rho'$ via the equation of state. To determine $\rho'$, we start from the Boltzmann equation for the phase-space distribution $f(p,t)$ of either of the hidden sector particle species:
\begin{equation}\label{eq:boltzmannf}
   \left(\partial_t-H\,p\,\,\partial_p\right)f(p,t) =\frac{1}{E(p)}C[f(p,t)]\,.
\end{equation}
Here we have used that, by isotropy and homogeneity, $f(p,t)$ can only depend on the modulus $p$ of its 3-momentum  and on time. $C[f]$ is the collision operator and $E$ is the energy. With the energy density given by eq.~\eqref{eq:MB}, we can integrate eq.~\eqref{eq:boltzmannf}, writing
\begin{align}
\label{eq:generalETBE}
    \frac{\partial\rho'}{\partial t}+3H\,\left(\rho'+P'\right)=&\int\frac{d^3 p}{(2\pi)^3}C[f(p,t)]
\end{align}
where we have integrated by parts and used $P'=\tfrac{1}{3}\langle p\tfrac{\partial E}{\partial p}\rangle$. The integrated collision operator for a $1\;2\into 3\;4$ process is
\begin{equation}
\int\frac{d^3 p}{(2\pi)^3}C[f]=g_1 g_2 \int \frac{d^3p_1}{(2\pi)^3}\frac{d^3p_2}{(2\pi)^3}f_1(p_1)f_2(p_2)v_\text{M\o l}\,\mathcal{E}(\vec{p}_1,\vec{p}_2)
\label{eq:coll1}
\end{equation}
where  $v_{\text{M\o l}}$ is the usual M\o ller velocity and the energy transfer rate $\cal E$ is, in terms of the matrix element $\mathcal{M}$ and the transferred energy $\Delta E_{tr}$, 
\begin{align}\hspace{-.4cm}
\label{eq:Entransfer}\mathcal{E}(\vec{p}_1,\vec{p}_2)=&\frac{1}{2 E_1 2E_2 v_\text{M\o l}}\int\prod_{i=3,4} \frac{d^3 p_i}{(2\pi)^3}\frac{1}{2 E_i} |\mathcal{M}|^2(2\pi)^4\delta^{(4)}(p_1+p_2-p_3-p_4)\Delta E_{tr}\,.
\end{align}
Solving eq.~\eqref{eq:generalETBE} enables us to obtain $\rho'$ as a function of the temperature of the visible sector $T$.
In the reannihilation region described in ref.~\cite{4ways}, eq.~\eqref{eq:generalETBE} was applied to a model with fermionic dark matter coupled to both the SM photon and a hidden-sector photon.~\footnote{Note that while our DFO region is very similar to the reannihilation region of ref.~\cite{4ways}, it differs from the latter in two ways: the first is that in our case the energy transfer from the SM to the dark sector proceeds via the SM to ALP transition rather than SM to DM, see figure~\ref{fig:evolutionDFO}. The second is that during freeze-out, the production of DM from the SM does not become the dominant production mechanism, and as a result we do not observe a bump in the DM comoving number density as opposed to the case in ref.~\cite{4ways}.
} In that case, the hidden sector is populated predominantly via $f\bar f\into \chi\bar\chi$ processes, and the mass degeneracies of the particles involved allow to analytically simplify the integrated collision operator along the lines of  \cite{cosmicabundances}, leaving a single integral to be evaluated numerically. In our model, the $2\to 2$ processes $f\bar{f}\to a\gamma$, $\gamma f\to a f$, $q\bar{q}\to ag$, $gq\to aq$ and hermitian conjugates must also be taken into account, as well as (inverse) decay processes $f\bar{f}\to a$. To treat the $2\to 2$ processes, some work is needed to cast the integrated collision operator into a form amenable to numerical integration. This is detailed and justified in Appendix \ref{app:collision}.

Given the hidden-sector energy density from eq.~\eqref{eq:generalETBE}, we obtain $\Tp$ from the equation of state of the hidden sector. The total HS energy density and pressure are given by the sum of the ALP and DM contributions,
\begin{equation}
\rho^\prime+P^\prime=\rho_a+\rho_{\chi}+P_a+P_{\chi} \; .
\end{equation}
Initially the ALPs and the DM will be ultra-relativistic, $P'=\rho'/3$, and the universe will be radiation-dominated with most of its energy density in the visible sector, $\rho\propto T^4$. Changing variables using $\frac{\partial}{\partial t}\approx-HT\frac{\partial}{\partial T}$,  eq.~\eqref{eq:generalETBE} becomes
\begin{equation}\label{eq:EBEinT}
    \frac{\partial\rho'}{\partial t}+4H\,\rho'=-H\left(T\frac{\partial\rho'}{\partial T}-4\rho'\right)=-HT\rho\frac{\partial}{\partial T}\left(\frac{\rho'}{\rho}\right)=\int\frac{d^3 p}{(2\pi)^3}C[f(p,t)]\,.
\end{equation}
As long as the hidden sector particles are relativistic and interactions are rapid, solving this equation for $\rho'$ will provide us with the temperature of the hidden sector via eqs.~\eqref{eq:MB}: $\Tp$ is obtained by iteratively solving
\begin{equation}\label{eq:rhoprimeoverrho}
\frac{\rho^\prime}{\rho}=\frac{\rho_a^{\rm eq}(\Tp)+\rho_\chi^{\rm eq}(\Tp)}{\frac{\pi^2}{30}g_{\rm eff,SM}(T)T^4} \; .
\end{equation} 

As $T'$ decreases, the hidden-sector particles will eventually become non-relativistic, and their interactions will freeze out. In this regime, $T'$ must be determined together with the HS number densities $n_\chi$ and $n_a$. Kinetic equilibrium is maintained (due to efficient $\chi a\to\chi a$ scattering), hence
\begin{equation}
\rho_\chi = \frac{\rho_{\chi}^{\rm{eq}}(\Tp)}{n_{\chi}^{\rm{eq}}(\Tp)} \ n_\chi\,, \qquad
P_\chi = \frac{P_{\chi}^{\rm{eq}}(\Tp) }{n_{\chi}^{\rm{eq}}(\Tp) }\ n_\chi =T^\prime n_\chi \; ,
\end{equation}
and similarly for the ALP. The hidden-sector equation of state is then given by
\begin{equation}\label{eq:hseofs}
\rho^\prime+P^\prime= \frac{\rho_{\chi}^{\rm{eq}}(\Tp)}{n_{\chi}^{\rm{eq}}(\Tp)}\ n_\chi +\frac{\rho_{a}^{\rm{eq}}(\Tp)}{n_{a}^{\rm{eq}}(\Tp)} \ n_a+T^\prime \left(n_\chi +n_a\right) \; .
\end{equation}
The quantities $n_\chi$, $n_a$ and $T'$ are finally obtained as the solutions of a system of three coupled differential equations, namely eq.~\eqref{eq:generalETBE} written in the form
\begin{equation}\label{eq:Tprime}
z \frac{\d\rho^\prime}{\d T^\prime}\frac{\d T^\prime}{\d z}=-3(\rho^\prime+P^\prime)+\frac{1}{H}\int\frac{d^3 p}{(2\pi)^3}C[f(p,t)] 
\end{equation}
in conjunction with the equation of state eq.~\eqref{eq:hseofs}, and the Boltzmann equations 
\begin{align}\begin{split}\label{eq:BERchi}
H z \frac{\d n_{\chi}}{\d z}+3 H n_{\chi}=&\sum_f \left\langle\sigma_{\chi\bar\chi\rightarrow f \bar f}v\right\rangle(T)n_{\chi}^{\rm{eq}}(T)^2+\sum_{i,j,k} \braket{\sigma_{\chi\bar\chi i\rightarrow jk}v^2} n_i^{\rm eq}\left(n_\chi^{\rm eq}\right)^2\\&+\left\langle\sigma_{a a \rightarrow \chi\bar\chi } v\right\rangle(\Tp) n_{a}^{2}-\braket{\sigma_{\chi\bar\chi \rightarrow a a} v}(\Tp)n_{\chi}^2
\end{split}\\
\begin{split}\label{eq:BERa}
H z \frac{\d n_{a}}{\d z}+3 Hn_a=&\braket{\Gamma_{a}}n_{a}^{\rm{eq}}(T)+\sum_{i,j,k}\left\langle\sigma_{i a \rightarrow j k} v\right\rangle(T) n_{a}^{\rm{eq}}(T) n_{i}^{\rm{eq}}(T)\\
&-\braket{\sigma_{a a \rightarrow \chi\bar\chi} v}(\Tp)n_{a}^2+\braket{\sigma_{\chi\bar\chi \rightarrow a a} v}(\Tp)n_{\chi}^2
\end{split}
\end{align}
with the initial condition for $T'$ provided by the solution of eq.~\eqref{eq:rhoprimeoverrho}.
Here $i,j,k$ are SM particles involved in the $a$-number changing processes, see table~\ref{tab:processes+couplings}.
\subsection{Finite-temperature effects}
We incorporate finite-temperature corrections in our analysis in light of the observation in ref.~\cite{Belanger:2020npe} that these have a large impact on the relic density in the case of sequential freeze-in. Such corrections arise due to the fact that perturbation theory breaks down in the presence of an additional scale, the temperature of the plasma. Calculations therefore require the resummation of diagrams, which can be implemented in a simple way, following ref.~\cite{Belanger:2020npe}, on adopting the hard thermal-loop (HTL) approximation~\cite{Chu:2011be}, where only loops involving soft momenta $\sim gT\lesssim T$ are resummed. Within this approximation, it turns out that the masses of fermions and gauge bosons appearing in the dispersion relations can simply be replaced by temperature dependent quantities. For the gauge bosons this amounts to adding a mass to the propagators which is the thermal Debye mass (neglecting differences between the transverse and longitudinal modes), which for the case of gluons is given by~\cite{Bhattacharya:2016zcn}
\begin{equation}
m_{g}^{2}(T)=\frac{4 \pi \alpha_s(2\pi T) T^{2}}{3}\left(n_{c}+\frac{n_{f}(T)}{2}\right) \; ,
\end{equation}
with $n_f$ the number of active flavours in the plasma, and for photons~\cite{Bhattacharya:2016zcn}
\begin{equation}
m_{\gamma}^{2}(T)=\frac{4 \pi \alpha_{\rm{em}} T^{2} n_{\mathrm{ch}}(T)}{3} \; ,
\end{equation}
with $n_{\mathrm{ch}}(T)$ the number of electromagnetically charged particles in the plasma.

For the case of fermions we also add a thermal mass correction to the propagators, neglecting the difference between particle and hole states~\cite{Bijnens:1998fm}. For the leptons this takes the form \cite{PhysRevD.26.2789}
\begin{equation}
m_l^2(T)=\frac{4 \pi \alpha_{\rm{em}}(2\pi T)T^2 q_f^2}{8}
\end{equation}
and for quarks
\begin{equation}
m_q^2(T)=\frac{4 \pi T^2 (\alpha_s(2 \pi T)/6+\alpha_{\rm{em}}(2\pi T) q_f^2)}{8} \; .
\end{equation}
In order to capture the effect of thermal corrections on interaction vertices we renormalize the coupling constants at the scale of the first Matsubara mode, $\omega=2\pi T$, making use of the renormalization group equations in vacuum~\cite{Bijnens:1998fm}. Note that we perform this running for the electromagnetic, strong couplings and Yukawa couplings, although for the latter we found that this resulted in a negligible effect on the results.

\section{Constraints}
\label{sec:constraints}
In this section we review the most important existing and future experimental constraints on our model. ALPs can significantly impact astrophysical, cosmological and collider processes, and the various constraints on ALPs have widely been discussed in the literature (see \cite{revisedconstraints} for a review). In section~\ref{sec:results} we will explore the regions of parameter space of our model where the correct relic density is obtained via different production mechanisms. We will therefore focus on those collider, astrophysical and cosmological constraints which are relevant in these regions.

While the discussion has been general so far, to discuss the constraints and for the subsequent numerical analysis we concentrate on the simplest case of strictly flavor-universal axion-fermion couplings, i.e.~universal values of $C_f$ and $g_{aff}$ in eqs.~\eqref{eq:lagrangian} and \eqref{eq:gaff}.

\subsection{Collider constraints}

\subsubsection{SLAC E137 electron beam dump experiment}
\begin{figure}
\begin{center}
\parbox[c][3cm]{4cm}{\centering \includegraphics[scale=0.7]{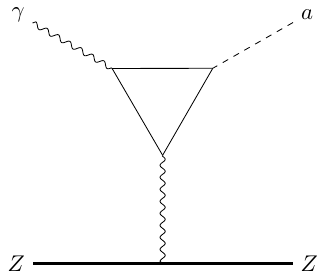}}
\parbox[c][3cm]{4cm}{ \vspace{1.35em}\hspace{0.4em}\includegraphics[scale=0.7]{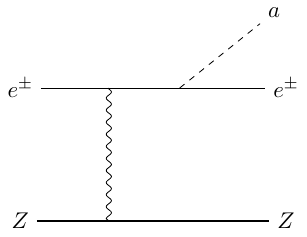}}
\parbox[c][3cm]{3.2cm}{\centering \includegraphics[scale=0.8]{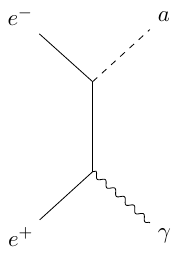}}
\parbox[c][3cm]{3.2cm}{\centering \includegraphics[scale=0.8]{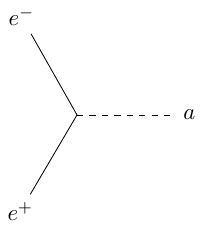}}\\
\parbox[c][2cm]{4cm}{\centering(a)}\parbox[c][2cm]{4.2cm}{\centering(b)}\parbox[c][2cm]{3.2cm}{\centering(c)}\parbox[c][2cm]{3.2cm}{\centering(d)}
\captionof{figure}{Feynman diagrams for the four different production mechanisms of ALPs in our model in the E137 beam dump experiment: (a) Primakoff production (b) bremsstrahlung (both off a target nucleus of atomic number Z) (c) non-resonant $e^+e^-$ annihilation (d) resonant $e^+e^-$ annihilation.}\label{fig:feynmanbeamdump}
\end{center}
\end{figure}

Electron and proton beam-dump experiments are among the accelerator-based experiments where the ALPs in our model could be produced. Within the parameter region of interest a strong constraint comes from the E137 beam dump experiment at SLAC \cite{E137}. Here ALPs could be produced via Primakoff, bremsstrahlung or positron annihilation (see figure \ref{fig:feynmanbeamdump}) and if these decay within the detector such events might be observed.
\paragraph{Calculating the number of events}
In the SLAC E137 electron beam dump experiment no events were detected \cite{E137}. This puts an experimental upper bound on the event rates $\mathcal{N}_s$
\begin{align}
\mathcal{N}_s<\mathcal{N}_s^{\text{up}}=2.996 \text{ for } \mathcal{N}_{s,\text{det}}=0 \text{ at 95\% CL}\,. 
\end{align}
Following \cite{E137,revisedconstraints,ALPtraum} we calculate the theoretically expected event rate to obtain a bound on the ALP's coupling and mass. The theoretical number of events can be obtained by an integral over all possible production cross sections folded with the probability to detect the ALP and the differential track-length distribution of the shower particles in the beam, multiplied by the total number of electrons dumped. 
To obtain the theoretical number of events in a specific decay channel we additionally multiply by the corresponding branching ratio, yielding
\begin{align}
\mathcal{N}_{\rm theory}(a\to XX)=\mathcal{N}_{e,\text{ inc}}\sum_{i,Z} P_Z \int \mrm{d}E \ T_i(E) \, p(E) \, \sigma_Z^{i\to a k}(E)  \ \mathcal{B}(a\to XX)\, . \label{eq:Ntheo}
\end{align}
$\mathcal{N}_{e,\text{ inc}}$ is the total number of electrons dumped ($\sim$ 30 Coulomb),  $P_Z$ is the number density of atoms per $\rm{cm}^{3}$ of atomic number Z and $ \sigma_Z^{i\to a k}$  are the cross sections for the different production processes with $i$ the initial particle. $T_i$ is the corresponding track-length distribution of the initial incoming particles, $i \in \{ \gamma, e^+, e^-\}$. $p(E)$ is the probability for the ALP to decay visibly. The differential track-length distributions for the secondary shower particles can be estimated in the Weizs{\"a}cker-Williams approximation which allows one to relate the incident electrons in the initial state to the distributions of photons, electrons and positrons which are produced in electromagnetic shower cascades. Intuitively, the differential track-length $T(E)$ is the length a particle with given energy $E$ will travel. One then directly considers the interactions of these particles with the target \cite{BUDNEV1975181}. We extract the track-length distributions for photons, electrons and positrons in the beam dump (figure~14 of reference \cite{E137}) which have been calculated using the SLAC EGS program \cite{EGS}. This spares us from finding an analytic expression for the Weizs{\"a}cker-Williams approximation. 
Through interactions of the shower particles with the target material, the ALPs in our model are predominantly produced by the Primakoff process ($Z \gamma \to Z a$), by bremsstrahlung ($Z e^{\pm} \to Z e^{\pm} a$) and by (non-)resonant positron annihilation ($e^+ e^- \to \gamma a$ and $e^+ e^- \to a$). The Feynman diagrams for these processes are shown in figure~\ref{fig:feynmanbeamdump}. We can sum over the production channels since all other final state particles produced in the process are absorbed in the shielding. The ALP subsequently decays and its decay products can be detected in the electromagnetic calorimeter. Therefore, in order to obtain the theoretically expected number of events, we include the branching fraction for the decay channel $a\to \gamma \gamma$ (loop induced) and $a\to e^+ e^-$, which opens up for $m_a>2m_e$. The various decay channels of the ALP in our model are summarised in appendix~\ref{app:decaywidths}.

Following ref.~\cite{ALPtraum} we include the probability $p(E)$ for the ALP to be detected. Taking the experimental layout into account, we can envisage at least five scenarios where the ALP could have decayed invisibly (see appendix~\ref{app:SLACproba}): Either none or only one of the decay particles reaches the detector or both of them reach the detector but the opening angle is too small and the two final state particles are indistinguishable. For instance, if the ALP is highly boosted and its lifetime is long, it will certainly decay behind the detector. We will thus have to take the ALP decay probability into account to obtain a final result for the theoretically expected number of events. Details about the calculation of the probability for the ALP to decay visibly can be found in appendix~\ref{app:SLACproba}. We assume that most of the ALPs are produced in the cooling water \cite{revisedconstraints}.
\paragraph{Details about the ALP production processes}
As mentioned above, the relevant production processes for our model in electron beam dump experiments are the Primakoff process, bremsstrahlung and (non-)resonant positron annihilation (see figure~\ref{fig:feynmanbeamdump}). In the Primakoff process, depicted in figure~\ref{fig:feynmanbeamdump}a), a secondary photon emits an ALP in the vicinity of the nucleus. Since the ALP in our model does not couple to photons at tree level, the process is induced at one-loop level, with a loop function depending on the momentum transfer. Details about the cross section and the loop calculation relevant for this process can be found in appendices~\ref{app:Primakoff} and \ref{app:loopdetails}, respectively. In bremsstrahlung (see figure~\ref{fig:feynmanbeamdump}b), the electron is scattered off the mass shell of the nucleus and returns to the mass shell by emission of an ALP. Lastly, the ALP can be produced by (non-)resonant positron annihilation, see figures~\ref{fig:feynmanbeamdump}c) and~\ref{fig:feynmanbeamdump}d). The cross sections for bremsstrahlung and positron annihilation can be found in appendices~\ref{app:bremsstrahlung} and~\ref{app:positronannihilation}.
\begin{figure}
\begin{center}
\includegraphics[scale=1]{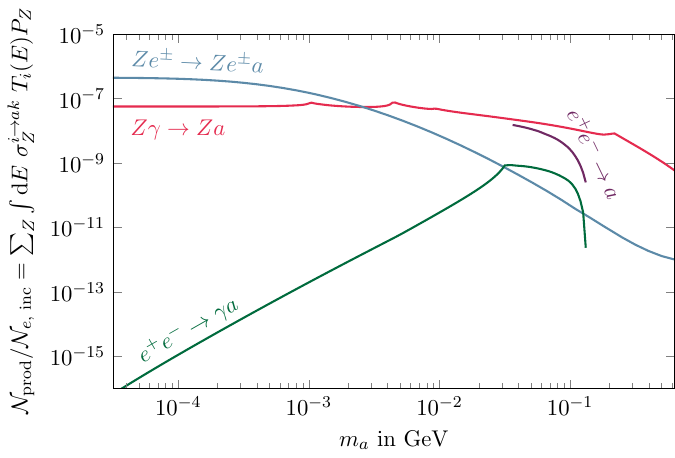}
\captionof{figure}{ALP yield per incident electron $\mathcal{N}_\text{prod}/\mathcal{N}_{e,\text{ inc}}=\sum_Z \int \mathrm{d}E \ \sigma_Z^{i\rightarrow ak}\ T_i(E) P_Z$ for the various production processes of the ALP in our model in electron beam dump experiments, namely Primakoff production $(Z\gamma \to Z a)$, bremsstrahlung $(Ze^\pm \to Ze^\pm a)$, non-resonant $(e^+e^- \to \gamma a)$ and resonant $(e^+e^- \to a)$ positron annihilation.}\label{fig:SLACxsecs}
\end{center}
\end{figure}
It is interesting to look at the interplay of the different contributions as a function of the ALP mass. From eqs.~\eqref{eq:primakoffSLAC} and \eqref{eq:BremsstrahlungSLAC} we see that for models where the ALP couples to both electrons and photons at tree level the Primakoff and the bremsstrahlung processes scale approximately as
\begin{align}
\sigma^{\gamma \to a}_Z \equiv\sigma(Z \gamma \to Z a) &\sim \alpha_{\rm{em}} g_{a\gamma \gamma}^2 \quad \mbox{and}\\
\sigma^{e \to a e}_Z \equiv\sigma(Z e^{\pm} \to Z e^{\pm} a) & \sim \alpha_{\rm{em}}^2 g_{aff}^2 \frac{m_e^2}{m_a^2} \; ,
\end{align}
for  $m_e\ll m_a\ll E_i$, with $E_i$ the energy of the initial particle.  This means that bremsstrahlung is suppressed by a factor $\sigma^{e \to a e}_Z/\sigma^{\gamma \to a}_Z \sim \alpha_{\rm{em}}m_e^2/m_a^2$ (see also ref.~\cite{2012.07894}). Interestingly, in our model, due to the loop suppression in the Primakoff process, see eq.~\eqref{eq:loop_coupling}, the cross sections scale as
\begin{align}
\sigma^{e \to a e}_Z/\sigma^{\gamma \to a}_Z \sim m_e^2/(m_a^2 \alpha_{\rm{em}})
\end{align}
and the Primakoff process becomes dominant only for larger ALP masses. This can be observed in figure \ref{fig:SLACxsecs} where we plot the ALP yield per incident electron from the various production processes depicted in figure~\ref{fig:feynmanbeamdump},
\begin{align}
\mathcal{N}_\text{prod}/\mathcal{N}_{e,\text{ inc}}=\sum_Z \int \mathrm{d}E \ \sigma_Z^{i\rightarrow ak} \ T_i(E) \ P_Z\; ,
\end{align}
as a function of the ALP mass $m_a$, for an ALP-electron coupling $g_{aff}=1$~GeV$^{-1}$ and ignoring the detection probability of the ALP. The range of accessible positron energies limits the range of producible ALPs to those with masses $m_{a,\min}\sim 35$~MeV for the resonant case and $m_{a,\max}\sim 135$~MeV for production by both resonant and non-resonant positron annihilation. 
For small ALP masses non-resonant positron annihilation scales approximately as $\sigma^{e\to a \gamma}\sim m_a^2$. While in the original analysis \cite{E137}, ALP production by non-resonant positron annihilation was included, in later analyses positron annihilation is often neglected altogether since the positrons arise as secondary particles in the beam. Yet, the number of secondary positrons can be large enough for positron annihilation to become important for large ALP masses. To be precise, it turns out that it indeed dominates over the often considered bremsstrahlung for ALP masses around $50-100$~MeV and clearly cannot be neglected in models where the Primakoff process is highly suppressed, see also ref.~\cite{SLACdarkphoton} for a related discussion in the case of dark photons. In our case, however, the ALP coupling to photons arises from fermion loops and the suppression of photoproduction is such that positron annihilation is non-negligible, but not large enough for it to become dominant.

\subsubsection{Exotic Higgs decays at the LHC}
Production of the ALP in Higgs decays would be mainly via $h \rightarrow a a$ (for $ m_a \leq m_h/2$) or $h \rightarrow Z a$ (for $m_a < m_h - m_Z \simeq 85$~GeV).  While it is possible to write extra effective operators for the $haa$ as well as the $hZa$ coupling, we assume for this work that only new terms are those in the Lagrangian in eq.~\eqref{eq:lagrangian}.  Therefore both these Higgs decay channels are mediated by a fermion loop and dominated by the top-quark contribution.

Higgs exotic decay searches at the LHC currently include $h\rightarrow Z(a \rightarrow gg)$~\cite{ATLAS:2020pcy} and $h\rightarrow a a$ with (1) both $a\rightarrow b \bar b$~\cite{ATLAS:2020ahi}, (2) both $a\rightarrow \ell^+ \ell^-$~\cite{ATLAS:2021ldb}, and (3) $aa \rightarrow b \bar b \mu^+ \mu-$~\cite{ATLAS:2018emt}.  A detailed study of these reveals (see appendix~\ref{app:higgsdecays} for further details on the sensitivity of each of these) that current searches are not sensitive to coupling regimes that would correspond to $g_{aff} \lesssim 1/$GeV.  We therefore can safely ignore limits from these searches.  

\subsubsection{Production in decays of mesons}
For intermediate mass ranges ($2 m_e \lesssim m_a \lesssim 5$ ~GeV), the best chance for detecting an ALP could be via decays of heavy mesons.  Dedicated experiments provide the possibility to obtain rather precise measurements of branching fractions and thereby constrain couplings to very small values.
We consider constraints on the emission of an ALP in rare $B$ and $K$ flavour-changing neutral current processes, proceeding at the quark level via $b\to s a$ or $s\to d a$. Depending on the decay mode and the lifetime of the ALP, such decays could be constrained by $B\to K^{(*)}\ell^+\ell^-$ and $K\to \pi\ell^+\ell^-$ or $B\to K^{(*)}\nu\bar{\nu}$ and  $K\to \pi\nu\bar{\nu}$. 
For ALP masses smaller than $2m_e$, the only decay possible is to photons, and the lifetime of the ALP is long enough that it decays outside the detector volume of current experiments.  In this case, it would show up as an invisible decay mode of the said meson with the accompanying products identifiable.

The main experimental constraints that we consider in this work are therefore:
\begin{itemize}
    \item $B^+ \to K^+ X (\to\mu^+ \mu^-)$ for long-lived scalar $X$ (LHCb~\cite{LHCb:2016awg}), the $95\%$ C.L.~upper limits on the branching ratio are given as a function of the lifetime of $X$ in the range 0.1 to 1000 ps.
     \item $B^0 \to K^{*0} X (\to\mu^+ \mu^-)$ where $X$ is a scalar particle with mass in the range 214 to 4350 MeV (LHCb~\cite{LHCb:2015nkv}), the $95\%$ C.L.~upper limits on the branching ratio are given as a function of the mass and lifetime of $X$. The limit is of the order $10^{-9}$ over the majority of this range.
         \item $B^0 \to K^{(*)} X (\to\mu^+ \mu^-)$ at fixed target experiments, limits can be extracted from CHARM results as described in ref.~\cite{Dobrich:2018jyi}. 
    \item $K^+ \to \pi^+\nu\bar{\nu}$  from NA62~\cite{NA62:2021zjw}, where $90\%$ C.L.~upper limits are given for the  $K^+ \rightarrow \pi^+X$ branching ratio, where $X$ is a long-lived scalar or pseudoscalar particle decaying outside the detector, for lifetimes longer than $100$ ps.
     \item $K^\pm \rightarrow \pi^\pm e^+e^-$, where NA48/2~\cite{NA482:2009pfe} provides a $90\%$ C.L.~upper limit (here we assume the lifetime should be less than $10$ ns).
    \item $K^\pm \rightarrow \pi^\pm X(\to\mu^+ \mu^-)$   for long-lived $X$ (NA48/2~\cite{NA482:2016sfh}) , the $90\%$ C.L.~upper limits on the branching ratio are given as a function of the lifetime of $X$ in the range 100 ps to 100 ns.
\end{itemize}
We currently do not apply the $K \rightarrow \pi \gamma \gamma$ (which only dominates near 100 MeV) and  $B_s \rightarrow \mu^+ \mu^-$ (which only dominates for non-minimal flavour violation and for $m_a \sim 3-10$ GeV). Further the $B \rightarrow K \nu\bar\nu$ from BaBar~\cite{BaBar:2013npw} is not included as the relevant parameter space is covered by the $K^+ \rightarrow \pi^+\nu\bar{\nu}$ search.

In order to obtain the constraints on our parameter space, we first calculate the $B\to K^{(*)} a$ and $K\to\pi a$ branching ratios. For the former, the expression can be found in refs.~\cite{Gavela:2019wzg} and \cite{MartinCamalich:2020dfe}. Form factors for the $B\to K$ transition are taken from ref.~\cite{Bharucha:2010im} and for $B\to K^*$ from ref.~\cite{Bharucha:2015bzk}. For the Kaon decay  and $K\to\pi a$, we follow ref.~\cite{Alves:2017avw}, taking into account the octet enhancement in non-leptonic Kaon decays.

For each point in parameter space, we then verify the ALP lifetime, which allows us to impose the appropriate limit, and then for the charged lepton channels multiply by the branching ratio for the necessary ALP decay, as given in appendix~\ref{app:decaywidths}. 
For the $B \to K^{(*)} X (\to\mu^+ \mu^-)$, $K^+ \rightarrow \pi^+X$ (where $X$ is long lived) and $K^\pm \rightarrow \pi^\pm X(\to\mu^+ \mu^-)$ channels, the experimental constraints are directly given for the contribution of a new particle $X$, as a function of the mass and lifetime of $X$, such that we do not require the SM branching ratio. These limits can easily be translated from the $m_a-\tau_a$ to the $(m_a,g_{aff})$-plane. For $K^+\to\pi^+e^+e^-$ however we need to include the SM contribution, for which we adopt the chiral perturbation theory result from ref.~\cite{Friot:2004yr}, taking into account the associated theoretical uncertainty.
The constraints from the CHARM experiment~\cite{CHARM:1985anb,CHARM:1983ayi} are extracted in ref.~\cite{Dobrich:2018jyi}, by first calculating the $B$ meson production spectra and then convoluting it with the branching ratio for $B\to K^{(*)}a$. In order to be detected, the ALP produced must then decay to muons within the detector volume. Making use of the probability of this occurring, a limit on the ALP coupling to fermions $g_{aff}$ was obtained for a given mass $m_a$, for further details about this procedure see ref.~\cite{Dobrich:2018jyi}.\footnote{We are very grateful to the authors of ref.~\cite{Dobrich:2018jyi} for providing us with the bounds they obtained in the $m_a-g_{aff}$ plane via private communication.}
\subsection{Astrophysical constraints}
\subsubsection{Supernova SN1987A}\label{sec:SN1987A}
The SN1987A neutrino observations in 1987 in the Large Magellanic cloud provide us with another strong constraint. The neutrino flux coming from the core collapse, which lasted a few seconds, was measured. The presence of weakly-coupled particles in supernovae would provide an additional cooling mechanism. By comparing to data from SN1987A, bounds on the ALP coupling can therefore be derived. 
Here, following ref.~\cite{Chang:2016ntp, Chang:2018rso} we will adopt the ``Raffelt criterion'' to obtain these bounds, i.e.~we demand that the luminosity emitted due to the ALP is less than that due to the neutrinos, $L_\nu=3\cdot 10^{52}$ ergs/s~\cite{StarsLaboratories}. The cooling time should also be in agreement with the data.
The high temperatures and density provided by SN1987A creates a conducive environment for the production of weakly-interacting particles such as the ALP. As the core temperature is around 30 MeV, and taking the Boltzmann tail into consideration, we could imagine that ALPs of masses up to $\mathcal{O}(100~\mathrm{MeV})$ could be generated. 
As our ALPs only interact with fermions at tree level, the dominant production mechanism of the ALPs is via bremsstrahlung. In order to calculate the production rate, we first need to relate the quark Lagrangian to the nuclear Lagrangian:
\begin{equation}
\mathcal{L}\supset -i \sum_N \frac{m_N\,C_N}{f_a}a\bar{N}\gamma^\mu\gamma_5N \; ,
\end{equation}
where we sum over the proton and neutron, $N=p,n$ with masses $m_N$ and coupling to the ALP $C_N$ respectively. Fortunately, the relation between the ALP-quark couplings and the ALP-nucleon couplings are well known, and the state of the art results can be found in ref.~\cite{diCortona:2015ldu}. We then obtain $C_p=C_n=0.43\,C_f$. Corrections to the diagrammatic calculation of the nuclear-scattering bremsstrahlung cross section are obtained using the results for the spin-flip current at N$^3$LO in chiral perturbation theory. Three multiplicative factors reproducing these corrections were provided in ref.~\cite{Chang:2018rso}, and are included in our analysis:
\begin{itemize}
    \item The factor $\gamma_f=1/(1+\tfrac{n_B\sigma_{np\pi}}{2\omega})$ acts as a cut-off preventing scattering at arbitrarily low energies, where $n_B$ is the baryon number density, $\omega$ is the energy, and $\sigma_{np\pi}\simeq 15$ is the nucleon-nucleon scattering cross section when the pion is massless.
    \item The factor $\gamma_p$ accounts for the change in phase space due to the non-zero pion mass, and is obtained from $s$ defined in Eq.~(49) of ref.~\cite{Hannestad:1997gc}, which we multiply by a factor $1/(1 - e^{-x})$ to account for the detailed-balance condition, following  ref.~\cite{Chang:2018rso}.
    \item Finally $\gamma_h$ contains the ratio of the mean free path of the weakly-coupled particle calculated at higher order in $\chi$PT to that at Born level, and is given by  $r_{Y_e}$ given in Eq.~(5) of ref.~\cite{Bartl:2016iok}.
\end{itemize}
Putting these factors together with the Born-level expression, the decay rate is given by:
\begin{equation}
\Gamma_a=\sum_N \frac{Y_N^2\,C_N^2}{8f_a^2}\,\frac{n_B^2\,\sigma_{np\pi}}{\omega} \, \gamma_f\gamma_p\gamma_h\, \sqrt{1-\frac{m_a^2}{\omega^2}} \; ,
\end{equation}
where $Y_N$ is the mass fraction of the nucleon $N$ (we adopt $Y_p=0.3$ ~\cite{Chang:2016ntp}).
Making use of this decay width, on integrating over the volume $V$ and the ALP phase-space we obtain the luminosity using~\cite{Chang:2018rso}
\begin{equation}
L_a=\int_0^{R_\nu} dV\int \frac{d^3k_a}{(2\pi)^3}\,\omega \,e^{-\omega/T}\,\Gamma_a\, \mathrm{exp}\left(-\int_0^{R_{\rm far}}dr \left(\Gamma_a+\Gamma_{all}^{\rm abs}\right)\right)
\end{equation}
where  $k_a$ is the ALP momentum. The integration limits involve two radii: $R_\nu$ is the neutrinosphere radius, $\simeq 40$ km, beyond which most neutrinos free stream until arriving at Earth, and the far radius is given by $R_{\rm far}\simeq$100 km. Further the contribution to the absorptive width of the decay of the ALP to leptons is included explicitly via
\begin{equation}
    \Gamma_{a\ell\ell}^{\rm abs}=\frac{m_a}{8\pi}\,\sum_\ell \,\Theta(m_a-2m_\ell)\,\frac{m_\ell^2}{f_a^2}\,\sqrt{1-\frac{4m_\ell^2}{m_a^2}}\;.
\end{equation}
In order to calculate the luminosity, we further require the temperature and density profile of the proto-neutron star. These profiles are subject to large uncertainties, which have been estimated in the past by comparing the different profiles proposed in the literature. We adopt the profiles introduced in ref.~\cite{Chang:2016ntp}, where a detailed study and comparison with other models for the hidden photon case is available. The comparison of the luminosity in the QCD axion and ALP case can be found in ref.~\cite{Chang:2018rso}.
\subsubsection{Horizontal branch stars}\label{sec:HBstars}
The conditions inside horizontal branch (HB) stars in globular clusters would allow a sizeable production of ALPs inside the stellar core. HB stars lie in the horizontal branch in the Hertzsprung-Russell diagram, i.e.~they contain little metal and are in the state of a stable helium burning core and a hydrogen burning shell. Red Giants (RG) evolve to HB stars after their core becomes so hot and dense that the helium ignites. The interaction of ALPs with particles in the core of HB stars could alter the stellar evolution of these stars and in order to avoid conflicts with observations, constraints must be placed on the ALP parameter space. The constraints coming from HB stars are two-fold:  If ALPs interact very weakly, they mostly escape freely, draining energy from the star. This puts an upper bound on the coupling. On the other hand, for larger couplings, they will be trapped inside the source and radiate energy which provides a lower bound.
\paragraph{Energy Loss Argument}
The first scenario we envisage is a very weakly interacting ALP, which would stream out freely of the hot core and accelerate the cooling of the star. A branch in the Hertzsprung-Russell diagram corresponds to a certain stage in the stellar evolution, depending on the type of nuclear fuel being burnt. Stars inside globular clusters travel along  the different branches of the Hertzsprung-Russell diagram during their stellar evolution. Hence, the number of stars inside a certain branch is proportional to the time the star ``lives'' in a certain branch.  The $R$-parameter which is the ratio of the number of HB stars over RG stars,
\begin{align}
R=\frac{N_\text{HB}}{N_\text{RG}}=\frac{\tau_\text{HB}}{\tau_\text{RG}}\; ,
\end{align}
is a suitable quantity to assess the impact of new particles on HB stars inside globular clusters: An additional energy loss mechanism by ALPs would manifest itself in a contraction and heating of the core, which would mainly have an effect on the nuclear fuel consumption (helium), hence reducing the star's Helium burning lifetime and therefore the number of HB stars. The Helium burning lifetime is related to the energy emitted per unit time and mass averaged over a typical HB core, $\langle \varepsilon \rangle$, via its luminosity. One usually assumes that the energy emission caused by additional particles should not exceed the energy emission from Helium burning, $\langle \varepsilon_a \rangle \lesssim \langle \varepsilon_{3\alpha} \rangle\approx 100  \ \text{erg} \ \rm{g}^{-1}\ \rm{s}^{-1}$ \cite{kevmass}. The dominant production mechanism for ALPs with a coupling to electrons in HB stars is the Compton process. Here, we consider three possible ALP production mechanisms, the Compton process, bremsstrahlung and the loop induced Primakoff process. For the calculation of the energy emitted by ALPs we use eq.~(12) in ref.~\cite{BosonicSuperWIMPS} for the Compton process, expression~(56c) given in ref.~\cite{screeningeffects} for the bremsstrahlung process, and eq.~(A2) in ref.~\cite{cosmobounds} for the Primakoff process where we replace the axion-photon coupling by expression \ref{eq:loop_coupling}, making the approximation of zero momentum transfer.
\paragraph{ALP opacity}
While in most models ALPs have very weak couplings, one could on the other hand imagine that the new particle has such a strong coupling that it is trapped inside the stellar core. In this scenario the ALP will scatter and decay inside the star, transporting heat between different regions and thereby contributing to radiative energy transfer. Again, observations indicate that models of stellar structure work well and that a new source of radiative energy transfer should therefore not exceed that due to photons, parametrized by the so-called Rosseland mean opacity. The photon opacity in HB stars is typically $\kappa_\gamma \sim 0.5$~cm$^2$/g. We include four sources for the ALP's effective opacity, namely (loop induced) decay into photons and the inverse Primakoff process (using eqs.~(A.2) and (A.6) from ref.~\cite{cosmobounds}, replacing $g_{a\gamma\gamma}$ by expression \ref{eq:loop_coupling}), the Compton process and bremsstrahlung (using eq.~(3.8) in ref.~\cite{kevmass}).
\subsection{Cosmological constraints}
If a substantial number of ALPs are produced in the very early universe, they can affect the successful predictions for big bang nucleosynthesis (see for example \cite{cosmobounds, Millea:2015qra,Depta:2020wmr}). Whether or not this is a concern depends on the ALP abundance, mass, and lifetime. 

If the ALPs and the DM are still relativistic during the time of BBN  (occuring at a photon temperature between $T \sim 1$~MeV and $T \sim 10$~keV), these additional relativistic degrees of freedom will contribute to the energy density of the universe, cf.~eq.~\eqref{eq:rhoSM}, which increases the speed at which it expands (see eq.~\eqref{eq:Friedmann}). A faster expansion causes an earlier freeze-out of neutrons, manifesting itself in a change of the neutron-to-proton ratio and changing the abundances of Helium-4 and Deuterium. The number of additional relativistic degrees of freedom around the time of BBN, $\Delta N^\text{BBN}_\text{eff}$, is thus constrained by measurements of these light element abundances \cite{Fields:2019pfx}.
In most of the scenarios we envisage, however, the hidden sector particles are frozen out before the QCD phase transition, and the hidden sector temperature is generally lower than the photon temperature.

Heavier ALPs with masses larger than $\sim 1-10$ MeV can still be constrained if they are abundant and long-lived. Out-of-equilibrium decays of ALPs after neutrino decoupling will heat up the plasma and therefore decrease the effective number of neutrinos $\Delta N^\text{CMB}_\text{eff}$ observed in the CMB. Since neutrino decoupling takes place just before BBN, comparably strong constraints on the ALP lifetime can also be inferred from $\Delta N^\text{BBN}_\text{eff}$ \cite{Millea:2015qra}. 

The combined constraints from $\Delta N_\text{eff}$ rule out axion-like particles with either masses below ${\cal O}$(10 MeV) or with lifetimes above ${\cal O}$(0.01 s) when assuming that the ALPs couple predominantly to photons \cite{Millea:2015qra, Depta:2020wmr}. However, this is not the case in our model.

Finally, for heavier ALPs, electromagnetic showers produced in ALP decays during or after BBN can destroy the newly created nuclei and thus directly alter the light element abundances, see \cite{Kawasaki:2020qxm,Depta:2020zbh} for recent studies. For ALP masses above the GeV scale, hadronic showers give rise to additional constraints, excluding abundant hadronically decaying particles with lifetimes down to about  ${\cal O}$(0.1 s). This is because cascade hadrons can scatter off background protons, which once again increases the neutron-to-proton ratio \cite{Reno:1987qw}; see \cite{Kawasaki:2017bqm} for a recent numerical analysis.

\subsection{Constraints on the DM}
\subsubsection{DM relic density constraint}\label{sec:measuredRelicDensity}
Our model should explain the DM relic density measured today and thus has to satisfy the relic density constraint. The DM relic density today measured by Planck is \cite{Planck}
\begin{align}\label{eq:relicDensity}
\Omega_\text{DM}h^2=\frac{\rho_\text{DM}}{\rho_\text{crit}/h^2}=0.120(1) \; .
\end{align}
Throughout the following sections, when calculating the DM relic densities for the different DM production mechanisms, we solve the Boltzmann equations until the comoving $\chi$ number density $Y_\chi=n_\chi/s$ stays constant. We then solve for the couplings which lead to the observed DM relic density today, where we assume that $\chi$ and its antiparticle $\bar \chi$ make up all the DM. Once the number changing interactions stop, the DM number density will become redshifted due to the ongoing expansion of the universe. 
Using entropy conservation in a comoving volume, we relate the number densities after freeze-out (of either the DM, see section~\ref{sec:reannihilation} or the bath particles annihilating into DM, see section ~\ref{sec:FI}) and the measured density parameter of DM today by
\begin{align}
\Omega_\chi h^2\equiv\frac{\rho_\chi}{\rho_{\rm{crit}}/h^2}=\frac{m_\chi n_{\chi,1}}{\rho_\text{crit}/h^2}\frac{s_0}{s_1}=\frac{m_\chi Y_\chi s_0}{\rho_\text{crit}/h^2} \; ,
\end{align}
with $\rho_{\text{crit}}/h^2=1.053672(24) \times 10^{-5} \ \rm{GeV}\ \rm{cm}^{-3}$ and $s_0=2891.2 \ \text{cm}^{-3}$ \cite{PDG} and where we have used the subscript ``$0$'' for quantities today and the subscript ``$1$'' for quantities after freeze-out, when we stop the simulation. Note that for our model $\Omega_\chi=\Omega_\text{DM}/2$ (see section~\ref{sec:model}).
\subsubsection{(No) constraints from dark matter phenomenology}

We have verified explicitly that, as one would expect, the couplings between the DM particle and the SM are too small to lead to an observable signal at current and near future direct or indirect detection experiments, in both the freeze-in and the DFO region. In particular, the present limits from AMS02 on the DM annihilation cross section into SM particles \cite{Bergstrom:2013jra, Cuoco:2017iax} would need to improve by several orders of magnitude to become relevant for our model. The same is true for present limits on dark matter-nucleon interactions in effective field theory for direct detection experiments \cite{Gelmini:2018ogy}.

Since our DM candidate has self-interactions via the axion couplings, it is interesting to ask whether it could resolve any of the small-scale structure problems plaguing the standard $\Lambda$CDM scenario. Along with possible hints for self-interacting DM come a variety of constraints from merging clusters (among them the Bullet cluster) or halo shapes, see ref.~\cite{1705.02358} for a summary. Most of the possible hints suggest a self-interaction cross section per DM mass of the order of $\sigma/m_\chi \sim 0.1-1 \ \rm{cm}^2/g .$
Interestingly, existing constraints turn out to be weaker or roughly of the same order as the cross sections needed to alleviate the tensions.
Since the hidden sector particles are in equilibrium in the DFO region, the interaction strength might become large enough to lead to an observable signal in our model. 

To quantify the self interaction strength in our model, we consider the momentum transfer cross section as proposed by \cite{1308.3419}.\footnote{Alternatively, the viscosity cross section $\sigma_{\mathrm{V}}=\int \mathrm{d} \Omega \sin ^{2} \theta \mathrm{d} \sigma/\mathrm{d} \Omega$ weighting forward and backward scattering equally is sometimes used.}
\begin{equation}
\sigma_{\mathrm{T}}=\int \mathrm{d} \Omega(1-|\cos \theta|) \frac{\mathrm{d} \sigma}{\mathrm{d} \Omega} \; .
\end{equation}
We find that, in the weakly coupled regime and at small velocities where the Born approximation is valid, the cross sections are too small to lead to sizeable interactions. 
In the non-relativistic strongly coupled regime for very light mediators, i.e.~for 
$$
\frac{g_{a\chi\chi}^2 m_\chi^3}{4 \pi m_a}>1 \; ,
$$ 
the cross section has been claimed to be Sommerfeld-enhanced \cite{0907.0235,1305.5438,flavourconstraints}. However, as we will see in section~\ref{sec:DFOresults}, the largest possible DM-ALP coupling in the DFO region is of the order of $g_{a\chi\chi}\sim 2.5\cdot 10^{-2}$~GeV$^{-1}$, such that if we keep the ratio $m_\chi/m_a=10$ fixed, this condition is never met.  
Additionally, as shown recently in ref.~\cite{2003.00021} (see also ref.~\cite{1704.02149}), the potential arising from a pseudoscalar exchange does not lead to any significant Sommerfeld enhancement at small velocities. Hence, the DM particles in our model do not self-interact sufficiently to explain the discrepancies in structure formation observed at small scales.
\section{Results}\label{sec:results}
In this section we will first present the results of the relevant constraints on our model, providing bounds on the ALP-fermion coupling, $g_{aff}$, as a function of the ALP mass, $m_a$. As introduced in section \ref{sec:constraints}, we focus on constraints from the electron beam dump experiment E137 at SLAC, from the supernova SN1987A, from horizontal branch stars and flavour constraints from heavy meson decays. 
Armed with the tools discussed in section~\ref{sec:model} we subsequently study the generation of DM in the early universe, calculating the dark matter relic density generated for a large range of ALP-SM and ALP-DM couplings for a specific set of masses. This requires solving the Boltzmann equations in various different production regimes. We focus on the regions of parameter space needed to reproduce the observed DM abundance in the freeze-in and in the DFO regions, i.e.~in DM genesis scenarios where the DM and the SM remain out of equilibrium, for a large range of ALP and DM masses. Finally, we compare these regions to the constraints on the ALP to study the phenomenological implications of our model. Throughout this section, when calculating the DM relic density, we fix the ratio $m_\chi/m_a=10$.
\subsection{Collider and astrophysical constraints}
\begin{figure}[t]
\begin{center}
	\includegraphics[scale=0.8]{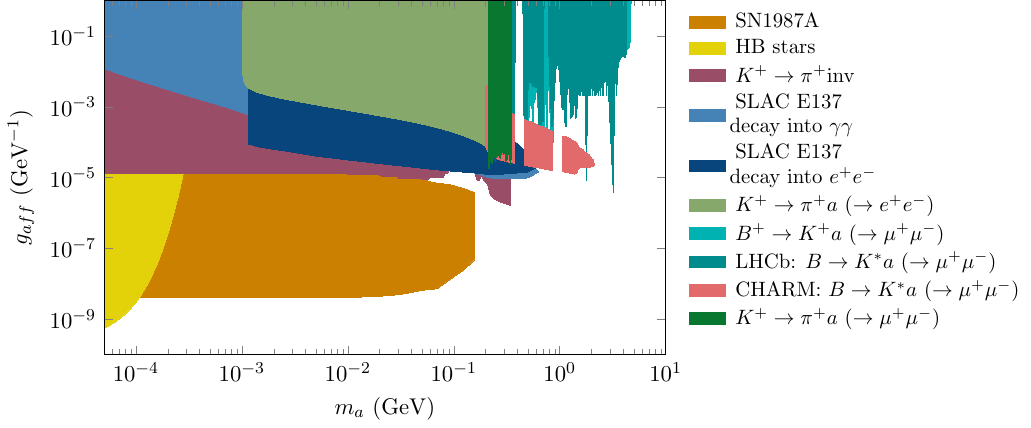}
	\captionof{figure}{Full constraint plot on the ALP-fermion coupling $g_{aff}$ as a function of the ALP mass $m_a$ for an ALP which only couples to SM fermions at tree level. We show collider and astrophysical constraints discussed in section ~\ref{sec:constraints}.} 
	\label{fig:constraints}
\end{center}
\end{figure}
Figure~\ref{fig:constraints} summarises the experimental bounds on ALPs discussed in section~\ref{sec:constraints} we obtained for the model considered in this work, see eq.~\eqref{eq:lagrangian}, where the ALP only couples to SM fermions at tree level.

We have considered a number of collider constraints coming from SLAC E137, NA62, NA48, LHCb and CHARM.
For the electron-beam dump at SLAC we solve $\mathcal{N}_\text{theory}(m_a, g_{aff})=\mathcal{N}_s^\text{up}=2.996$, with $\mathcal{N}_\text{theory}$ given by eq.~\eqref{eq:Ntheo}, for different values of $m_a$. The ALP can be produced by the Primakoff process, by bremsstrahlung or by electron-positron annihilation. It can decay into two photons or, for $m_a>2m_e$, into an electron-positron pair. For even larger ALP masses, various other decay channels open up, cf.~figure~\ref{fig:Branchingfraction}. The original analysis \cite{E137} considers separately 1) Primakoff production with decay into either two photons or into $e^+e^-$, 2) bremsstrahlung production with decay into $e^+e^-$ and 3) non-resonant positron annihilation with subsequent decay into $e^+e^-$. In the present work, we provide a complete analysis without making assumptions on dominant production and/or decay channels. In order to provide a combined bound, we sum over all possible production processes, since the only observable final-state particles are the decay products of the ALP (additional final-state particles from the production process are absorbed in the beam dump absorber). In figure~\ref{fig:constraints} we depict in light and dark blue the constraints from a decay of the ALP into two photons and an electron-positron pair, respectively. 
As explained in more detail in section~\ref{sec:constraints}, a too large ALP-fermion coupling would lead to an overproduction of ALPs, whereas ALPs with too small couplings are long-lived and would escape the detector. The upper bound thus stems from the fact that the detection probability of ALPs decaying too early becomes exponentially suppressed, leading to the typical nose-like shape in both SLAC beam dump constraints. Comparing our results to figure~19 of ref.~\cite{colliderprobes}, we find that the combination of several production processes excludes a slightly larger region of parameter space. This stems from the dominant Primakoff production process for larger ALP masses, cf.~figure~\ref{fig:SLACxsecs}. Since the loop interaction is kinematically suppressed, the bounds on the ALP-fermion coupling $g_{aff}$ for an ALP decaying into two photons are shifted towards larger couplings compared to the constraints on a tree-level ALP-photon coupling, see for instance figure~2 in ref.~\cite{revisedconstraints}.

Flavour constraints also have an important impact on the parameter space considered. The calculation of the bounds from rare $B$ and $K$ decays shown in figure~\ref{fig:constraints} was discussed in section~\ref{sec:constraints}. Rare $B$ decays provide access to the parameter space for masses up to $m_a\sim m_B-m_K^{(*)}$. The LHCb bounds from $B^+ \to K^+ X (\to\mu^+ \mu^-)$ are shown in cyan, and from  $B^0 \to K^{*0} X (\to\mu^+ \mu^-)$ in dark cyan, for a long-lived scalar $X$, as indicated. We further show the CHARM bound coming from the $B^0 \to K^{(*)} X (\to\mu^+ \mu^-)$ analysis in pink. Together, these form a powerful probe of $g_{aff}$, excluding almost completely couplings down to $\mathcal{O}(10^{-5})$ GeV$^{-1}$, as pointed out in ref.~\cite{Dobrich:2018jyi}. Coming to rare $K$ decays, we show the NA48/2 limits on $K^\pm \rightarrow \pi^\pm e^+e^-$ in light green and from $K^\pm \rightarrow \pi^\pm X(\to\mu^+ \mu^-)$ for long-lived $X$ in dark green as well as the NA62 limit from $K^+ \to \pi^+\nu\bar{\nu}$ in purple. These three bounds are highly powerful and complementary probes of the parameter space, providing complete coverage down to $g_{aff}\sim 10^{-5}$ for $m_a\lesssim m_K-m_\pi$.  Note that the uneven nature of certain constraints from $B$ and $K$ meson decays is a direct artefact of the experimental limits, and the gaps in the constraints are due to the fact that in regions where $m_a$ corresponds to the mass of certain mesons, a reliable limit cannot be obtained.

The astrophysical constraints from SN1987A and from HB stars in globular clusters are depicted in  orange and yellow respectively. 
The most stringent constraint relevant for the freeze-in region is from SN1987A. This restricts the coupling $g_{aff}$ to be less than $\sim 4\cdot 10^{-9}$~GeV$^{-1}$ for ALP masses up to $\mathcal{O}(100)$ MeV. The only constraint which probes lower couplings is that from HB stars, but this is only relevant for very small values of $m_a$. As described in section~\ref{sec:SN1987A}, the bound is obtained, following ref.~\cite{Chang:2016ntp, Chang:2018rso}, by demanding that the luminosity emitted due to the ALP is less than that due to the neutrinos, $L_\nu=3\cdot 10^{52}$~ergs (i.e.~the ``Raffelt condition''). The upper line in the constrained region comes from the fact that if the coupling is too large the ALPs would not escape from the Supernova, and the lower line comes from the fact that if the coupling is too low, fewer ALPs would be produced. Note that our bound is obtained by following the calculation in  ref.~\cite{Chang:2016ntp, Chang:2018rso}, which involves several updated nuclear physics calculations, as well as including the energy dependence of the optical depth such that the energy loss at large couplings is correctly accounted for, and the bound therefore differs by an order of magnitude from previous work. We have verified that multiplying or dividing this luminosity by a factor two would result in a change in the limit on the coupling of less than 50\%.

The constraint coming from HB stars excludes a wide region of parameter space for small ALP masses ($1 \ {\rm eV} <m_a<10$~keV) and small couplings $g_{aff} \lesssim 3\times 10^{-9} \text{ GeV}^{-1}$ (yellow), by requiring $\langle \epsilon_a(m_a,g_{aff}) \rangle \lesssim 100\ \rm{erg} \ \rm{g}^{-1}\ \rm{s}^{-1}$. Not surprisingly, since the dominant production mechanism in HB stars for ALPs with a coupling to fermions is the Compton process and since the Primakoff process in our model is loop-suppressed, we essentially obtain the same bound as in figure~4 in ref.~\cite{kevmass}, which the authors calculated considering the Compton process only. It was pointed out in ref.~\cite{kevmass} that radiation is not the dominant mechanism for heat transfer inside HB stars and that indeed convection is more efficient. Let us stress that this analysis is a crude estimate of the bound for our model and therefore most likely subject to large uncertainties. Making dedicated simulations is beyond the scope of this paper, in particular since we are mainly interested in ALPs with masses ranging in the MeV-GeV range for which the number densities are Boltzmann-suppressed in HB stars. For more recent analyses of constraints on the ALP-photon coupling from HB stars see for instance refs.~\cite{CARENZA2020135709,PhysRevLett.125.131804,PhysRevLett.113.191302}. A global analysis from  HB stars, red giants and white dwarfs suggests a non-zero ALP-SM coupling (\emph{HB-hint}) \cite{Giannotti_2017}. Other strong astrophysical constraints on an ALP-electron coupling come from the delay of Helium ignition in red giants, see for example \cite{refId0}. 

It is worthwhile mentioning that we have not made any assumptions about the branching ratios or preferred decay modes for the calculation of the bounds on our model. Since the ALP only couples to fermions at tree level, the branching ratios for the various production and decay modes follow naturally.
\subsection{Phase diagram}
\begin{figure}[t]
\begin{center}
\includegraphics[scale=0.6]{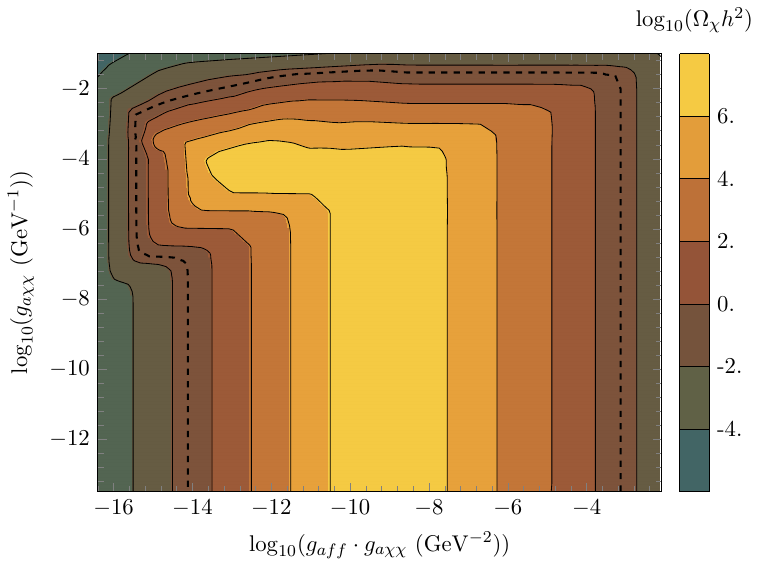}
	\includegraphics[scale=0.6]{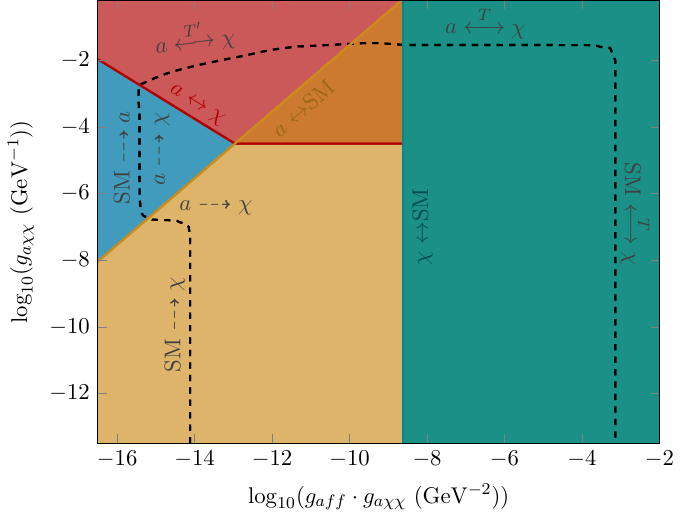}
	\captionof{figure}{Hidden sector coupling $g_{a\chi\chi}$ as a function of the product of the hidden sector and the ALP-fermion coupling, $g_{a\chi\chi}\cdot g_{aff}$, for $m_\chi=10$~GeV and $m_a=1$~GeV. The dashed line corresponds to the combination of couplings which give the correct DM relic density $\Omega_\text{DM}h^2=0.12$ as measured by the Planck telescope. \textit{Left:} Contour plot of the DM relic density as a function of $g_{a\chi\chi}\cdot g_{aff}$ and $g_{aff}$, giving the so-called ``Mesa'' phase diagram: For our choice of masses six different DM genesis scenarios are possible, namely freeze-in from the SM, freeze-in from ALPs, sequential freeze-in, decoupled freeze-out (DFO), freeze-out from the hidden sector and freeze-out from the SM. \textit{Right:} Different phases of the diagram, corresponding to equilibrium relations between the three sectors. The equilibrium relations essentially determine the
production mechanism.}\label{fig:phase}
\end{center}
\end{figure}
As described in detail in section~\ref{sec:model}, the interplay between the three different sectors in our model, namely the SM particles, the ALP and the DM, gives rise to various DM production mechanisms. Solving the Boltzmann equations in a large range of hidden sector and connector couplings and plotting the final abundance as a function of these couplings gives what in \cite{4ways} (and later in \cite{Hambye:2019dwd} for the massive mediator case) was dubbed a ``Mesa'' phase diagram.
For our model, this phase diagram is shown in the left panel of figure~\ref{fig:phase} for an ALP with mass $m_a=1$~GeV and the DM with mass $m_\chi=10$~GeV. The dotted line indicates the combination of couplings $g_{a\chi \chi}$ and $g_{aff}$ such that the final abundance corresponds to the observed DM density.\footnote{Note that we plot the abundance as a function of $g_{a\chi\chi}$ and the product of $g_{a\chi\chi}\cdot g_{aff}$ as these are the proportionality factors for the competing DM production processes, cf.~table~\ref{tab:processes+couplings}.} To aid the understanding of what follows, we also plot in the right panel of figure~\ref{fig:phase} the different phases of the diagram, corresponding to equilibrium relations between the three sectors, as these essentially determine the production mechanism. Let us have a look at the various regimes in more detail:\footnote{This is extensively discussed in ref.~\cite{Hambye:2019dwd}, and we will therefore only briefly comment on the different shapes of the regions.}\\
\vspace*{0.8em}\\
\parbox[t]{0.15\textwidth}{\vspace*{-2.3em}\begin{align*}
\text{SM} &\dashrightarrow \chi\\
(a &\stackrel{T}{\longleftrightarrow} SM)
\end{align*}}\hspace*{2em} \parbox[t]{0.8\textwidth}{\textbf{Freeze-in from SM.} Starting from the bottom-left (small $g_{a\chi\chi}$ and $g_{a\chi\chi}\cdot g_{aff}$) of the phase diagram in figure~\ref{fig:phase}, the dominant process is $f \bar f \to \chi\bar \chi $ and, for reheating temperatures above $T_{RH}\gtrsim 200$~GeV, ultraviolet-dominated $2 \into 3$ interactions. DM is produced from scattering of SM fermions in the thermal bath without ever reaching equilibrium. From equation~\eqref{eq:freezein} we realise that the final relic density is proportional
 to $(g_{a\chi\chi}\cdot g_{aff})^2$ (cf.~table~\ref{tab:processes+couplings}). The size of each term individually is a free parameter, however, $g_{aff}$ is large enough to ensure $a$-SM equilibrium. Hence, we expect a vertical line in the $(g_{a\chi\chi} \cdot g_{aff},g_{a\chi\chi})$ phase diagram.}\\
 \vspace*{0.8em}\\
\parbox[t]{0.15\textwidth}{\vspace*{-2.3em}\begin{align*} a &\dashrightarrow \chi\\
(a &\stackrel{T}{\longleftrightarrow} SM)
\end{align*}}\hspace*{2em} \parbox[t]{0.8\textwidth}{\textbf{Freeze-in from ALPs.} Going upwards in the phase diagram, i.e.~increasing the value of the hidden sector coupling, while still keeping $g_{aff}$ large enough for efficient ALP-SM scattering, collisions of thermal ALPs become more likely than collisions of thermal SM particles. Following the same arguments as above, $Y_\chi$ scales like $g_{a\chi\chi}^4$ and is therefore independent of $g_{aff}$. Hence, we expect a horizontal line in the $(g_{a\chi\chi} \cdot g_{aff},g_{a\chi\chi})$ phase diagram, appearing as a plateau in figure~\ref{fig:phase}.}\\
\vspace*{0.8em}\\
\parbox[t]{0.15\textwidth}{\vspace*{-2.3em}\begin{align*} \text{SM} &\dashrightarrow a \\
a &\dashrightarrow \chi \\
(a  &\cdots\chi)\end{align*}}\hspace*{2em} \parbox[t]{0.8\textwidth}{\textbf{Sequential freeze-in.} The relic abundance is obtained by a chain of sequential reactions $\sum_{i,j,k} ij \to ak$ (with $\{i,j,k\}$ SM particles) followed by $aa \to \chi \bar \chi$. Hence, $Y_\chi \propto \braket{\sigma_{aa \to \chi \bar \chi} v}n_a^2$, i.e.~$Y_\chi \propto (g_{a\chi\chi})^4$. Similarly, $n_a \propto \sum_{i,j,k} \braket{\sigma_{i a \rightarrow j k} v}n_{a}^{\rm{eq}}(T) n_{i}^{\rm{eq}}(T)$, i.e.~$n_a \propto (g_{aff})^2$. Finally, $Y_\chi \propto (g_{aff}\cdot g_{a\chi\chi})^4$ and we expect a vertical line, but shifted to the left because the ALPs and the DM are now both out of equilibrium. Note that our calculation in this region is subject to uncertainties (see section~\ref{sec:seqFI}). On performing a full analysis as in ref.~\cite{Belanger:2020npe} we would expect the line in figure~\ref{fig:phase} to move to the right since as our approximation may overestimate the production of ALPs up to a factor of two. Consequently, a larger ALP-fermion coupling would be needed in order to produce the observed DM relic density.} \\
\vspace*{0.8em}\\
\parbox[t]{0.15\textwidth}{\vspace*{-2.3em}\begin{align*}a &\stackrel{\Tp}{\longleftrightarrow} \chi\\
(\text{SM} &\dashrightarrow a)\\
(\text{SM} &\dashrightarrow \chi)
\end{align*}}
\hspace*{2em} \parbox[t]{0.8\textwidth}{\textbf{DFO.} As explained in section~\ref{sec:reannihilation}, the ALP-SM and the SM-DM processes enter the final relic density only weakly via $\Tp$ (see also section~\ref{sec:DFOresults} for an extended discussion). The dependence on $g_{aff}$ is mild, hence the almost horizontal line in the phase diagram. For increasing $g_{aff}$, more hidden sector particles are produced and the strength among the hidden sector particles has to slightly increase to ensure a later freeze-out of $\chi-$ particles.}\\ 
\vspace*{0.8em}\\
\parbox[t]{0.15\textwidth}{\vspace*{-2.3em}\begin{align*}a &\stackrel{T}{\longleftrightarrow} \chi\\
(a &\stackrel{T}{\longleftrightarrow} SM)
\end{align*}}\hspace*{2em} \parbox[t]{0.8\textwidth}{ \textbf{Thermal freeze-out from ALPs.} 
Going further to the right in the phase diagram, i.e.~further increasing $g_{aff}$, the ALPs and the SM and then also the SM and DM will equilibrate and all three sectors will share a common temperature. In the right half of the phase diagram DM will therefore be produced by freeze-out. However, $g_{aff}$ is still small enough for $aa\leftrightarrow \chi \bar \chi$ to be the more efficient process. The behaviour is as in DFO except that the weak dependence on $g_{aff}$ disappears completely, since the hidden and the visible sector share the same temperature.}\\
\vspace*{0.8em}\\
\parbox[t]{0.15\textwidth}{\vspace*{-2.3em}\begin{align*} \text{SM}&\stackrel{T}{\longleftrightarrow} \chi\\
(a &\stackrel{T}{\longleftrightarrow} SM)
\end{align*}}\hspace*{2em} \parbox[t]{0.8\textwidth}{\textbf{Thermal freeze-out from SM.} This is the usual thermal freeze-out from $2 \to 2$ DM-SM scattering which in our model is set by $f \bar f \leftrightarrow \chi \bar \chi$ scattering and hence by the product $(g_{a\chi\chi}\cdot g_{aff})^2$. As in freeze-in from SM we expect a vertical line, except that the final relic abundance depends now inversely on $(g_{a\chi\chi}\cdot g_{aff})^2$.}

\subsection{Freeze-in region}\label{sec:FIresults}
\begin{figure}[t]
\begin{center}
\includegraphics[scale=0.34]{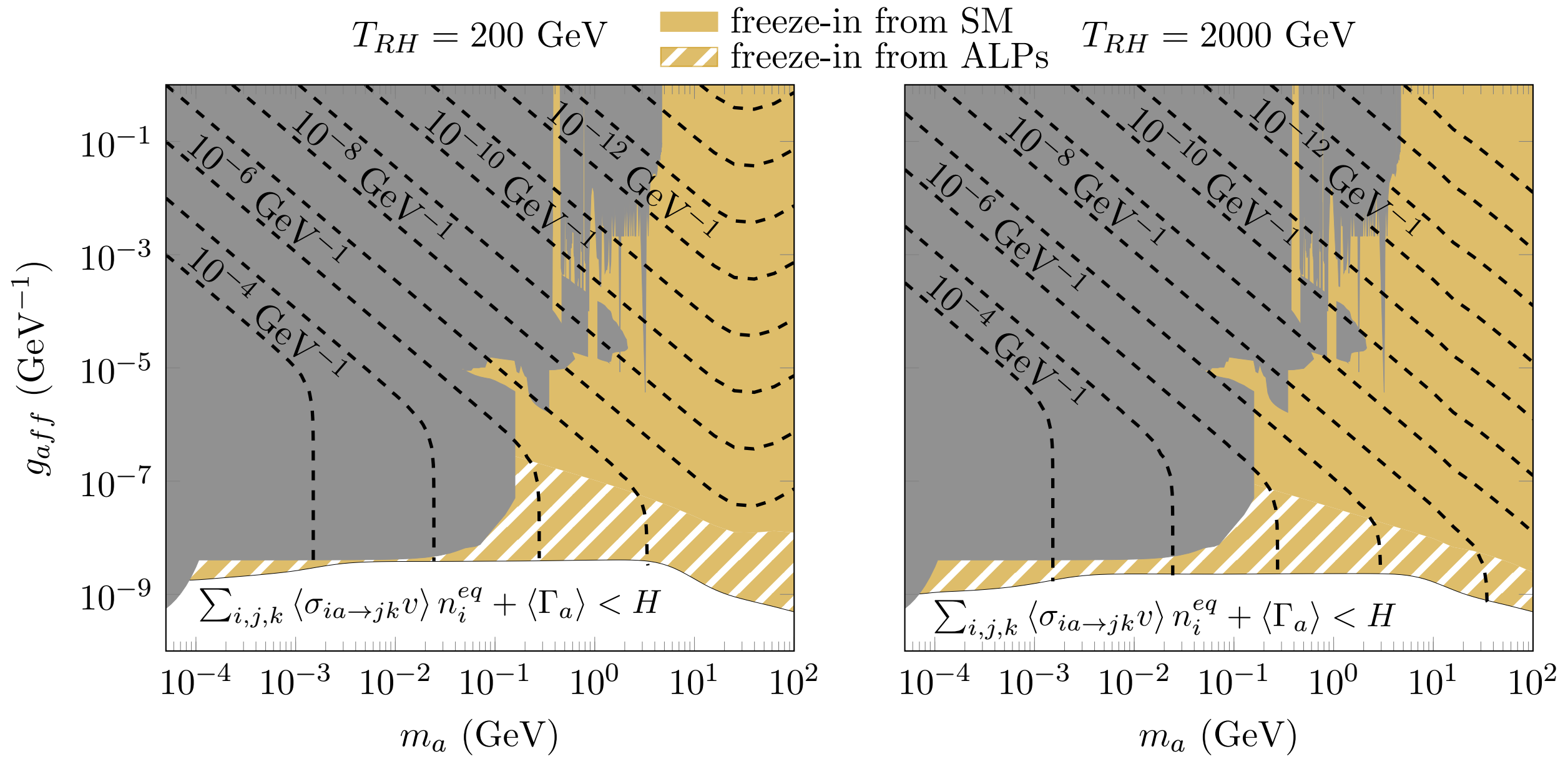}
	\captionof{figure}{ALP-fermion coupling $g_{aff}$ as a function of $m_a$ for lines of constant hidden-sector coupling $g_{a\chi\chi}$ (plotted in dashed black) which reproduce the observed DM relic density via freeze-in from SM and freeze-in from ALPs as indicated for reheating temperatures $T_{RH}=200$~GeV (left) and $T_{RH}=2000$~GeV (right). We fixed the ratio $m_\chi/m_a=10$. The lower line indicates the value of $g_{aff}$ for which the ALPs and the SM reach equilbrium. In grey, we have included the relevant constraints on our ALP model on the connector coupling $g_{aff}$ in this parameter region (cf.~figure~\ref{fig:constraints}).}\label{fig:constraintsplusfreezein}	
\end{center}
\end{figure}
\begin{figure}[t]
\begin{center}
\includegraphics[scale=0.85]{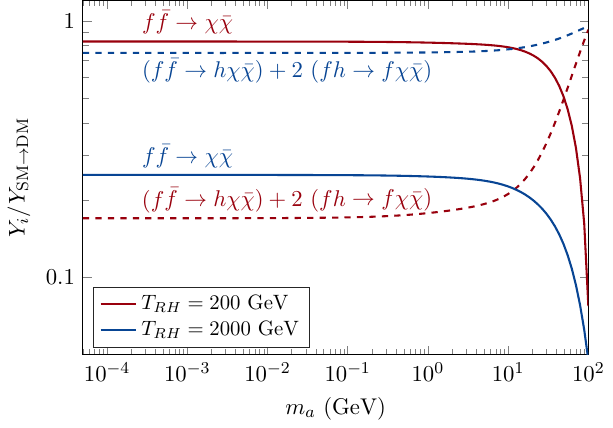}
	\captionof{figure}{Relative contributions of infrared-dominated interaction $f \bar f \to \chi \bar \chi$ (solid) and ultraviolet-dominated $2 \into 3$ interactions $(f \bar f \to h \chi \bar \chi)+2 (f h \to f \chi \bar \chi)$ (dashed) to the final $\chi$-abundance as a function of the ALP mass in the freeze-in from SM regime for reheating temperatures $T_{RH}=200$~GeV (red) and $T_{RH}=2000$~GeV (blue). We fixed the ratio $m_\chi/m_a=10$.}\label{fig:UVvsIR}
\end{center}
\end{figure}
Having discussed collider and astrophysical constraints on the mediator in our model in the previous section, we now investigate the phenomenological implications of a feebly interacting DM particle. The freeze-in region is defined by the couplings of the DM to both the ALPs and the visible sector being so tiny that the DM particles never reach thermal equilibrium. Their initial abundance being zero or negligible, the DM particles are gradually produced by out-of-equilibrium scattering of particles in the thermal bath. The ALPs, on the other hand, are in thermal equilibrium with the SM in the conventional freeze-in scenario. (We do not consider DM production via sequential freeze-in as its description is subject to theoretical uncertainties, as described in section~\ref{sec:seqFI}.)
 Hence the reaction rates satisfy 
\begin{align}
\nonumber 
\sum_{i,j,k}\left\langle\sigma_{i a \rightarrow j k} v\right\rangle n_{i}^{\rm{eq}}(m_\chi) +\braket{\Gamma_{a}}\,>& \,H(m_\chi) 
\qquad \,\,\text{(equilibrium of ALP-SM)}\, ,\\
 	\begin{array}{l}
 	{\displaystyle \sum_{i,j,k} \left\langle\sigma_{\chi\bar\chi[k]\rightarrow i j}v^{[2]}\right\rangle [n^{\rm{eq}}_k(m_\chi)] \ n^{\rm{eq}}_\chi(m_\chi)}\\
 {\displaystyle	 \qquad \left\langle\sigma_{\chi\bar\chi\rightarrow a a}v\right\rangle n^{\rm{eq}}_\chi(m_\chi)\,}
 	\end{array}
 	\Bigg\}  <&\, H(m_\chi) 
 	\qquad\quad \text{(DM out of equilibrium)} \label{eq:eqconditions} \; ,
 \end{align}
 	where, in the second line, the symbols in square brackets pertain to $2 \into 3$ interactions.
Depending on the relative size of the couplings $g_{aff}$ and $g_{a\chi\chi}$, DM generation by either 
$i j \to \chi \bar \chi [k]$ (where $i,j,k$ are SM particles) or $aa \to \chi \bar \chi $ scattering will be more efficient. This results in two distinct regimes which we label \emph{freeze-in from SM particles} (cf.~section~\ref{sec:FISM}) and \emph{freeze-in from ALPs} (cf.~section~\ref{sec:FIALP}). In these regimes, the DM relic abundance is given by eqs.~\eqref{eq:freezein} and \eqref{eq:freezeinfromALPs}, respectively. Since the relic density is proportional to the couplings entering the matrix element squared of the dominant process, it suffices to factorize the couplings and solve for those couplings satisfying the relic density constraint in eq.~\eqref{eq:relicDensity}. 

As discussed in section~\ref{sec:FISM}, ultraviolet-dominated contributions to the DM relic density from $a$-mediated $2\into 3$ scattering processes $f\bar f\into h\chi\bar\chi$, $fh\into f\chi\bar\chi$ and $\bar f h\into\bar f\chi\bar\chi$ become important for reheating temperatures above a few hundred GeV. These contributions introduce a dependence on the reheating temperature since the final relic abundance scales with $T_{RH}$, see eq.~\eqref{eq:maxTRH}. Here we consider two representative scenarios: For $T_{RH}= 200$~GeV, UV-dominated processes are practically negligible; the DM abundance will therefore be independent of the exact value of $T_{RH}$ if it is chosen even lower (but above the temperature of top quark freeze-out).  By contrast, for $T_{RH}\sim 2000$~GeV, the result is fully dominated by the UV-dominated processes. This behaviour can be observed in figure~\ref{fig:UVvsIR} where we compare the relative contribution of the IR- and UV-dominated processes to the final DM abundance as a function of the ALP mass in the freeze-in from SM scenario, i.e.~where $aa \to \chi \bar \chi$ scattering is negligible. We keep the ratio $m_\chi/m_a=10$ fixed. 
 Indeed, for $T_{RH}=200$~GeV, the DM abundance is set by $f\bar f \to \chi \bar \chi$ scattering, except for large ALP masses where the $2 \into 3$ interactions always dominate. By contrast, for reheating temperatures of the order of a few TeV, it is the ultraviolet contributions that set the final DM abundance.

In figure~\ref{fig:constraintsplusfreezein} we depict the results for these two scenarios in a large range of ALP masses, together with the constraints derived in the previous section.  We plot the value for the ALP-SM coupling $g_{aff}$ as a function of the ALP mass $m_a$ for lines of constant hidden sector coupling $g_{a\chi\chi}$ (dashed black lines). Inside the yellow region DM production from SM fermion scattering is more efficient (freeze-in from SM particles), and inside the yellow hatched region production from ALP scattering (freeze-in from ALPs). These freeze-in regimes span a vast region of parameter space in the $(m_a,g_{aff})$-plane. We cut the plot at the upper theoretical bound $C_f\lesssim 4\pi$ which results in $g_{aff}=C_f/f_a\lesssim 1$~GeV$^{-1}$ since $f_a\gg v$ with $v$ the electroweak scale, cf.~appendix~\ref{app:model}. Below the lower boundary the $\chi$-particles are produced via sequential freeze-in or decoupled freeze-out (DFO) because the ALPs and the visible sector cease to share the same temperature. 

The point where the ALPs equilibrate with the SM plasma is affected by $\mathcal{O}(1)$ finite-temperature corrections and by UV-dominated processes entering the equilibrium conditions eq.~\eqref{eq:eqconditions}, notably $f\bar f\into ah$. These can enhance ALP production at earlier times. As a consequence, the simple conditions in eq.~\eqref{eq:eqconditions} are only approximately true and in fact, equilibration can be attained for some $T\gtrsim m_\chi$. To obtain the lower boundary in figure~\ref{fig:constraintsplusfreezein}, we solve the Boltzmann equation for the ALP number density numerically and determine the couplings for which $n_a(T)=n_a^\text{eq}(T)$ for some $T_{RH}\gtrsim T\gtrsim m_\chi$.

In general, the ALP-SM interaction strength $g_{aff}$ must be comparably large if thermal equilibrium is to be reached, and is therefore likely in reach of upcoming experiments. For $g_{aff}  \gtrsim 10^{-6}-10^{-5}$~GeV$^{-1}$ and $m_a$ below a few GeV, the parameter space is ruled out by constraints from rare $B$ and $K$ decays and the electron beam dump experiment at SLAC. For masses below $\sim 0.2$~GeV, the remaining parameter space is covered by the SN constraint, see also figure~\ref{fig:constraints}. However, we remark that the relic density constraint inhibits large hidden sector couplings: $g_{a\chi\chi}$ has to be as small as $\sim 10^{-15}$~GeV$^{-1}$ for the largest connector couplings and for large ALP masses as in this parameter region DM production is controlled by $f\bar f \to \chi \bar \chi$ + ($2 \into 3$ terms) and therefore by the product of couplings $g_{aff}\cdot g_{a\chi\chi}$ (see table~\ref{tab:processes+couplings}). 

When the DM is dominantly produced by freeze-in from ALPs, we expect the contours in constant $g_{a\chi\chi}$ in the $(m_a,g_{aff})$-plane to be vertical because the relic density is independent of the connector coupling and proportional to $(g_{a\chi\chi} m_\chi)^4$. In the freeze-in from SM scenario the relic density is both proportional to $m_\chi^2\propto m_a^2$ and to $g_{aff}^2$. For $T_{RH}=200$~GeV and DM masses above the mass of the top quark, this dependence is weakened by the Boltzmann suppression of the particles predominantly producing the DM via $t\bar t\to\chi\bar\chi$, leading to an upwards bending of lines of constant hidden sector coupling. Here, interactions have to become stronger; between reheating and freeze-out of the top there is barely enough time to produce the heavy DM particle. 
For the ultraviolet-dominated case, the lines of constant $g_{a\chi\chi}$ get shifted towards smaller values of $g_{aff}$ since more DM is produced at earlier times. Not surprisingly, the freeze-in from the mediator scenario is independent of the two reheating temperatures since $aa \to \chi \bar \chi$ is infrared dominated and the ALPs belong to the SM bath.
We would like to add that, depending on the temperature, finite-temperature corrections can influence the DM production from fermion scattering by up to $\mathcal{O}(100\%)$, resulting in a change in the final relic density of $\sim 10\%$ for $T_{RH}=200$~GeV and $\sim 5\%$ for $T_{RH}=2000$~GeV.

Note that, in principle, ALPs with masses $m_a\lesssim$~MeV which are still relativistic during big bang nucleosynthesis are severely constrained by cosmology, see section~\ref{sec:constraints}. We do not include these bounds in our analysis since the relevant parameter space is already excluded by collider experiments or astrophysics. For larger masses, lifetimes above $\tau\sim 0.01$~s can be excluded~\cite{Millea:2015qra}; yet in the freeze-in regime the ALP-fermion coupling strength is large such that lifetimes are $< 0.01$~s.

In summary, we find that for feebly interacting dark matter, which has tiny couplings to the mediator, the correct relic density can be obtained in a large region of unexcluded parameter space, likely within the reach of future experiments.

\subsection{DFO region and cosmological constraints}\label{sec:DFOresults}
As the hidden sector coupling $g_{a\chi\chi}$ increases, the interactions among the ALPs and the DM become frequent enough to  bring them in thermal equilibrium, however at a temperature $\Tp$ which is distinct from that of the photons, $T$. These now constitute a dark sector which is thermally decoupled from the visible sector. Energy is gradually transferred to the hidden sector by rare scatterings of SM particles into both the ALPs and the DM. The DM relic density is set by $aa\leftrightarrow \chi\bar \chi$ interactions and the mechanism resembles ordinary freeze-out but occurs at a different temperature, i.e.~by \emph{decoupled freeze-out (DFO)}. To obtain the relic density in the DFO regime we proceed as outlined in section~\ref{sec:reannihilation} and section~\ref{sec:measuredRelicDensity}. The calculation is more involved than for freeze-in, since now a stiff system of coupled differential equations \eqref{eq:Tprime}, \eqref{eq:BERchi} and \eqref{eq:BERa} has to be solved, which we have accomplished using the stiff solver package dvode \cite{dvode}. We track the evolution of $Y_\chi\equiv n_\chi/s$ until the DM particles freeze out and $Y$ stays constant (more precisely, we require the relative change in $Y_\chi$ to be smaller than $\varepsilon=5\cdot 10^{-4}$).
\begin{figure}[t]
\begin{center}
	\hspace*{-0.6em}\includegraphics[scale=0.77]{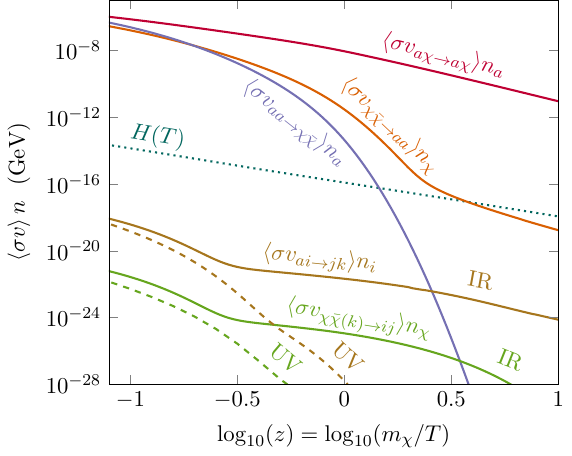}\includegraphics[scale=0.77]{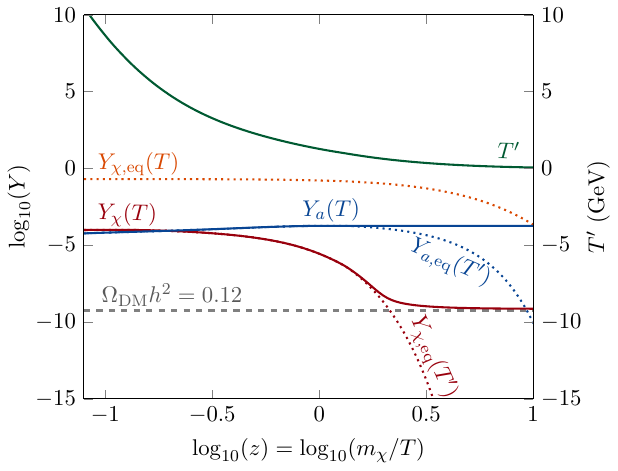}
	\captionof{figure}{Illustrative example of the evolution of the number densities and the reaction rates in the DFO scenario for $m_\chi=10$~GeV and $m_a=1$~GeV and for a set of couplings which give the measured dark matter relic abundance: $g_{aff}=9.44\cdot 10^{-12}$~GeV$^{-1}$ and $g_{a\chi\chi}=1.06\cdot 10^{-2}$~GeV$^{-1}$. \emph{Left:} Evolution of the different terms competing on the right hand side of the Boltzmann equations for $n_\chi$ and $n_a$ as a function of $\log_{10}(z)=\log_{10}(m_\chi/T)$. We also plot the evolution of the reaction rate for elastic scattering among the hidden sector particles, $a\chi \to a \chi$. \emph{Right:} The evolution of the comoving number density of $\chi$ (red) and $a$ (blue) as a function of $\log_{10}(z)=\log_{10}(m_\chi/T)$. The equilibrium distributions $Y_{\chi,\rm{eq}}(\Tp)$ (dotted red) and $Y_{a,\rm{eq}}(\Tp)$ (dotted blue) and $Y_{\chi,\rm{eq}}(T)$ (dotted orange) are also shown for reference. We plot along the evolution of the hidden sector temperature $\Tp$ (green).}\label{fig:evolutionDFO}
\end{center}
\end{figure}
As in the case of freeze-in, here too, UV contributions to SM-DM and SM-ALP interactions can introduce a dependence on the reheating temperature. However, these only enter indirectly via the energy transfer from the SM. We have checked this explicitly and find that for reheating temperatures below $\mathcal{O}$(TeV)  they do not significantly influence the final DM relic abundance. To be more specific, we find that a variation of $T_{RH}=200$~GeV and $T_{RH}=2000$~GeV leads to a change of $\sim 22 \%$ in the final DM relic abundance and a change $\sim 7 \%$ in the required value of $g_{a\chi\chi}$. Hence, the DFO scenario retains very little sensitivity to the ultraviolet for reheating temperatures below a few TeV, and in the following we can neglect these contributions. However, we do add finite-temperature corrections to both ALP and DM production from the SM, which result in $\mathcal{O}(1)$ corrections, mainly originating from thermal mass effects.

In the left panel of figure~\ref{fig:evolutionDFO} we present an illustrative example of the evolution of the various reaction rates at play in the DFO regime, together with the Hubble rate, for $m_\chi=10$~GeV and $m_a=1$~GeV and for a set of couplings which give the measured dark matter relic abundance ($g_{aff}=9.44\cdot 10^{-12}$~GeV$^{-1}$ and $g_{a\chi\chi}=1.06\cdot 10^{-2}$~GeV$^{-1}$). For reactions faster than the Hubble rate, i.e.~lines above the turquoise dotted Hubble line, the particles involved achieve equilibrium. Once the reaction rates drop below the Hubble rate, the particles become so dilute that they no longer collide sufficiently often and the process shuts off. We also distinguish between the infrared- (solid) and ultraviolet- (dashed) dominated SM-ALP and SM-DM interactions. A few observations can be made: Firstly, $aa\leftrightarrow \chi \bar{\chi}$ interactions are the dominant DM number changing interactions throughout the evolution. The mechanism which sets the relic abundance is therefore conceptually very similar to the simpler freeze-out scenario, however governed by $\Tp$ rather than $T$. Secondly, neither the SM-ALP nor the SM-DM reaction rates ever exceed the Hubble rate, and the hidden sector is indeed decoupled. Nevertheless, the SM slowly transfers energy to the hidden sector both via SM-ALP and SM-DM interactions, with the dominant contribution coming from SM-ALP interactions. Lastly, elastic scattering $a\chi (\bar \chi) \leftrightarrow a \chi (\bar \chi)$ stays active until after freeze-out of the $\chi$-particles, as assumed in section~\ref{sec:reannihilation}. For the same parameters, in the right panel of figure~\ref{fig:evolutionDFO} the evolution of the comoving number densities of the DM and the ALP are shown, as well as the hidden sector temperature $\Tp$.  We also plot the equilibrium distributions $Y_{\chi,\rm{eq}}(\Tp)$ and $Y_{a,\rm{eq}}(\Tp)$ as dotted lines, and, for comparison, $Y_{\chi,\rm{eq}}(T)$. As in the simpler thermal freeze-out scenario, $Y_a$ and $Y_\chi$ follow the equilibrium distributions, $Y_{a,\rm{eq}}(\Tp)$ and $Y_{\chi,\rm{eq}}(\Tp)$, respectively. For rapid interactions the right hand side of the Boltzmann equation vanishes and $Y_\chi$ stays constant.\footnote{Due to the relation between $T$ and $T'$, affected by the rate of energy transfer to the hidden sector, $Y_\chi$ and $Y_a$ ostensibly increase in the plot; this is because the plot shows the evolution of the comoving number densities as a function of the SM temperature $T$.} The ALPs and the DM separate from the equilibrium distributions at the same time, around $\Tp\simeq m_\chi/10$, since no other particles are around to maintain frequent number changing interactions. Since $\Tp <T$ the inverse SM freeze-out temperature can be as small as $z\sim 2-3$.

\begin{figure}[t]
\begin{center}
	\includegraphics[scale=1]{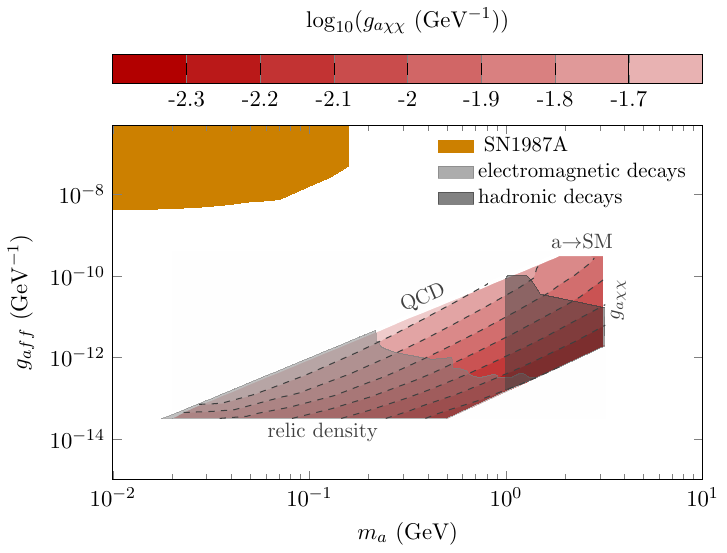}
		\captionof{figure}{Contour plot of the hidden-sector couplings $g_{a\chi\chi}$ in the $(m_a,g_{aff})$-plane which give the observed DM relic density via decoupled freeze-out (DFO). We fixed the ratio $m_\chi/m_a=10$. We have included the relevant constraints on our ALP model on the connector coupling $g_{aff}$ in this parameter region (cf.~figure~\ref{fig:constraints}). We additionally plot the region excluded by cosmological constraints of additional particles decaying electromagnetically (light grey) and decaying hadronically (dark grey). For explanations about the boundaries of the DFO region and the constraints from electromagnetic and hadronic decays see the main text.}\label{fig:constraints_DFO}
\end{center}
\end{figure}

In figure~\ref{fig:constraints_DFO} we show the region of parameter space where the correct relic density is obtained for a large range of masses and couplings, again keeping the ratio $m_\chi/m_a=10$ fixed, together with the constraints derived in the previous section. We have checked that the observations made in the previous paragraph hold for the whole set of parameters depicted in figure~\ref{fig:constraints_DFO}. The region in the $(m_a,g_{aff})$-plane is shaped by a number of conditions:
\begin{itemize}
\item \textbf{Relic density} For connector couplings, $g_{aff}$, which are too small, the hidden sector does not become sufficiently populated. Although $g_{a\chi\chi}$ might be large enough to establish equilibrium between the hidden sector particles, $n_\chi^{\rm{eq}}(\Tp)$ can never reach the amount of DM density observed today. This is indicated by the lower boundary.
\item \textbf{\textit{a} $\leftrightarrow$ SM} On the other hand, if the connector coupling $g_{aff}$ is too large, the interactions between the hidden sector and the SM become strong enough to establish thermal equilibrium. Depending on the hidden sector coupling, the DM is then either produced by freeze-in (see section~\ref{sec:FIresults}) or thermal freeze-out. We remark that the numerical solution close to the transition between the freeze-in and the DFO regime is challenging and we chose this upper boundary conservatively.
\item \textbf{\textit{g}$_{a\chi\chi}$} The DM-mediator interaction is (cf.~eq.~\eqref{eq:lagrangian})
\begin{align}
C_\chi \frac{m_\chi}{f_a} a\bar \chi \gamma_5 \chi \equiv g_{a\chi\chi} m_\chi a \bar{\chi} \gamma_5 \chi\,.
\end{align}
Our effective theory is valid only below the scale $f_a$. Thus, the reheating temperature $T_{RH}$ should be below this scale to safely ignore UV contributions. On the other hand, $T_{RH}$ has to be higher than $m_\chi$. We consequently need a hierarchy $f_a \gtrsim T_{RH} \gtrsim m_\chi$, i.e.~small $g_{a\chi\chi}=C_\chi/f_a$, and this is the reason why the DM (the ALP for a fixed mass ratio) should not be too heavy. 
\item \textbf{QCD} Finally, we employ a perturbative description of the strong interactions, only convergent at high enough energies. The DM abundance should therefore be set by interactions happening at temperatures before the QCD phase transition. In practice, we set the upper boundary labelled ``QCD'' by requiring that the $\chi$-particles freeze out at temperatures above the threshold $T_\text{pert}=600$~MeV.
\end{itemize}
The shape of the contours of constant $g_{a\chi\chi}$ can be understood as follows. By increasing the strength of the ALP-fermion coupling, more energy will be transferred to the hidden sector and heat up the hidden sector particles. This will increase the amount of DM particles, $n^{\rm{eq}}_\chi(\Tp)$. In order to satisfy the relic density constraint~\eqref{eq:relicDensity}, the interaction strength among the hidden sector particles has to increase to keep $\chi\bar\chi \rightarrow aa $ active for longer, such that the $\chi$-particles can continue to annihilate. The DM-ALP interaction is proportional to the product $(g_{a\chi\chi}\cdot m_\chi)^4$. For increasing $m_a$, $g_{a\chi\chi}$ has to decrease to compensate the effect of increasing $m_\chi$. We find that in this scenario $g_{aff}$ enters only via SM $\to a$ processes because the energy transfer and DM production from the connector process $f \bar f \rightarrow \chi \bar \chi$ is always subdominant. The dependence on $g_{aff}$ is therefore very mild. Furthermore, we observe that $g_{a\chi\chi}$ has to be large compared to the freeze-in scenario to maintain equilibrium among the hidden sector particles. 

It turns out that it is extremely challenging to test the DFO region with collider searches for ALPs: in this scenario the visible sector and the mediator barely talk to each other, such that the sensitivity of experiments probing the relevant mass range, in particular the SN bound and the bounds from rare $B$ and $K$ decays, would have to improve by several orders of magnitude. However, the tiny coupling between the SM particles and the ALP makes the ALP relatively long lived, and since the ALPs are abundantly produced along with the DM particles (cf.~figure~\ref{fig:evolutionDFO}) their decay can have important implications for the cosmological history during and after big bang nucleosynthesis, as briefly discussed in section~\ref{sec:constraints}. Of course, the imprint ALPs will leave on cosmological observables strongly depends on its decay products and its lifetime. In the range of ALP masses we consider, the dominant decay channel varies, see figure~\ref{fig:Branchingfraction}. For ALPs with masses $m_a\lesssim 2m_\mu$, i.e.~ALPs which dominantly decay into photons and $e^+e^-$, the constraints are very similar. Here, we apply the bounds from ref.~\cite{Kawasaki:2020qxm} on very long-lived ALPs in the sub-GeV mass range. Generically speaking, they exclude sufficiently abundant particles decaying electromagnetically with lifetimes $\tau_a\sim 10^3-10^5$~s. The bound labelled ``electromagnetic decays'' in figure~\ref{fig:constraints_DFO} was obtained by applying the bounds on the ALP's lifetime from figures~4 and~5 in ref.~\cite{Kawasaki:2020qxm}, interpolating between the mass of the decaying particle and the dominant decay channel of the ALP. For masses in the range $2 m_\mu \lesssim m_a \lesssim 1$~GeV, the ALP dominantly decays into muons. In principle, the applicable constraints are the electromagnetic ones here -- see also the discussion in ref.~\cite{cosmobounds} -- and the lifetime is short enough for them not to matter. For hadronic decays the bounds become more severe, cf.~section~\ref{sec:constraints}, and lifetimes above $\tau \sim 0.1$~s can be excluded. In ref.~\cite{Kawasaki:2017bqm} bounds are provided on hadronically decaying particles with masses in the GeV-TeV range. The smallest mass for which results are available is 30 GeV. We apply the corresponding bound to our model, extrapolating from the given shapes that the bounds will remain approximately constant for lower masses.
We have checked that, taking into account the branching ratios of the various decay channels, the energy injection from ALPs is sufficient in the DFO region for these constraints to apply. However, as outlined in section~\ref{sec:constraints}, photo-dissociation is clearly not the only possible scenario. For instance, ALPs decaying into photons can re-equilibrate with the SM, a scenario which was studied in depth in ref.~\cite{Millea:2015qra}, excluding much shorter lifetimes. However, as studied in ref.~\cite{Depta:2020zbh}, since the temperature of our hidden sector $\Tp$ is in general well below $T$, we expect these bounds to be alleviated in our case. In order to obtain precise cosmological bounds a detailed analysis of the evolution of the ALP number density and its imprint on cosmological observables would hence be necessary, yet beyond the scope of this paper.

We conclude this section by remarking that, as mentioned above, the upper left boundary labelled ``QCD'' arises due to a theoretical limitation. It might be interesting to describe the processes involved at low energies using non-perturbative techniques in order to explore the remaining parameter space. In doing so, we should be able to close the gap between the ``QCD'' line and ALP-SM equilibrium condition in figure~\ref{fig:constraintsplusfreezein}.

\section{Conclusion}
\label{sec:conclusion}
To conclude, we study a fermionic DM model with a pseudoscalar (ALP) mediator coupled to SM and DM fermions at tree level. 
In this model, different production mechanisms can set the relic abundance, depending on which interactions are most important.
In fact, we find that, similar to the case of a fermionic DM model with a massive dark photon mediator studied in ref.~\cite{Hambye:2019dwd}, six different production mechanisms are possible for the set of ALP and DM masses we have considered, namely freeze-in from SM particles, freeze-in from mediators, sequential freeze-in, decoupled freeze-out (DFO), freeze-out from mediators and freeze-out from the SM.
We numerically solve the general set of coupled Boltzmann equations for a range of ALP-fermion and ALP-DM couplings to obtain the phase diagram for a fixed ALP and DM mass ($m_a=1$~GeV, $m_\chi=10$~GeV).
We focus on two cases where the hidden sector, comprised of the DM and ALP, does not thermalise with the visible sector: freeze-in (from the SM or from ALPs) and DFO.

In the freeze-in and DFO regimes, we determine the relic density by solving the relevant Boltzmann equations.
This is most challenging in the DFO region, as discussed in sec.~\ref{sec:reannihilation}, where a set of three coupled differential equations must be solved (the Boltzmann equations for the ALP and DM number densities and the Boltzmann equation governing the energy transfer from the SM to the dark sector), and a variety of processes contribute to ALP and dark matter production.
We derive for the first time the energy transfer Boltzmann equation for the case where there is a single hidden sector particle in the final state, and present a new strategy for determining the hidden sector temperature from the evolution of the energy density.
Throughout our work we take into account thermal effects by implementing temperature-dependent corrections to the SM masses and couplings, as these can alter the results at the $\mathcal{O}(1)$ level, in particular in the case of the production of ALPs. The dominant contribution stems from the increased masses of the in medium SM particles. 

In a model where the ALP-fermion couplings arise from an extended Higgs sector in the UV, both infrared-dominated and ultraviolet-dominated processes contribute to freeze-in, and either of them may dominate depending on the hierarchy of the reheating temperature and the electroweak scale. 
In particular, we find that UV-dominated $2 \into 3$ interactions play a role for the production via freeze-in from the SM for reheating temperatures $T_{RH}\gtrsim 200$~GeV. We therefore choose to study one typical infrared-dominated ($T_{RH}= 200$~GeV) and one ultraviolet-dominated scenario ($T_{RH}= 2$~TeV). 
In the DFO region all ultraviolet-dominated processes enter only at the subleading level via the energy transfer to the hidden sector, such that the final DM abundance is largely insensitive to the ultraviolet contributions for reheating temperatures below a few TeV, and these are therefore neglected.

We also pay particular attention to calculating the experimental bounds on the relevant parameter space, including collider, astrophysics, cosmology and DM constraints.
Some of the most powerful constraints come from collider experiments, particularly beam dump constraints coming from SLAC E137 and flavour constraints notably from NA62 and LHCb, together constraining masses up to the $B$ meson mass and the $g_{aff}$ coupling down to $\mathcal{O}(10^{-6}~\mathrm{GeV}^{-1})$.
For the beam dump experiment we improve previous analyses by including all dominant ALP production processes, namely the Primakoff process, bremsstrahlung and (non-)resonant positron annihilation. 
In particular, we show that positron annihilation can become the dominant process for larger ALP masses in models where the ALP coupling to photons is highly suppressed and in this case clearly cannot be neglected.
For the scenario considered, where the only tree-level couplings of the ALP are to fermions, we added the different production scenarios since they cannot be distinguished by experiment.
Our analysis excludes a large part of the parameter space in the $(m_a,g_{aff})$-plane and puts upper and lower bounds on the coupling.
For the flavour constraints, we implement the state-of-the-art bounds for the most constraining channels from NA62, NA48/2, LHCb and CHARM, which together rule out a large section of parameter space, down to $g_{aff}\sim\mathcal{O}(10^{-5}$ GeV$^{-1})$ for masses up to $m_a\sim m_{K^+}-m_{\pi^+}$, and in the range $g_{aff}\sim\mathcal{O}(10^{-4}$~GeV$^{-1})$ to $\mathcal{O}(10^{-2}$~GeV$^{-1})$ up to $m_a\sim m_{B}-m_{K^{(*)}}$. 

Out of all the constraints considered, astrophysics provides the most stringent bounds on our parameter space.
In particular SN1987A and HB stars strongly probe the parameter space for $m_a\lesssim\mathcal{O}$(0.1~GeV), extending down to $g_{aff}\sim\mathcal{O}(10^{-9}$~GeV$^{-1})$.
For the former constraint, we adapt the state-of-the-art analysis~\cite{Chang:2016ntp, Chang:2018rso} to the scenario we consider, including N$^3$LO corrections in chiral perturbation theory.
While we do consider DM direct detection, indirect detection and self-interactions, due to the pseudoscalar nature of the mediator, as expected, none of these provide relevant constraints.

 The tiny coupling between the SM and the mediator which is required to reproduce the observed DM relic density in the DFO region is responsible for making the new sector so difficult to probe. It therefore seems that current and near future experiments cannot access this region of parameter space, yet cosmology can provide a powerful tool to constrain secluded dark sectors. Here we carry out an estimate of the consequences that the presence of ALPs would have on cosmology for the DFO region: Applying the constraints from refs.~\cite{Kawasaki:2020qxm} and \cite{Kawasaki:2017bqm}, we exclude very long lived ($\tau_a\gtrsim 10^3-10^5$~s) electromagnetically decaying ALPs and short lived ($\tau_a\gtrsim 0.1$~s) hadronically decaying ALPs as their decay products would photodissociate light elements.

In sections~\ref{sec:FIresults} and \ref{sec:DFOresults} we present the regions of parameter space in the $(m_a,g_{aff})$-plane where the correct relic density is obtained, with contour lines showing the required value of $g_{a\chi\chi}$.
For freeze-in, as seen in figure~\ref{fig:constraintsplusfreezein}, we find a large unexcluded region, extending to $g_{aff}\sim 10^{-9}~\mathrm{GeV}^{-1}$ for $10^{-4}$~GeV~$<m_a<10^2$ GeV. 
For larger values of $g_{aff}$ and $m_a$, smaller values of $g_{a\chi\chi}$ are required.
Probing this region, for the largest values of $g_{a\chi\chi}$, would require the flavour constraints to improve substantially, or  low-mass resonance searches at ATLAS or CMS.
For the DFO regime, as seen in figure~\ref{fig:constraints_DFO}, the correct density is obtained for relatively low values of $g_{aff}\sim 10^{-13} - 10^{-10}~\mathrm{GeV}^{-1}$ and large values of $g_{a\chi\chi}$ ($\sim 10^{-2}~\mathrm{GeV}^{-1}$).
Again it appears that cosmology offers the only means of probing such small values of $g_{aff}$. Moreover, it turns out the DFO region is disfavoured by the standard BBN scenario.

There are several possible improvements to our analysis that could be envisaged. As discussed in Sec.~\ref{sec:seqFI}, a thorough study of the sequential freeze-in region requires solving the unintegrated Boltzmann equations. 
It would also be interesting to compare our results in the DFO region to those obtained using the unintegrated Boltzmann equations. Furthermore, it might be worthwhile using non-perturbative methods to investigate the DFO mechanism in the region above the QCD boundary.
In this work we focus on a Dirac fermion dark matter particle, but, as previously mentioned, a Majorana fermion is an equally well-motivated possibility which could merit a detailed study, although one should expect qualitatively similar results.
Finally, the potential sensitivity of future experiments is of utmost interest, and in this view it should be worthwhile  to assess the sensitivity of planned future flavour experiments and upgrades, beam dump experiments and cosmology measurements. 

\acknowledgments
The authors thank Rupert Coy,  Sacha Davidson, Fatih Ertas, Rouven Essig, Thomas Hambye, Felix Kahlh\"ofer, Sam McDermott, Vivian Poulin, Christopher Smith, Michel Tytgat and Wei Xue for useful discussions and correspondence. Further we thank Babette D\"obrich for providing the CHARM constraint in numeric form, and Florian Domingo for sharing with us his code for the decay width of pseudoscalars in the NMSSM.
The project leading to this publication has received funding from Excellence Initiative of Aix-Marseille University -- A*MIDEX, a French ``Investissements d'Avenir'' programme (AMX-19-IET-008 - IPhU and ANR-11-IDEX-0001-02), and from the OCEVU Labex (ANR-11-LABX-0060).

\appendix

\section{Effective ALP-DM Lagrangian from a two Higgs doublet model}\label{app:model}

The effective Lagrangian eq.~\eqref{eq:lagrangian} can be UV completed by a renormalizable model containing additional Higgs doublets and singlets, with the flavour-diagonal couplings of the ALP being a consequence of the Higgs coupling structure. Here we present a simple example for pedagogical purposes.

Our starting point is the Lagrangian of a type-I two Higgs doublet model with an extra scalar singlet,
 \begin{align}\hspace{-.7cm}
\nonumber{\cal L}=&\;{\cal L}_{\rm kin}-\left(y^d_{ij} \bar Q_{Li} \Phi_2 D_{Rj}+y^e_{ij} \bar L_{Li} \Phi_2 E_R+y^u_{ij} \bar Q_{Li}\widetilde\Phi_2 U_{Rj}\hc\right)-\tilde m_1^2\,|\Phi_1|^2\\
\nonumber &-\tilde m_2^2\,|\Phi_2|^2+M^2\,|\phi|^2-\frac{\tilde\lambda_1}{2}|\Phi_1|^4-\frac{\tilde\lambda_2}{2}|\Phi_2|^4-\frac{\lambda_\phi}{2}|\phi|^4- \tilde\lambda_3\,|\Phi_1|^2\,|\Phi_2|^2\\
 &-\tilde\lambda_4\,(\Phi_1^\dag \Phi_2)(\Phi_2^\dag \Phi_1)-\kappa_1\,|\Phi_1|^2\,|\phi|^2-\kappa_2\,|\Phi_2|^2|\phi|^2-\tilde\lambda_{12}\left(\Phi_1^\dag \Phi_2 \phi^2\hc\right)\,.
 \end{align}
 Here ${\cal L}_{\rm kin}$ contains all the gauge-kinetic terms for the SM gauge and fermion fields as well as for the scalars $\Phi_1$, $\Phi_2$ and $\phi$. $\Phi_1$ and $\Phi_2$ are two Higgs doublets with the gauge quantum numbers of the SM Higgs, and $\phi$ is a complex scalar which is neutral under the SM. This is the most general renormalizable Lagrangian allowed by a $\U{1}_{\rm PQ}$ global symmetry under which $\Phi_2$ and the SM fermions are neutral, while $\Phi_1$ and $\phi$ are charged. Note that this symmetry is not a PQ symmetry in the stricter sense of being anomalous with respect to QCD, and therefore unsuitable for solving the strong CP problem.

 For $M^2>0$, $\phi$ will take a VEV $\vev{\phi}=f_a/\sqrt{2}$, which can be taken real and positive. We define real scalar fields $\sigma$ and $a$ by
 \begin{align}
 \phi(x)=\frac{1}{\sqrt{2}}\left(f_a+\sigma(x)\right)e^{i\frac{a(x)}{f_a}}\,.
 \end{align}
 We anticipate that $\Phi_1$ and $\Phi_2$ will also take vacuum expectation values of the order $v\ll f_a$. This parameterization is useful to extract leading-order effects in $v/f_a$; to leading order we have $f_a=\sqrt{\frac{2}{\lambda_\phi}}M$. The fields $\sigma$ and $a$ are approximate mass eigenstates, up to mixing with the $\Phi$-like fields with mixing angles of order $v/f_a$.
 
 At low energies, $\sigma$ (whose VEV is zero to leading order) can be integrated out. The effective $\Phi_{1,2}$ mass parameters and quartic couplings at low energies will be modified according to suitable matching conditions from replacing $|\phi|^2\into f_a^2/2$ in the above Lagrangian and from four-point interactions with $\sigma$ exchange, such that $\tilde m_i^2\into m_i^2$ and $\tilde\lambda_k\into\lambda_k$. Up to higher-dimensional operators, the effective Lagrangian is now that of a type-I two-Higgs doublet model with an additional field $a(x)$,
 \begin{align}\nonumber
 {\cal L}=&\;{\cal L}_{\rm kin}-\left(y^d_{ij} \bar Q_{Li} \Phi_2 D_{Rj}+y^e_{ij} \bar L_{Li} \Phi_2 E_R+y^u_{ij} \bar Q_{Li}\widetilde\Phi_2 U_{Rj}\hc\right)-m_1^2\,|\Phi_1|^2\\
 &\nonumber- m_2^2\,|\Phi_2|^2 -\frac{\lambda_1}{2}|\Phi_1|^4-\frac{\lambda_2}{2}|\Phi_2|^4-\lambda_3\,|\Phi_1|^2\,|\Phi_2|^2\\
 &-\lambda_4\,(\Phi_1^\dag \Phi_2)(\Phi_2^\dag\Phi_1)-\frac{\lambda_{12}\,f_a^2}{2}\left(\Phi_1^\dag \Phi_2\,e^{2ia/f_a}\hc\right)\,.
 \end{align}
 With the $a$-dependent field redefinition $\Phi_1(x)\into e^{2ia(x)/f_a}\Phi_1(x)$,
 $a$ disappears from the potential. However, the $\Phi_1$ kinetic term is not invariant,
 \begin{align}
     \nonumber\left(D_\mu \Phi_1\right)^\dag D^\mu \Phi_1\into \left(D_\mu \Phi_1\right)^\dag D^\mu \Phi_1-2i\frac{\partial_\mu a}{f_a}\,\Phi_1^\dag\overleftrightarrow D^\mu \Phi_1+{\cal O}\left(\frac{a^2}{f_a^2}\right)\,.
 \end{align}
 
The ``Higgs basis'' in the space of $\Phi_1$ and $\Phi_2$ is defined by rotating $(\Phi_1, \Phi_2)\into (H, \Phi)$ by an angle $\beta$ such that the Higgs VEV is contained in one doublet $H$ only, $\tan\beta=\vev{\Phi_2}/\vev{\Phi_1}$. In general, $\beta$ is independent of the mixing angle $\alpha$ which parameterizes the mixing between the mass eigenstates. However, in the decoupling limit, these two angles are aligned and the fields $H$ and $\Phi$ contain the mass eigenstates as
 \begin{align}
 H(x)=\left(\begin{array}{c} G^+(x) \\ \frac{v+h(x)+i G^0(x)}{\sqrt{2}}\end{array}\right)\,,\qquad \Phi(x)=\left(\begin{array}{c} H^+(x) \\ \frac{H^0(x)+i A^0(x)}{\sqrt{2}}\end{array}\right)\,.
 \end{align}
 We assume that the $\Phi$-like states are also heavy, so they can be integrated out as well. (This step can in principle be interchanged with integrating out $\sigma$ or performed simultaneously, depending on the mass hierarchies.) The effective Lagrangian becomes
 \begin{align}\label{eq:EFTL}
 \nonumber
 {\cal L}=&\;{\cal L}_{\rm kin}-\Big(Y^d_{ij}\bar Q_{Li} H D_{Rj}+Y^e_{ij} \bar L_{Li} H E_{Rj}+Y^u_{ij} \bar Q_{Li}\widetilde H U_{Rj}\hc \Big)\\&\;+m^2\,|H|^2-\frac{\lambda}{2}|H|^4
 -2i\frac{\partial_\mu a}{f_a}\,\cos^2\beta\,H^\dag\overleftrightarrow D^\mu H
 \end{align}
 where $Y^{u,d,e}=y^{u,d,e}\sin\beta$.
The $(\partial a)\,H^\dag\overleftrightarrow D H$ term would cause $a$ to mix with the $Z$ boson after electroweak symmetry breaking, which is avoided by another field redefinition:
\begin{align}
H(x)\into H(x)e^{-2i\cos^2\beta\,a(x)/f_a}\,.
\end{align}
The Yukawa terms will shift accordingly, and one obtains for the final axion couplings in this field basis at leading order in $1/f_a$
\begin{align}\label{eq:Leff}
\nonumber\hspace{-.8cm}
{\cal L}=&\;{\cal L}_{\rm kin}+m^2\,|H|^2-\frac{\lambda}{2}|H|^4-\Big(Y^d_{ij}\;\bar Q_{Li} H D_{Rj}+Y^e_{ij}\;\bar L_{Li} H E_{Rj}+Y^u_{ij}\; \bar Q_{Li}\widetilde H U_{Rj}\hc\Big)\\
 &-2\cos^2\beta\frac{a}{f_a}\Big(i\,Y^d_{ij}\;\bar Q_{Li} H D_{Rj}+i\,Y^e_{ij}\;\bar L_{Li} H E_{Rj}-i\,Y^u_{ij}\; \bar Q_{Li}\widetilde H U_{Rj}\hc\Big) \; .
\end{align}
The second line corresponds to the ALP-fermion couplings in eq.~\eqref{eq:lagrangian} with $C_{d,l}=-C_u=2\cos^2\beta$, after switching to four-spinor notation. These are, for our purposes, flavour-universal couplings (the minus sign in the up-type quark coupling is immaterial for the processes we consider). They can be regarded as PQ charges of a suitably redefined $\U{1}_{\rm PQ}$ under which $\Phi_2$ carries charge $2\cos^2\beta$.

Other models where an ALP is obtained by more general extensions of the Higgs sector will induce different axion-fermion couplings. For example, choosing a type-II two-Higgs doublet model (as in the standard DFSZ model for QCD axions \cite{Dine:1981rt}) would yield different couplings for up-type and down-type quarks and leptons, unless $\tan\beta=1$.

It is possible to transform the dimension-5 couplings in eq.~\eqref{eq:Leff} into $\partial_\mu a\, j_{\rm PQ}^\mu$ terms, where $j_{\rm PQ}^\mu$ denotes the fermionic $\U{1}_{\rm PQ}$ current. In models where $\U{1}_{\rm PQ}$ is anomalous, this will induce also $aF\widetilde F$ couplings. The $\partial_\mu a\, j_{\rm PQ}^\mu$ are usually eliminated again after electroweak symmetry breaking, using the equations of motion to transform them into the Yukawa-like couplings of eq.~\eqref{eq:lagrangian}, which involves precisely the same anomaly term. We emphasize that additional $aF\widetilde F$ terms are never induced at high scales, unless there are heavy fermions with charges under both PQ and the Standard Model (such as in the KSVZ model for a QCD axion).

To obtain our dark matter model, we now add two more features.
The first is a mass term for $a$ which explicitly breaks the PQ symmetry; we assume $m_a\ll f_a$ and write
\begin{align}
\Delta{\cal L}_{\rm mass}= \frac{m_a^2}{4}\left(\phi-\phi^*\right)^2\,.
\end{align}

The second is the dark matter candidate $\chi$, for which there are several possibilities. In this paper we focus on a Dirac dark matter candidate $\chi=\chi_L+\chi_R$ with PQ charges allowing for a coupling
\begin{align}
\Delta{\cal L}_\chi= -\left(y_\chi\,\phi\overline\chi_L\chi_R\hc\right)
\end{align}
which, after PQ breaking, gives rise to a Dirac mass $m_\chi=y_\chi f_a/\sqrt{2}$ and an axionic coupling:
\begin{align}
\Delta{\cal L}_{\rm eff}=-m_\chi\;\overline\chi\chi-i\,\frac{m_\chi}{f_a}\;a\,\overline\chi\gamma^5\chi\,.
\end{align}
In this simple model, the reduced ALP-fermion couplings $g_{aff}$ tend to be of the same order of magnitude as the reduced ALP-DM couplings $g_{a\chi\chi}$, unless $\cos\beta\ll 1$. As we show in the main text, hierarchically different couplings can lead to interesting effects. They are well-motivated in UV models with multiple axions and a clockwork-like structure for the couplings, leading to exponentially different effective decay constants for different sectors of the theory \cite{Kaplan:2015fuy}. One might speculate, for instance, that the DM sector and the SM are localized on different lattice sites in theory space, giving rise to a large hierarchy $g_{aff}\ll g_{a\chi\chi}$.

Moreover, this model predicts $C_\chi=1$ and therefore $g_{a\chi\chi}=\frac{1}{f_a}$. The DFO phase as discussed in sections \ref{sec:reannihilation} and \ref{sec:DFOresults} will therefore not be realized in this simple scenario, since DFO requires significant connector couplings $g_{a\chi\chi}\sim (100\text{ GeV})^{-1}$. A mass scale $f_a\sim 100$ GeV is of course ruled out by observation. We emphasize nevertheless that there is no obstacle in principle to raising $C_\chi$, although doing so with the particle content of the present minimal model would require fine-tuning (by adding an additional explicitly PQ breaking term, namely a bare Dirac mass term for $\chi$ which could partially cancel the Dirac mass from spontaneous PQ breaking).

We finally remark that another possibility for the dark matter candidate would be a single Weyl dark matter candidate $\chi=\chi_L$, coupled to $\phi$ according to
\begin{align}
\Delta{\cal L}_\chi=-\frac{1}{2}\left(y_\chi\,\phi \chi_L^T C \chi_L\hc\right)\,.
\end{align}
This coupling translates into a Majorana mass after PQ breaking. A third possibility would be a model with two Weyl fermions and a see-saw-like structure where both a Dirac mass and a Majorana mass are allowed, with the dark matter phenomenology depending on the hierarchy between the two.
\section{Collision term for the energy transfer Boltzmann equation}\label{app:collision}
For a process $1\,2\to 3\,4$, neglecting the back-reaction, the integrated collision operator is given in eq.~\eqref{eq:coll1} in terms of the energy transfer rate $\mathcal{E}(\vec{p}_1,\vec{p}_2)$ for an initial state with momenta $\vec{p}_1$ and $\vec{p}_2$, and of the M\o ller velocity
\begin{equation}
v_\text{M\o l}=\frac{\sqrt{(s-(m_1+m_2)^2)(s-(m_1-m_2)^2)}}{2 E_1 E_2}\equiv \frac{F}{E_1E_2}.
\end{equation}
Assuming the momentum distributions to be Maxwell-Boltzmann-like, the collision term becomes
\begin{align}\label{eq:ETBE}
\int\frac{d^3 p}{(2\pi)^3}C[f]=&\,g_1 g_2 \int \frac{d^3p_1}{(2\pi)^3}\frac{d^3p_2}{(2\pi)^3}e^{-(E_1+E_2)/T}v_\text{M\o l}\,\mathcal{E}(\vec{p}_1,\vec{p}_2)\\
\nonumber=&\,\frac{g_1 g_2}{32 \pi^4} \int dE_+ dE_- ds\, e^{-E_+/T}F\,\mathcal{E}(\vec{p}_1,\vec{p}_2)
\end{align}
where, following ref.~\cite{cosmicabundances},
in the second line we have replaced $E_+=E_1+E_2$, $E_-=E_1-E_2$ and $s=(E_1+E_2)^2-(\vec{p}_1+\vec{p}_2)^2$. 
The energy transfer rate $\mathcal{E}(\vec{p}_1,\vec{p}_2)$ can be written as in eq.~\eqref{eq:Entransfer}. For the transferred energy $\Delta E_{tr}$, there are two cases which we need to distinguish: For $f\bar{f}\to\chi\bar{\chi}$, $\Delta E_{tr}=E_+$, and for $f\bar{f}\to a\gamma$, $\gamma f\to a f$, $q\bar{q}\to ag$, $gq\to aq$ (and the corresponding hermitian conjugate processes) we have $\Delta E_{tr}=E_3$. It turns out that in the former case, as studied in ref.~\cite{4ways}, the identical particles in the initial state and $\Delta E_{tr}$ being independent of the final state result in several simplifications in $\mathcal{E}(\vec{p}_1,\vec{p}_2)$, absent for the latter. 

In order to demonstrate the difference between these two cases let us now perform the integration in $p_4$ (where in the following discussion we will use the notation $p_i=|\vec{p}_i|$):
\begin{align}\hspace{-.4cm}
\nonumber\mathcal{E}(\vec{p}_1,\vec{p}_2)=&\,\frac{1}{4 F}\int \frac{d^3 p_3}{(2\pi)^3}\frac{1}{2 E_3 2 E_4} |i\mathcal{M}|^2(2\pi)\delta(E_1+E_2-E_3-E_4)\Delta E_{tr}
\end{align}
where $E_4=(m_4^2+|\vec{p}_1+\vec{p}_2-\vec{p}_3|^2)^{1/2}$. The next steps depend on the value of $\Delta E_{\rm tr}$. Clearly if $\Delta E_{\rm tr}=E_+$, it does not affect the integration over $p_3$ and $\Omega$, resulting in the simple relation $\mathcal{E}(\vec{p}_1,\vec{p}_2)=\sigma(\vec{p}_1,\vec{p}_2) E_+$. However, when $\Delta E_{\rm tr}=E_3$ integrating in $p_3$ and $\Omega$ is not so straightforward. For the latter case, let us consider the argument of the delta function which we will call $g(p_3)\equiv E_1+E_2-E_3-E_4$:
\begin{align}
\nonumber g(p_3)=\,E_+-E_3-(m_4^2+E_+^2-s+p_3^2-2p_1 p_3\, c_{13}-2 p_2 p_3\, c_{23})^{1/2}
\end{align}
where $c_{ij}=\cos\theta_{ij}$, and we have defined $\theta_{ij}$ to be the angle between the momenta of particles $i$ and $j$.
Since we perform the integration in $c_{13}$, we need to replace $\theta_{23}$ by $\theta_{13}+\theta_{12}$, where $c_{12}=(E_+^2-s-p_1^2-p_2^2)/(2 p_1 p_2)$. In order to obtain the solution $p_3^0$ of the equation $g(p_3)=0$ analytically, and to speed up the computation, we are obliged to make the assumption $E_3=p_3$. This assumption is easily justifiable for the dominantly contributing  case $f=t$ (and results in an effect of at most $\sim 1.4\%$), and we have checked that $q\bar q\into ga$ and $f\bar f\into \gamma a$ dominate over $f\gamma \into f a$ and $q g \into q a$. We then find
\begin{align}\label{eq:Etransfer}
\mathcal{E}(\vec{p}_1,\vec{p}_2)=\frac{1}{8\pi F}\int p_3^2d p_3dc_{13}\frac{1}{4 E_4} |i\mathcal{M}|^2\frac{\delta(p_3-p_3^0)}{|g'(p_3)|_{p_3\to p_3^0}}
\end{align}
where we have replaced $\Delta E_{\rm tr}$ by $E_3$. The result for the integrated collision operator is then obtained by inserting eq.~\eqref{eq:Etransfer} in eq.~\eqref{eq:ETBE} and performing the integration numerically, setting $E_3=p_3$. 

For the case of $f\bar f\to\chi\bar \chi$, and $\mathcal{E}(\vec{p}_1,\vec{p}_2)=\sigma(\vec{p}_1,\vec{p}_2) E_+$, the integrated collision operator can easily be simplified to~\cite{4ways}:
\begin{align}\label{eq:ETBE-chichi}
\int\frac{d^3 p_3}{(2\pi)^3}C[f_3]=&\frac{g_1 g_2}{32 \pi^4} \int ds\, \sigma(s)\,(s-4 m^2) \,s\, T K_2\left(\frac{\sqrt{s}}{T}\right)
\end{align}
where we have adopted the limits of integration~\cite{cosmicabundances,4ways}: $s>(m_1+m_s)^2$, $E_+>\sqrt{s}$ and $ -\tfrac{2F}{s}\sqrt{E_+^2-s}<E_-+E_+ \tfrac{m_1^2-m_2^2}{s}<\tfrac{2F}{s}\sqrt{E_+^2-s}$.
\vspace{.5cm}\\
The case of inverse decays, $1\,2\to X$, is not identical, and deserves to be studied as it has not previously been required, starting with 
\begin{align}
\nonumber\mathcal{E}(\vec{p}_1,\vec{p}_2)=&\frac{1}{2 E_1 2E_2 v_{Mol}}\int\frac{d^3 p_X}{(2\pi)^3}\frac{1}{2 E_X} |i\mathcal{M}|^2(2\pi)^4\delta^{(4)}(p_1+p_2-p_X)\Delta E_{tr}\\
=&\frac{1}{2 E_1 2E_2 v_{Mol}}\frac{1}{2 E_X} |i\mathcal{M}|^2(2\pi)\delta(E_1+E_2-E_X)E_+ \; .
\end{align}
We can then insert this expression in eq.~\eqref{eq:ETBE}, and find for $m_1=m_2=m$
\begin{align}
\nonumber \int\frac{d^3 p_3}{(2\pi)^3}C[f_3]=&\frac{g_1 g_2}{32 \pi^3} \int dE_+\, e^{-E_+/T}\left|i\mathcal{M}\right|^2\, \frac{ 2 F\,E_+}{m_X^2}\sqrt{E_+^2-m_X^2}\,\Theta(m_X^2-4m^2)\\
=&T\frac{g_1 g_2}{32 \pi^3}m_X\sqrt{m_X^2-4 m^2}\,\left|i\mathcal{M}\right|^2\,K_2\left(\frac{m_X}{T}\right)\Theta(m_X^2-4m^2)
\end{align}
which in terms of the decay width of particle $X$ can be expressed as
\begin{align}
\nonumber \int\frac{d^3 p_3}{(2\pi)^3}C[f_3]=&T\,\frac{g_1 g_2}{2 \pi^2}\, \Gamma_X\,m_X^3\,K_2\left(\frac{m_X}{T}\right)\Theta(m_X^2-4m^2) \; .
\end{align}

\section{Cross sections and input for Boltzmann equations}\label{app:xsecs}
\renewcommand{\arraystretch}{1.4}
\begin{table}[t!]
\begin{center}
\begin{tabular}{|c|c|}
\hline
\textbf{Process} & \textbf{Feynman-diagram} \\
\hline
 \parbox[c][2.7cm]{4cm}{\centering $\chi\bar\chi\to a a$\\eq.~\eqref{eq:xsecxxaa}}&\parbox[c][2.7cm]{3cm}{\centering\includegraphics[scale=0.6]{FeynmanDiagrams/xxaa.pdf}}\parbox[c][2.7cm]{3cm}{\centering\includegraphics[scale=0.6]{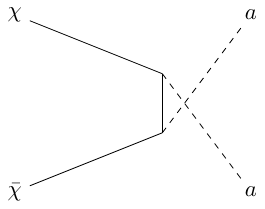}} \\
\hline
 \parbox[c][2.7cm]{4cm}{\centering $\chi\bar\chi\to f\bar f$\\eq.~\eqref{eq:xsecxxff}}&\parbox[c][2.7cm]{3cm}{\centering\includegraphics[scale=0.6]{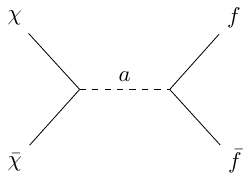}} \\
\hline
\parbox[c][2.7cm]{4cm}{\centering $a \gamma \to f\bar f$, $ag\to q\bar q$\\eqs.~\eqref{eq:xsecagammaff}, \eqref{eq:xsecagqq}}&\parbox[c][2.7cm]{3cm}{\centering \includegraphics[scale=0.6]{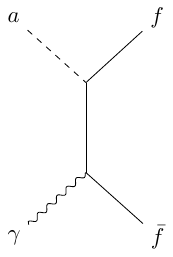}}\parbox[c][2.7cm]{3cm}{\includegraphics[scale=0.6]{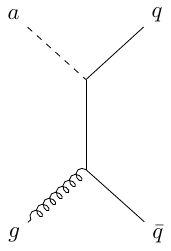}}\\
\hline
\parbox[c][2.5cm]{4cm}{\centering $a f\to \gamma f$, $a q \to g q$ \\eqs.~\eqref{eq:xsecafgammaf}, \eqref{eq:xsecaqgq}}&
\parbox[c][2.5cm]{3cm}{\includegraphics[scale=0.6]{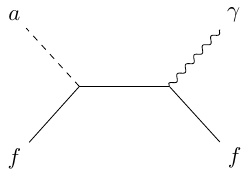}}\parbox[c][2.5cm]{3cm}{\centering\includegraphics[scale=0.6]{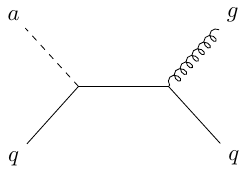}}\\
\hline
\parbox[c][2.5cm]{4cm}{\centering $a h\to f \bar f$ \\eq.~\eqref{eq:xsecahff}}&
\parbox[c][2.5cm]{3cm}{\includegraphics[scale=0.6]{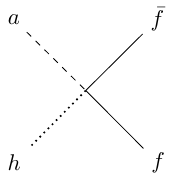}}\\
\hline
\parbox[c][2.5cm]{4cm}{\centering $f \bar f \to h \chi \bar \chi$, $f h \to f \chi \bar \chi$, $\bar f h \to \bar f \chi \bar \chi$ \\eq.~\eqref{eq:xsecffhxbarx}}&
\parbox[c][2.5cm]{3.5cm}{\includegraphics[scale=0.6]{FeynmanDiagrams/ffbartohxxbar.pdf}}\parbox[c][2.5cm]{3.5cm}{\includegraphics[scale=0.6]{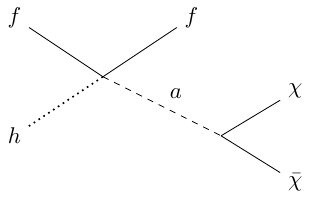}}\parbox[c][2.5cm]{3.5cm}{\includegraphics[scale=0.6]{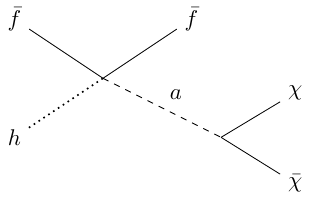}}\\
\hline
\parbox[c][2.5cm]{4cm}{\centering $ a \to f\bar f$\\ eq.~\eqref{eq:decayaxionleptons}} &\parbox[c][2.5cm]{3cm}{\includegraphics[scale=0.6]{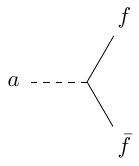}}\\
\hline
\end{tabular}
\captionof{table}{Feynman-diagrams of the annihilation of the DM via the hidden sector and connector processes, as well as the co-annihilation of ALPs at tree level.}\label{tab:xsec}
\end{center}
\end{table}
\renewcommand{\arraystretch}{1}
In table \ref{tab:xsec} we collect the different annihilation processes for the DM and the ALPs which govern the Boltzmann equations in eq.~\eqref{eq:generalBoltzmann}. In this section we provide expressions for the cross sections for the required $2\to 2$ processes and the collision term for the relevant $2\to 3$ processes.
\subsection{Hidden sector process}
For the annihilation of the DM particles to ALPs, i.e.~annihilation in the hidden sector (for which the Feynman diagrams at tree level are shown in table~\ref{tab:xsec}), the total unpolarized cross section in the centre of mass frame is given by
\begin{eqnarray}\label{eq:xsecxxaa}
\hspace{-1cm}\nonumber\sigma(\chi\bar{\chi}\rightarrow a a)&=&\frac{(g_{a\chi\chi}m_\chi)^4}{16\pi s}\frac{\sqrt{s-4 m_a^2}}{\sqrt{s-4 m_\chi^2}} \Bigg( -2
-\frac{m_a^4}{m_a^4-4 m_a^2 m_\chi^2+m_\chi^2 s}\\
&&
\frac{6m_a^4-4m_a^2s+s^2}{(s-2m_a^2)\sqrt{(s-4m_a^2)(s-4m_\chi^2)}}\,\text{ln}\tfrac{s-2m_a^2+\sqrt{(s-4m_a^2)(s-4m_\chi^2)}}{s-2m_a^2-\sqrt{(s-4m_a^2)(s-4m_\chi^2)}}
\Bigg)
\label{eq:HSsection}
\end{eqnarray}
\subsection{Connector processes}
The Feynman diagram of the $\chi \bar{\chi} \rightarrow f \bar{f} $ connector process is depicted in table~\ref{tab:xsec}. For the total cross section in the centre of mass frame we find at tree level
\begin{align}\label{eq:xsecxxff}
\sigma(\chi \bar{\chi} \rightarrow f \bar{f})=\frac{(g_{a \chi \chi}m_\chi)^2 (g_{aff}m_f)^2 n^c_f s \sqrt{s-4 m_f^2}}{16 \pi  \left(m_a^2-s\right)^2 \sqrt{s-4 m_\chi^2}}\, ,
\end{align}
 with $n^c_f$ the number of colour degrees of freedom of the fermion.
For reheating temperatures above a few hundred GeV contributions from the ultraviolet-dominant $2 \to 3$ processes $f \bar f \to h \chi \bar \chi$, $f h \to f \chi \bar \chi$ and $\bar f h \to \bar f \chi \bar \chi$ become important.
Following ref.~\cite{Elahi:2014fsa}, but calculating the matrix element explicitly and including the final state masses one finds (the fact that not all involved particles are scalars does not change the integral because we neglect Fermi/Bose factors)
\begin{align}\nonumber
\dot{n}_{\chi}+3 H n_{\chi} =\,\,&\frac{(g_{a \chi \chi}m_\chi)^2 (g_{aff}m_f)^2}{v^2}\frac{n^c_f\ T}{(4 \pi)^{7}}\,\times\\& \quad\int_{s_{\text{min}}}^{\infty} \mathrm{d} s \int_{0}^{1} \mathrm{~d} x_{2} \int_{x_{2}}^{1} \mathrm{d} x_{1}\,s^{3/2}\,K_1\left(\frac{\sqrt{s}}{T}\right) \frac{4p_a^2 s}{\left(p_a^2-m_a^2\right)^2}\label{eq:xsecffhxbarx}
\end{align}
with $p_a^2=(x_1-x_2)s+m_h^2$ and $s_{\text{min}}=4 m_f^2(T)$ for $f \bar f \to h \chi \bar \chi$ and $p_a^2=(x_1-x_2)s+m_f^2$ and $s_{\text{min}}=(m_f(T)+m_h)^2$ for $f h \to f \chi \bar \chi$, $\bar f h \to \bar f \chi \bar \chi$.
\subsection{SM-ALP processes}
For the co-annihilation of ALPs to SM particles, the Feynman diagrams at tree level are shown in table~\ref{tab:xsec}. Here we collect the results for the total unpolarized cross section in the centre of mass frame:
\begin{align}
\hspace{-2cm}\nonumber\sigma(a\gamma \rightarrow f \bar f)=&\frac{(g_{aff}m_f)^2 \ \alpha_{\rm{em}} \ q_f^2\ n^c_f}{m_a^2+m_\gamma^2-s}\sqrt{\frac{\lambda(s,m_f^2,m_f^2)}{\lambda(s,m_a^2,m_\gamma^2)}}\\
\nonumber &\Bigg[\frac{m_a^4-4 m_a^2 m_f^2+\left(m_\gamma^2-s\right)^2}{\sqrt{\lambda(s,m_a^2,m_\gamma^2)(s(s-4 m_f^2)}} \ln \left(\tfrac{\sqrt{s} \left(m_a^2+m_\gamma^2-s\right)+\sqrt{\lambda(s,m_a^2,m_\gamma^2)} \sqrt{s-4 m_f^2}}{\sqrt{s} \left(m_a^2+m_\gamma^2-s\right)-\sqrt{\lambda(s,m_a^2,m_\gamma^2)} \sqrt{s-4 m_f^2}}\right)\qquad\\
\label{eq:xsecagammaff}&\qquad-\frac{m_a^2 \left(m_\gamma^2+2 m_f^2\right) \left(m_a^2+m_\gamma^2-s\right)}{(m_a^2-m_\gamma^2)^2m_f^2+m_a^2 m_\gamma^2s-2(m_a^2+m_\gamma^2)m_f^2 s+m_f^2 s^2}\Bigg]\; ,\\
\label{eq:xsecagqq}\sigma(ag \rightarrow q \bar q)=&\frac{1}{6} \  \sigma(a \gamma \rightarrow f\bar{f})(\alpha_{\rm{em}} \rightarrow \alpha_s,\ q_f^2\rightarrow 1,\ m_\gamma\to m_g)\; ,\\
\nonumber\sigma(af \rightarrow \gamma f)=& \frac{(g_{aff} m_f)^2 \ \alpha_{\rm{em}}\ q_f^2}{4 s}\sqrt{\frac{\lambda(s,m_f^2,m_\gamma^2)}{\lambda(s,m_f^2,m_a^2)}}\int_{-1}^{1} \mathrm{d} \cos(\theta) \frac{1}{\left(s-m_f^2\right)^2 \left(t-m_f^2\right)^2}
\\
\nonumber &\,   \Big(m_a^2\left(m_\gamma^2 \left(2 m_f^2 (s+t)-2 m_f^4-s^2-t^2\right)-2 \left(s+t-2 m_f^2\right) \left(s\,t-m_f^4\right)\right)\\
\label{eq:xsecafgammaf}&\,\,+ \left(s-m_f^2\right) \left(t-m_f^2\right) \left((s+t-2 m_f^2)^2+2 m_a^4\right)\Big)\; ,\\
\sigma(aq \rightarrow g q)=&\frac{4}{3}\ \sigma(a f \to \gamma f)(\alpha_{\rm{em}}\to \alpha_{s},\ q_f^2\rightarrow 1,\ m_\gamma\to m_g) \; ,\label{eq:xsecaqgq}
 \end{align}
 with $n^c_f$ the number of colour degrees of freedom of the fermion and $q_f$ the fermion's electric charge.
Note that expressions~\eqref{eq:xsecafgammaf} and~\eqref{eq:xsecaqgq} are equivalent to the ones for $\bar f a \to \bar f \gamma$ and $\bar q a \to \bar q g$.

 For reheating temperatures above a few hundred GeV contributions from the ultraviolet-dominant process $a h \to f\bar f$ become important. In this case we find for the cross section in the centre of mass frame
 \begin{equation}\label{eq:xsecahff}
 \sigma(ah \rightarrow f\bar f)=\frac{(g_{aff}m_f)^2 n^c_f}{8 \pi v^2}\sqrt{\frac{\lambda(s,m_f^2,mf^2)}{\lambda(s,m_a^2,m_h^2)}}\; .
 \end{equation}
\section{Decay width of the ALP}\label{app:decaywidths}
\begin{figure}[t]
\begin{center}
\includegraphics[scale=1]{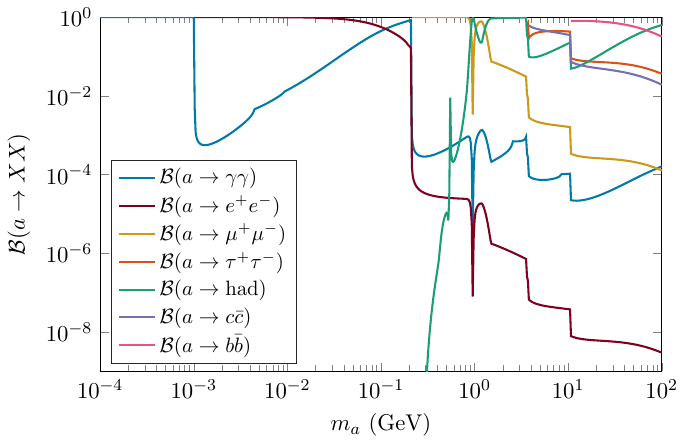}
	\captionof{figure}{Branching fractions of the ALP into photons, leptons and hadrons as indicated, as a function of the ALP mass.}\label{fig:Branchingfraction}	
\end{center}
\end{figure}
The total decay width of the ALP plays an important role at various points in our work, particularly for the collider and cosmological constraints. For example, for the collider constraints it is important because the ALP lifetime determines whether it can be triggered on, or if it decays within the detector or outside (leading to invisible decay). At tree level, the ALP decays into leptons or quarks, however at loop level it may decay into photons or gluons. Our calculations for the leading partial decay widths are summarised below.
In the perturbative regime the decay width of an ALP decaying to fermions is
\begin{align}
\Gamma(a\rightarrow f\bar f)=\frac{(g_{aff}m_f)^2 n^c_f m_a}{8\pi}\sqrt{1-\frac{4m_f^2}{m_a^2}}\,, \label{eq:decayaxionleptons}
\end{align}
with $n^c_f$ the number of colour degrees of freedom of the fermion. This expression is used for leptons and for decays into heavy quarks $c\bar{c}$ and $b\bar{b}$, provided $m_a$ is above the threshold for the decay into a pair of $D$ or $B$ mesons respectively. For the contribution of the light quarks, we take into account the decays of ALPs to hadrons via
\begin{itemize}
    \item If the mass of the ALP is less than 1.2 GeV, the decay width is calculated using code obtained by private communication, created for ref.~\cite{domingo2017decays}. This code was written for the decay of an NMSSM CP-odd Higgs, but with an appropriate choice of parameters ($P_{11}=\sqrt{2} v g_{aff}$, $\tan\beta=1$ and decoupled neutralinos, charginos and heavy Higgs bosons) this particle can be identified with our ALP. 
    \item Above 1.5 GeV, the decay width into light quarks is given by~\cite{colliderprobes}
\begin{align}\label{eq:decayaxionhad}
\Gamma(a\to {\rm hadrons}) = g^2_{aff} \frac{9 m_a^3 \alpha_s^2}{32 \pi} \left(1 + \frac{83 \alpha_s}{4 \pi} \right), \end{align}
as in this regime the strong coupling is taken to be perturbative.
\item Between 1.2 and 1.5 GeV we interpolate between the above determinations of the decay widths, in order to obtain a smooth result for the total decay width of the ALP, important particularly for the cosmological constraints.
\end{itemize}
This concludes the discussion of the tree-level decays of the ALP.
In order to take into account experimental constraints on ALP decays to photons, we further require the decay width into two photons via a fermion loop. This is given by
 \begin{equation}
 \Gamma(a\to \gamma\gamma)_\text{loop}=\frac{ \alpha^2 |T(m_a^2)|^2}{4 \pi^3 m_a}\, ,
 \end{equation}
 where the triangle loop function is
 \begin{equation}
 T(s)=\sum_f g_{aff}\,m_f^2\,n^c_f\,q_f^2 \arcsin^2\left( \sqrt{s}/(2 m_f)\right) \; ,
 \end{equation}
 where $q_f$ is the electric charge of the fermion. 
In figure~\ref{fig:Branchingfraction}, we provide the branching ratios for all the above mentioned decays.
\section{Axion production cross sections in SLAC E137 experiment}\label{app:xsecsSLAC}
\subsection{Primakoff production}\label{app:Primakoff}
The differential cross section for the production of ALPs via the Primakoff mechanism (see figure~\ref{fig:feynmanbeamdump}a) is given by eq.~(A1) in ref.~\cite{E137},
\begin{align}
\frac{\mrm{d}\sigma_Z^{\gamma \to a}}{\mrm{d}\Omega}=8\alpha_{\rm{em}}\frac{\Gamma(a\to \gamma\gamma)}{m_a^3}\mid F(t) \mid^2\frac{\theta^2}{\left(\theta^2+\frac{1}{4}\left(\frac{m_a}{E_\gamma}\right)^4\right)^2}\,,\label{eq:primakoffSLAC}
\end{align}
where $\Gamma(a\to \gamma\gamma)$ is the loop induced ALP decay width to two photons (see app.~\ref{app:loopdetails} for details), $t= (p_a-p_\gamma)^2$ is the momentum exchange with $p_\gamma$ the four-momentum of the incoming electron and $p_a$ the four-momentum of the outgoing ALP. $\Theta$ is the scattering polar angle in the laboratory frame. $Z$ and $F(t)$ are the target atomic number and target form factor, respectively. For $t\leq 3\times 10^{-6}$~GeV$^2$ we use the atomic form factor, given by eqs.~(A5) -- (A6) in~\cite{E137},
\begin{align}
|F_\text{atomic}(t)|^2=Z^2\left(\frac{a^2|t|}{a^2|t|+1}\right)+Z\left(\frac{a^{\prime2}|t|}{a^{\prime 2}|t|+1}\right)^2
\end{align}
with the parameters
\begin{align}\label{eq:aparameters}
a  &= \left\{\begin{array}{cl}
122.8/m_{e} & \text { for hydrogen} \\
 111\ Z^{-1 / 3}/m_e & \text { for oxygen} \\
\end{array}\right.\\
a^{\prime}  &= \left\{\begin{array}{cl}
 282.4/m_e & \text { for hydrogen} \\
 773\ Z^{-2 / 3}/m_e & \text { for oxygen} \;. \\
\end{array}\right.
\end{align}
For $t> 3\times 10^{-6}$~GeV$^2$ the elastic scattering form factor of the nuclei is used. The dipole form factor for the proton and the elastic form factor for oxygen are given by \cite{Hofstadter}
\begin{align}
F_{\mathrm{di},H}(t)&=\frac{1}{(1+\frac{|t|}{q_0^2})^2} \\
F_{\mathrm{el},O}(t)&=Z\left(1-\frac{a_0^2\ t}{8}\right)e^{-a_0^2\ t/4} \; ,
\end{align}
with $q_0^2=0.71$~GeV$^2$ and $a_0=8.97$~GeV$^{-1}$.
To obtain an expression for the momentum exchange $t$ we need to find the relation between the energy of the incoming photon and the energy of the outgoing axion. Following~\cite{ALPtraum}, the momentum exchange can be expressed as
\begin{align}
 t &= (p_\gamma - p_a)^2 = -\frac{m_a^4}{4 \, E_a^2} - E_a^2 \, \theta^2 \; .
\end{align}
under the assumption that $\theta\ll 1$, $m_a\ll E_a,m_N$ and $p_t\approx 0$. However, since both the photon track-length distribution and our expression for the Primakoff cross section are given in terms of the photon energy, we have to find an expression for the transverse momentum squared $t$ in terms of $E_\gamma$. To first order approximation we have $E_\gamma \approx E_a$. As pointed out in~\cite{ALPtraum}, the term proportional to $m_a^4$ is negligible since we are considering small axion masses. Expanding to second order and inverting gives
\begin{align}
E_a \approx E_\gamma - \frac{(E_\gamma \theta)^2}{2 m_N}
\end{align}
and consequently
\begin{align}
t &= (p_\gamma - p_a)^2=-2m_N(E_\gamma-E_a)\approx (E_\gamma \theta)^2 \,.
\end{align}
For our largest $m_a=1$ GeV we get a relative error of $0.1 \%$ for the photon energy which is negligible considering the fact that we extract the track-length distribution from a graph. 
\subsection{Bremsstrahlung}\label{app:bremsstrahlung}
The differential cross section for axion production by bremsstrahlung (see figure~\ref{fig:feynmanbeamdump} b) is given by
\begin{equation}\label{eq:BremsstrahlungSLAC}
\begin{split}
\frac{\mrm{d}^2\sigma_Z^{e \to a e}}{\mrm{d}\Omega_a\mrm{d}E_a}=&\frac{\alpha_{\rm{em}}^2g_{aff}^2m_e^2}{4\pi^2}\frac{E_e}{U^2}\left( x^3-\frac{2m_a^2x^2(1-x)}{U}\right.\\
&+\left.\frac{2m_a^2}{U^2}[m_a^2x(1-x)^2+m_e^2x^3(1-x)]\right)\chi\,,
\end{split}
\end{equation}
with $U=E_e^2\theta_a^2 x+m_e^2x+m_a^2(1-x)/x$ and $x=E_a/E_e$. $\chi$ is the integrated form factor obtained from integrating $\chi=\int_{t_{\text{min}}}\mrm{d}t |F(t)|^2(t-t_{\text{min}})/t^2$~\cite{E137_2},
\begin{align}
\chi&=\chi_\text{elastic}+\chi_\text{inelastic} \nonumber \\
&= Z^2\left[\text{ln}\left(\frac{a^2m_e^2(1+l)^2}{a^2t_\text{min}+1}\right)-1\right]+Z\left[\text{ln}\left(\frac{a^{\prime 2}m_e^2(1+l)^2}{a^{\prime 2}t_\text{min}+1}\right)-1\right]\,,
\end{align}
where $t_\text{min}=\left(U/(2E_e(1-x))\right)^2$ is the minimal momentum transfer squared and $l=E_e^2\theta_a^2/m_e^2$ and $a$ and $a^\prime$ are to a reasonable approximation given by eqs.~\eqref{eq:aparameters}. $m_e$ is the mass of the electron, $E_e$ is the energy of the incoming electron and positron and $E_a$ the energy of the ALP in the laboratory frame.
\subsection{Positron annihilation}\label{app:positronannihilation}
\paragraph{Non-resonant case}
The non-resonant cross section for positron annihilation in the centre of mass frame is given by (see figure~\ref{fig:feynmanbeamdump}c)
\begin{align}
\frac{\mrm{d} \sigma^{e^+ \to a \gamma}}{\mrm{d} \ \mrm{cos}(\theta^*)}=\frac{g_{aff}^2 m_f^2 \alpha_{\rm{em}}}{2s}\left(\frac{m_a^2}{s-m_a^2}\right)\frac{1}{1-\beta^2\mrm{cos}(\theta^*)^2},
\end{align}
where $\theta^*$ is the scattering polar angle of the ALP in the centre of mass frame and $\beta=\sqrt{1-4m_e^2/s}$ is the speed of the electron target \cite{E137}. $s=2E_+m_e+2m_e^2$ is the centre of mass energy squared. The maximal opening angle for the ALP to be detected is given in the laboratory frame (see appendix \ref{app:SLACproba}). We therefore have to relate the ALP's scattering angle in the centre of mass frame, $\theta^*$, with $\theta^{\mrm{lab}}_\text{max}$, the angle in the laboratory frame. This can be achieved by expressing the Mandelstam variable $t$ which is Lorentz invariant in the laboratory frame and in the centre of mass frame. We find
\begin{align}
\cos(\theta^{\mrm{lab}})=\frac{E_+^{\mrm{lab}} E_a^{\mrm{lab}} -E_+^* E_a^{*}+\cos (\theta^* ) \sqrt{E_a^{*2}-m_a^2} \sqrt{E_+^{*2}-m_e^2}}{\sqrt{E_+^{\mrm{lab}2}-m_e^2} \sqrt{E_a^{\mrm{lab}2}-m_a^2}},
\end{align}
where the asterix denotes quantities in the centre of mass frame.
The ALP's energy in the the laboratory frame is
\begin{align}
E_a^{\mrm{lab}}=\frac{s+m_a^2+(s-m_a^2)\beta\cos(\theta^*)}{4m_e}.
\end{align}
In the centre of mass frame the energies of the ALP and of the positron are 
\begin{align}
E_a^*=\frac{4 E_+^{*2}+m_a^2}{4 E_+^*} \qquad E_+^*=\frac12\sqrt{s}=\frac12\sqrt{2(m_e^2+m_e E_+^{\mrm{lab}})}.
\end{align}
We have now all the ingredients to express the differential cross section in terms of the scattering angle and the positron energy in the laboratory frame over which we integrate. To deal with the infrared divergence in the cross section for $s \rightarrow m_a^2$, i.e.~small photon energies, we apply a cut on the centre of mass energy around $\sqrt{s}=3$~MeV as proposed in ref.~\cite{E137}.
\paragraph{Resonant production}
The resonant positron annihilation production cross section (see figure~\ref{fig:feynmanbeamdump}d) can be obtained from the Breit-Wigner formula and is in the centre of mass frame given by \cite{LuSLAC}
\begin{align}
\sigma^{e^+ \to a}(E_a^*) =\frac{\pi^2}{2E_a^{*2}} \frac{\Gamma(a\to e^+ e^-) \Gamma(a\to XX)}{(E_a^*-m_a)^2+\Gamma_a^2/4}\; ,
\end{align}
with $\Gamma_a$ the total decay width of the ALP and
 \begin{align}
\Gamma(a\to XX)= \left\{
	\begin{array}{ll}
	\Gamma_\text{loop}(a\to \gamma \gamma) & \text{for the ALP decaying into two photons } \\
	\Gamma(a\to e^+ e^-) & \text{for the ALP decaying into } e^+e^-.
	\end{array}
	\right.
\end{align}
Since the Breit-Wigner formula contains the decay width of the ALP which is strongly peaked compared to the step size in the positron energy the cross section is approximated by a delta function, i.e.
\begin{align}
\sigma^{e^+ \to a}(E_a^*)=\sigma^{e^+ \to a}_{\text{tot}} \delta(E_a^*-m_a)
\end{align}
where 
\begin{align}
\sigma^{e^+ \to a}_{\text{tot}} (a\to XX) = \int_{\infty}^\infty \mathrm{d}E_a^* \sigma^{e^+ \to a}(E_a^*)= \frac{\pi^2}{2m_a^{2}} \Gamma(a\to e^+ e^-) \mathcal{B}(a\to XX)
\end{align}
and hence independent of the positron energy. The number of events in a specific decay channel can therefore be approximated by
\begin{align}\hspace{-.8cm}
\mathcal{N}(a\to XX)&=\mathcal{N}_{e,\text{ inc}} \ \sigma^{e^+ \to a}_{\text{tot}}(a\to XX) \int\mrm{d}E_a^* \ T(E_+) \ p(E_+) \ \frac{\mrm{d}E_+}{\mrm{d}E_a^*}\,  \delta(E_a^*-m_a) \sum_{Z} P_Z \nonumber \\
&=\mathcal{N}_{e,\text{ inc}} \ \sigma^{e^+ \to a}_{\text{tot}}(a\to XX) \ T(E_+) \ p(E_+) \ \frac{m_a}{m_e} \ \sum_{Z} P_Z\,, 
\end{align}
where $T_+(E)$ is the track-length distribution of positrons. The energy of the positron and of the ALP are in the laboratory frame given by
\begin{align}
E_+^\text{lab}=\frac{m_a^2-m_e^2}{2m_e} \; , \qquad E_a^\text{lab}=E_+^\text{lab} + m_e \; 
\end{align}
respectively.
 \section{Details on the ALP's detection probability}\label{app:SLACproba}
 In this appendix we will give details about the probability for the ALP in the SLAC electron beam dump experiment to decay invisibly, i.e.~we consider the cases where the ALP could have been produced but not been detected. In these cases the non-observation of axions in the experiment does not allow us to set bounds on the ALP's coupling and mass. To determine whether the ALP decays inside or outside the decay volume we need its decay length. The decay length of the ALP in the laboratory frame is
 \begin{align}
 l_a=\beta \gamma_a \tau_a =\frac{\beta \gamma_a}{\Gamma_a} \approx \frac{E_a}{m_a \Gamma_a}
 \end{align}
 where we assume $E_a\gg m_a$~\cite{ALPtraum}. In the following discussion we consider the case where the axion decays into two photons.\footnote{In principle, the same arguments can be applied for the case where the ALP decays to an electron-positron pair. Some approximations may not be reasonable as the decay products are massive (concerning boost, typical opening angle). To derive the bounds, we have used the same expressions and obtain very good agreement with the results presented in~\cite{E137} (based on a Monte-Carlo simulation) and~\cite{colliderprobes}.}
 \begin{figure}
 \begin{center}
 \includegraphics[scale=0.8]{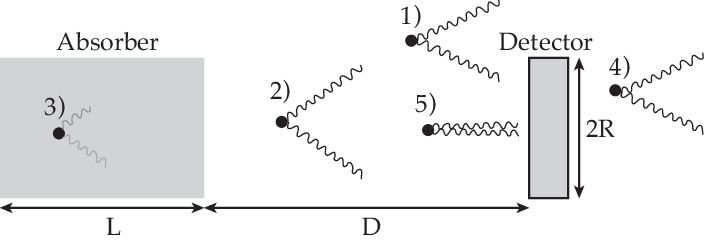}
 \captionof{figure}{Different scenarios for the ALPs to decay invisibly (here schematically shown for photons).}\label{fig:alpdecay}
 \end{center}
 \end{figure}
 Taking the experimental layout into account, the ALP cannot be detected if (the scenarios are summarised in figure~\ref{fig:alpdecay}):
 \begin{itemize}
 \item the opening angle of the ALP is too big, $p(\theta_{a}>\theta_{\text{max}})=0$. The detector which is approximated as cylindrical with radius $R=1.5$~m can detect photons with a maximal separation angle of $\text{sin}(\theta_{\text{max}})=R/(D+L)\approx\theta_{\text{max}}$ (see definition in figure~\ref{fig:alpdecay}), which for the dimensions of E137 corresponds to $\theta_{\text{E137, max}}=0.22$~deg. 
 \item the ALP decays too early and/or the opening angle between the photons is too large, so that both photons would miss the detector. The opening angle between the two photons is determined by the boost of the ALP. The minimal opening angle for high ALP boosts is given by $\theta_{\gamma,\text{ min}}\approx 2/\gamma_a$~\cite{ALPtraum, revisedconstraints}. The distribution of the number of photons $\mrm{d}N/\mrm{d}\theta_\gamma$ is peaked at this minimal $\theta_{\gamma,\text{ min}}$ and the typical separation between the two photons arriving at the detector is therefore given by
 \begin{align}
 d_{\gamma\gamma}=\text{sin}\left[\theta_{\gamma,\text{ min}}(D+L-l_a)\right]\approx\theta_{\gamma,\text{ min}}(D+L-l_a)=\frac{2(D+L-l_a)}{\gamma_a}\,,
 \end{align}
  where $D+L-l_a$ is the distance from the detector at which the ALP decays. $d_{\gamma\gamma}$ should not exceed the dimension of the detector, $p(d_{\gamma\gamma} > 2R)=p(D+L-l_a > RE_a/m_a)=0$.
 \item the ALP decays inside the absorber and the photons get absorbed or the ALP decays behind the detector. The probability that the ALP decays inside the decay volume is given by
 \begin{align}
 p(l_a)=p(l_a>D)-p(l_a>D+L)=e^{-D/l_a}-e^{-(D+L)/l_a}\,.
 \end{align}
 Indeed, we find that this significantly reduces the number of detectable particles.
 \item the ALP decays in front of the detector but the separation between the two photons is too small. The typical separation between the photons has to be larger than the minimal resolution of the detector, $R_{\text{E137, min}}\approx 3$ mm, which would lead to an indistinguishable signal, $p(d_{\gamma\gamma}<R_{\text{min}})=0$.  
 \end{itemize}

 To sum up, the probability to detect both photons is given by~\cite{ALPtraum}
 \begin{align}\label{eq:proba}
 p(l_a)&= \left\{
 	\begin{array}{ll}
 	e^{-D/l_a}-e^{-(D+L)/l_a} & \text{if } \ E_a<2m_a\cdot L/R_{\text{min}}, \\
 	 &E_a>(L+D)\cdot m_a/\left(R+1/\Gamma_a\right)\text{ and } \theta<\theta_\text{max}\\
 	0 & \text{otherwise} \,.
 	\end{array}
 	\right. 
 \end{align}

\section{Details on constraints from exotic Higgs decays}\label{app:higgsdecays}

To understand the sensitivity of the LHC exotic Higgs decay searches, it is first important to calculate the decay table for the axion.  A detailed calculation is available in the appendix and the branching fractions are plotted in figure~\ref{fig:Branchingfraction} for a range of ALP masses from 10 MeV to 100 GeV. What is relevant for our discussion here is that the branching fraction into two photons remains consistently smaller than $10^{-3}$ as soon as the ALP mass is large enough to allow $a \rightarrow \mu^+ \mu^-$.  Next, as expected, once each new massive fermion mode turns on, it quickly dominates as the width in that channel is proportional to $m_f^2$.  We also assume that as soon as the mass threshold is above 1.5 GeV ($\simeq 2 m_\rho$), decays into hadrons via loop-mediated $a \rightarrow g g$ open up. Therefore, once hadronic decays become allowed, any branching into muons becomes again vanishingly small.  Our calculation (assuming $a \rightarrow gg$ open only above 1 GeV) gives $\mathcal{B}(\mu^+ \mu^-) < 10^{-3}$ as soon as $\tau^+ \tau^-$ and $c \bar c$ modes open (i.e.~$m_a \gtrsim 1.8$~GeV).   Since all current exotic Higgs decays look for $m_a \gtrsim 10$~GeV only, we can safely assume that none of the muonic searches will be able to see the signatures of our model. We therefore need only think about the $h \rightarrow aa \rightarrow 4b$ channel. The limits on $h \rightarrow aa \rightarrow 4b$ are only able to exclude branching ratios less than one in the small mass range between 18-22.5 GeV.  However, this would still require a branching fraction of 0.75 of the SM Higgs into this one channel alone.  Given that in our model, these channels are dependent on 1-loop contributions only the decay width of the Higgs into these modes cannot be of comparable size to the usual tree-level 2-body SM decays.  We therefore conclude that none of these searches unfortunately have any sensitivity to our model.

The current limit~\cite{ATLAS:2020pcy} in $h\rightarrow Z(a \rightarrow gg)$ is in principle sensitive to our model in the very narrow range $0.5~\text{GeV} < m_a \lesssim 2.7~\text{GeV}$.  However, this is precisely the mass range where theoretical calculations are wildly unpredictable due to hadronic contributions and we choose not to apply these.

Lastly, the $h\rightarrow Z \gamma$ searches can 
be used in certain phase space regions where the two photons from $a$-decay are collimated, i.e.~$h\rightarrow Z(a \rightarrow \gamma \gamma) $.  The $h\rightarrow ZZ^* \rightarrow 4 \ell$ measurements can also be used to place a limit on new physics contributions as the off-shell Z decaying into two leptons can also be interpreted as a new particle that decays into two leptons.  A study of the collimated di-gamma decay was done in \cite{Brooijmans:2020yij}.  However, interpreted in our model, it corresponds only to coupling values $C_{aff} \gtrsim 10^2/$GeV.  Since this lies wildly outside the self-consistent EFT regime, we assume that the $h \rightarrow Z \gamma$ measurements are currently not sensitive enough to provide a useful constraint.  Similarly, the $4\ell$ decay mode of the Higgs currently carries the signal strength accuracy $\mu = 1.44 \pm 0.4 $~\cite{ATLAS:2015egz} which translates to an upper limit on the decay width $\Gamma(h\rightarrow Za)\times \mathcal{B}(a\rightarrow 2\ell) \lesssim 0.44 \times \Gamma(h\rightarrow ZZ^*\rightarrow 4\ell)$.  Again, due to the loop suppression in our decay, we end up with exclusions on $C_{aff} \gg 1$ only and there is no usable limit from this measurement. 
 \section{Loop diagrams relevant for constraint calculations}\label{app:loopdetails}
 The ALP in our model can only decay into photons via a fermion loop. Applying the constraints from the effective axion-photon coupling, $g_{a\gamma\gamma}$, to the axion-fermion coupling, $g_{aff}$, demands more theoretical groundwork. The main task is to calculate the Primakoff cross section for this coupling. The resulting cross section can then be inserted into the expression for the ALP yield for the beam dump constraint and the expressions for the total energy outflow and the ALP opacity for the HB stars constraints.
 \begin{figure}[t]
 \begin{center}
 \hspace{1.2cm}
 \begin{subfigure}{0.45\textwidth}
 \includegraphics[scale=0.9]{FeynmanDiagrams/primakoff.pdf}
 \end{subfigure}
\begin{subfigure}{0.45\textwidth}
\vspace{-.2cm}
\includegraphics[scale=0.9]{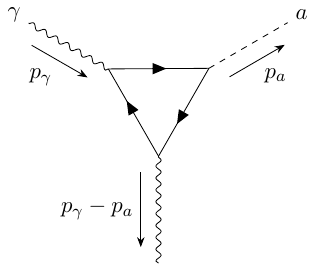}
\end{subfigure}
 \captionof{figure}{\emph{Left:} Feynman diagram of the Primakoff production process via a fermion loop. \emph{Right:} Feynman diagram of the amplitude we need to calculate to replace $g_{a\gamma \gamma}$ by $g_{a\gamma \gamma}^\text{loop}$ in the Primakoff production process.}\label{fig:primloopfeynman}
 \end{center}
 \end{figure}
 The approach we adopt is to replace the expression for the axion-photon coupling in the ``decay width'' in equations~\eqref{eq:primakoffSLAC} by an expression for the axion-photon coupling induced by a fermion loop. This coupling will now depend on the momentum transfer squared, $t=(p_\gamma - p_a)^2$. The Feynman diagram of the Primakoff process induced by a fermion loop is shown in figure~\ref{fig:primloopfeynman} on the left. It should be noticed that the decay width in equation~\eqref{eq:primakoffSLAC} has to be regarded as a definition of the axion-photon coupling rather than an actual physical decay width,
 \begin{align}
 \frac{\mrm{d}\sigma}{\mrm{d}\Omega}&=8\alpha_{\rm{em}}\frac{\Gamma}{m_a^3}\mid F(t) \mid^2\frac{\theta^2}{\left(\theta^2+\frac{1}{4}\left(\frac{m_a}{E_\gamma}\right)^4\right)^2}\nonumber\\
 &=\alpha_{\rm{em}}\frac{g_{a\gamma\gamma}^2}{8\pi}\mid F(t) \mid^2\frac{\theta^2}{\left(\theta^2+\frac{1}{4}\left(\frac{m_a}{E_\gamma}\right)^4\right)^2}\,.
 \end{align}
 By comparing the amplitude of the loop diagram and the tree-level effective diagram we will deduce what $g_{a\gamma \gamma}$ should be replaced by. The Feynman diagram of the amplitude we have to consider is depicted in figure~\ref{fig:primloopfeynman} on the right. We proceed as in ref.~\cite{axionanomalies} and find that the amplitude of the triangle graph is given by
 \begin{align}
 \mathcal{T}^{\alpha\beta} &  =\int\frac{d^{4}k}{(2\pi)^{4}%
 }(-1)\operatorname*{Tr}\left[  \frac{i}{\slashed k+\slashed p_{\gamma}-m_f}\gamma^{\alpha
 }\frac{i}{\slashed k-m_f}\gamma^{\beta}\frac{i}{\slashed k+\slashed p_{\gamma}-\slashed p_a-m_f}\gamma
 _{5}\right] \ \nonumber\\
 &  =-i\frac{1}{2\pi^{2}}\varepsilon^{\alpha\beta\rho\sigma}p_{a,\rho
 }p_{\gamma,\sigma}m_f C_{0}(p_{a}^{2},p_{\gamma}^{2},(p_{a}-p_{\gamma})^{2},m_f^{2},m_f^{2},m_f^{2})\,,
 \end{align}
 with the three-point scalar function $C_0$ defined as in ref.~\cite{feyncalc}. The matrix element squared of the diagram is given by (with $t=(p_a-p_\gamma)^2$ in eq.~\eqref{eq:primakoffSLAC})
 \begin{align}
 |\mca{M}|^2_\text{loop}=\frac{\alpha_{\rm{em}}^2 (m_a^2-t)^2 \left[\sum_f g_{aff} m_f^2 q_f^2 n^c_f  |C_0(m_a^2,0,t,m_f^2,m_f^2,m_f^2)|\right]^2}{2\pi^2}\,.
 \end{align}
 By calculating the same diagram but with an effective ALP-photon coupling $g_{a\gamma\gamma}$ at tree level, i.e.
 \begin{align}
 |\mca{M}|^2&=g_{a\gamma\gamma}^2 \epsilon_{\mu \nu \rho \sigma }\epsilon_{\alpha \beta \kappa \lambda }g^{\alpha \mu } \ g^{\beta \nu } \ p_\gamma^{\rho} \ p_\gamma^{\kappa} \ (p_\gamma-p_a)^{\sigma } \ (p_{\gamma}-p_a)^{\lambda }\nonumber\\
 &=2 \ g_{a\gamma\gamma}^2 \left((p_a\cdot p_\gamma)^2- (p_a\cdot p_a)(p_\gamma\cdot p_\gamma)\right)=\frac{g_{a\gamma \gamma}^2(m_a^2-t)^2}{2}
 \end{align}
 we obtain the replacement
 \begin{align}\label{eq:loop_coupling}
 g_{a\gamma \gamma}^2\rightarrow\frac{\alpha_{\rm{em}}^2 \left[\sum_f g_{aff} m_f^2 q_f^2 n^c_f  |C_0(m_a^2,0,t,m_f^2,m_f^2,m_f^2)|\right]^2}{\pi^2}
 \end{align}
 in the expression for the Primakoff production in the E137 SLAC experiment, eq.~\eqref{eq:primakoffSLAC}, and in the expression for the Primakoff production in horizontal branch stars. For the constraints from horizontal branch stars we make the approximation of zero momentum transfer, i.e.~$t \approx 0$.
\bibliographystyle{jhep}
\bibliography{literatur}
\end{document}